\documentclass[jcp,onecolumn,floatfix]{revtex4-1}

\usepackage{graphics} 
\usepackage{graphicx} 
\usepackage{color} 
\usepackage{bm} 
\usepackage[dvipsnames]{xcolor,colortbl}
\usepackage{amsmath}
\usepackage{amssymb}
\usepackage{amsfonts}

\definecolor{DarkYellow}{RGB}{80, 80, 0}

\begin{document} 
\title{The asymmetric Wigner bilayer}

\author{Moritz Antlanger$^{1, 2}$} 
\author{Gerhard Kahl$^1$} 
\author{Martial Mazars$^{2}$} 
\author{Ladislav \v{S}amaj$^{3}$} 
\author{Emmanuel Trizac$^4$} 

\affiliation{$^1$Institute for Theoretical Physics and Center for
  Computational Materials Science (CMS), TU Wien, Austria
  \\ $^2$Laboratoire de Physique Th\'eorique (UMR 8627),
  Universit{\'e} Paris-Sud, Universit\'e Paris-Saclay, and CNRS,
  Orsay, France \\ $^3$Institute of Physics, Slovak Academy of
  Sciences, Bratislava, Slovakia \\ $^4$LPTMS, CNRS, Univ. Paris-Sud,
  Universit\'e Paris-Saclay, Orsay, France}


\begin{abstract}
We present a comprehensive discussion of the so-called asymmetric
Wigner bilayer system, where mobile point charges, all of the same
sign, are immersed into the space left between two parallel,
homogeneously charged plates (with possibly different charge
densities). At vanishing temperatures, the particles are expelled from
the slab interior; they necessarily stick to one of the two plates,
and form there ordered sublattices. Using complementary tools
(analytic and numerical) we study systematically the self-assembly of
the point charges into ordered ground state configurations as the
inter-layer separation and the asymmetry in the charge densities are
varied. The overwhelming plethora of emerging Wigner bilayer ground
states can be understood in terms of the competition of two strategies
of the system: the desire to guarantee net charge neutrality on each
of the plates and the effort of the particles to self-organize into
commensurate sublattices. The emerging structures range from simple,
highly commensurate (and thus very stable) lattices (such as staggered
structures, built up by simple motives) to structures with a
complicated internal structure. The combined application of our two
approaches (whose results agree within remarkable accuracy) allows to
study on a quantitative level phenomena such as over- and
underpopulation of the plates by the mobile particles, the nature of
phase transitions between the emerging phases (which pertain to two
different universality classes), and the physical laws that govern the
long-range behaviour of the forces acting between the plates.
Extensive, complementary Monte Carlo simulations in
the canonical ensemble, which have been carried out at small, but
finite temperatures along selected, well-defined pathways in parameter
space confirm the analytical and numerical predictions within high
accuracy. The simple setup of the Wigner bilayer system offers an
attractive possibility to study and to control complex scenarios and
strategies of colloidal self-assembly, via the variation of two simple
system parameters. 
\end{abstract}

\date{\today}

\maketitle


\section{Introduction}
\label{sec:introduction}

In the 1930s, Eugene P. Wigner put forward the claim
\cite{Wigner:1934} that the (ordered) ground state configurations of
electrons in a metal are ``close packed lattice configurations'',
forming thereby a so-called Wigner crystal. Actually, such
configurations were -- at least so far -- never observed in
experiment: neither in a metal nor in any three-dimensional
system. Instead, the corresponding ordered configurations were
identified in two-dimensional systems where the Wigner crystal reduces
to a hexagonal monolayer lattice. Electrons which form at a He
interface a hexagonal lattice \cite{Grimes:1979} were presumably (and
more than 40 years after Wigner's claim) the first realization of a
two-dimensional Wigner crystal. Later on, two-dimensional Wigner
crystals were realized in semi-conductor hetero-structures
\cite{TsSG82,EiMD04,WZYE07,Wang:2012,ZHDK14}, graphene \cite{AbCh09},
or in quantum dots, trapped ionic plasmas {and other dusty
plasmas \cite{MoIv09}. Also Wigner crystals were reported to be
experimentally observed in colloidal systems \cite{Pert01}. A few
studies were dedicated to laterally confined two-dimensional systems
of charged particles, investigating if such systems crystallize at
sufficiently low temperatures also into Wigner crystals
\cite{Bedanov:1994,Lozovik:1987,Lozovik:1990,Lozovik:1990a,Lozovik:1992,Bolton:1993},
see also \cite{BaHa80}. Other highly ordered trapped ionic systems
have been studied with a distinct quantum computing perspective
\cite{Mart2016,Blatt2012}.


The extension of the two-dimensional monolayer problem to the
so-called symmetric bilayer Wigner problem was studied ever since the
1990s; it is now well understood
\cite{GoPe96,Weis:01,LoNe07,OgML09,Samaj12_1,Samaj12_2}. Classical
point charges confined between two parallel, oppositely charged plates
(both of them characterized by the same charge density) that are
separated by a distance $d$, self-assemble in five archetypical
structures, termed \textrm{I} to \textrm{V}; as they are throughout
staggered lattices of simple structural motives (such as triangles,
rectangles, squares, or rhombs), the sublattices formed on each of the
layers are commensurate and are -- in addition -- locally charge
neutral (with the plate charge compensated by those of the point
ions). While the results were initially quite controversial, a
quasi-exact analytic approach put forward by two of the authors
\cite{Samaj12_1,Samaj12_2} provided the following results for this
numerically delicate problem: (i) phase \textrm{I} is stable only for
$d = 0$; (ii) exact $d$-values where the transitions between adjacent
phases take place and the order of the respective phase transitions
could be specified. It should be mentioned that these structural
motives were identified in a number of experiments (see, e.g.,
\cite{Mitchell:98,Winkle:86,Eisenstein:04}).

In this contribution, we report about the natural generalization of
the aforementioned symmetric Wigner double-layer problem to the {\it
  asymmetric} case, i.e., when the two plates (with indices 1 and 2),
which are separated by a distance $d$, can carry different charge
densities ($\sigma_1$ and $\sigma_2$) \cite{rque10}. From an
experimental point of view, one can consider the parallel plates as
the surfaces of two sufficiently large colloidal particles, which are
separated by a minute distance; in the space left between these
particles, oppositely charged (therefore all of the same sign)
microscopic point charges are immersed. As in the symmetric case,
Earnshaw's theorem \cite{Earnshaw} constrains the energy-minimizing
configuration: the charges have to be located on either of the
plates. The interplate distance $d$ (which, for convenience is
replaced by a reduced, dimensionless distance $\eta$) and the charge
asymmetry parameter $A$ (defined as $A = \sigma_2/\sigma_1$ with $A
\in [0, 1]$), remain as the only parameters that specify our
system. Using two complementary tools (analytical and numerical) we
identify the (ordered) ground state configurations that the charged
particles are able to form on the plates at {\it vanishing}
temperature. Additional Monte Carlo (MC) simulations, carried out at
small, but {\it finite} temperatures provide evidence about the
thermal stabilities of the predicted lattice structures. In this
contribution, we thereby demonstrate that the system is able to
self-organize -- via subtle changes in the parameters $\eta$ and $A$
-- into a rich plethora of ordered structures.

The aforementioned {\it analytic} approach is an extension of the
Coulomb lattice summation method for periodic structures, introduced
in \cite{Samaj12_1,Samaj12_2}. Lattice Coulomb summations can be
transformed into rapidly converging series representations, which can
be calculated straightforwardly up to arbitrary numerical
accuracy. This unprecedented numerical accuracy is counteracted by the
limited applicability of the formalism: its complexity rapidly
increases with that of the involved structures (either via an
increasing number of particles or via distortions of ideal
lattices). The {\it numerical} approach is a highly specialized
optimization technique which relies on ideas of evolutionary
algorithms (EA) \cite{Gol89,Gottwald:05}. Our implementation of the
EA, which is mimetic in character (i.e., it combines global and local
search techniques), relies on a heavy use of Ewald summation
techniques (see \cite{Mazars:11} and references therein); it
guarantees a substantial reduction in computational costs. Due to
numerical restrictions, unit cells with up to 40 particles have been
considered. The robustness, the efficiency, the reliability, and the
capacity of our algorithmic implementation to cope in high dimensional
search spaces in problems characterized by minute energy differences
of competing structures has been tested in numerous cases (see, for
example,
\cite{Fornleitner:08,Fornleitner:08a,Pauschenwein:08,Doppelbauer:10,Doppelbauer:12}).
These attractive features are counteracted by the fact that no
guarantee can be given that the converged values corresponds indeed to
the ``true'' ground state configuration. The numerical and analytical
approaches are complementary in the sense that they compensate
mutually for their respective shortcomings.
As will be demonstrated here, the two approaches are able to provide
together a comprehensive picture of this intricate problem within a
remarkable degree of accuracy and consistency.

Extensive MC simulations have been carried out in selected regions
(that are specified in the body of the text) of the parameter space,
i.e., in the $(\eta, A)$-plane. These simulations have been performed
in the canonical ensemble, assuming a small, but finite temperature
and thus provide information about the thermal stability of the ground
state configurations predicted by the analytic and the numerical
approaches. A standard MC technique has been used
\cite{Frenkel:01,Allen:17} (featuring flexible cell shape and trial
particle moves from one plate to the other) and suitable Ewald
summation techniques \cite{Mazars:11} guarantee for efficient
simulations; ensembles typically contain $\sim$ 4000 particles.

In the numerical approach and in the simulations, the classification
of the emerging structures has been realized via suitably defined bond
orientational order parameters \cite{Steinhardt:83} and the occupation
index $x$ to be defined below.
The overwhelming complexity of the emerging diagram of states can be
understood in terms of the competition of two disparate strategies of
the system which cannot be reconciled in the asymmetric case: (i)
maintaining charge neutrality on each of the plates and (ii)
self-organizing into commensurate sublattices on the two plates. In
the {\it symmetric} case, these two principles are compatible, leading
to the five above mentioned archetypical structures: these are rather
simple, staggered (and thus commensurate) lattices, based on
triangles, square, rectangles, or rhombs. However, as soon as charge
asymmetry sets in (i.e., as soon as $A < 1$), the situation is
different: the system is not always able to guarantee both charge
neutrality {\it and} commensurability of the sublayers at the same
time.

In the symmetric case the hexagonal monolayer was stable only at $\eta
= 0$; in the asymmetric case, this phase \textrm{I} is stable in a
rather large portion of parameter space and represents the origin of
all bilayer configurations: they emerge from the monolayer as one
particle moves from layer 1 to layer 2, creating thereby the bilayer
structures \textrm{I}$_x$ (for intermediate and large $A$-values and
rather small $\eta$'s) and \textrm{V}$_x$ (for intermediate $A$'s and
rather large $\eta$-values); both transitions (\textrm{I}$_x$ $\to$
\textrm{I}$_x$ and \textrm{I} $\to$ \textrm{V}$_x$) are of second
order, characterized by a non-conventional set of critical
exponents. Starting off from the structures \textrm{I}$_x$ and
\textrm{V}$_x$, a rich plethora of ordered bilayer ground state
configurations emerges: the spectrum ranges from highly stable
structures (with strongly correlated sublattices on the layers and a
small number of particles per unit cell) to essentially uncorrelated
hexagonal sublattices at large $\eta$-values, covering thereby at
intermediate $\eta$-values highly complex structures, that carry
features of five-fold symmetry. Similar as in the symmetric case, the
identification of the ordered ground state configurations turned out
to be a particularly tricky task, as competing structures were
characterized by minute differences in energies; the complementarity
of the analytic and of numerical approaches proved valuable in this
analysis.

Violation of local charge neutrality was observed.  For the majority
of the state points and keeping in mind that we took the surface
charges $\sigma_1$, $\sigma_2$ positive while the ions bear a negative
charge, we encounter a phenomenon that we have termed
``undercharging'': layer 2 (which carries the smallest charge) carries
a net positive charge, i.e., this layer is -- as compared to its
charge density -- ``underpopulated'' by charges; only for $A$-values
close to unity the inverse effect (i.e., the ``overcharging''
phenomenon) is observed.
Finally we point out that the transitions between the structures
\textrm{I} to \textrm{V}, which occur for $A \gtrsim 0.9$, namely the
transitions $\textrm{I\kern -0.3ex I}\to\textrm{I\kern -0.3ex I\kern
  -0.3ex I}$ and $\textrm{I\kern -0.3ex I\kern -0.3ex
  I}\to\textrm{I\kern -0.3ex V}$ are of second order, now being
characterized by mean-field critical exponents.  Thus the system shows
a remarkable critical behaviour, with two second-order phase
transitions pertaining to different universality classes.

The rather extensive MC simulations confirm with remarkable accuracy
the theoretical predictions (i.e., structural features, regions of
stability of the different phases, etc.). 
Yet, open and still unanswered issues remain. One of the most
pertinent ones is the question of the system's ability to form
non-periodic, but ordered ground state configurations (as they are,
for instance, observed in quasi-crystalline particle arrangements) or
disordered structures (as they are, for instance found in systems
interacting via soft, bounded -- in this context termed ``stealthy''
-- interactions; see, for instance \cite{Zhang:15}). The former case
is not unlikely to occur, as the snub square particle arrangement or
ordered structures with features of five-fold symmetry can be
considered as precursors of quasi-crystalline lattices.
Finally, one should also investigate the phase diagram of the system
as we proceed to higher temperatures, i.e., towards melting of the
structures identified.

In the currently wide-spread investigations of self-assembly scenarios
and self-organization strategies in colloidal systems, one can observe
a trend towards an increasing complexity in the properties of the
system and/or in the internal architecture of the colloidal particles:
shape, surface decoration, the consideration of colloidal mixtures,
affecting the solvent through various additives, or by applying
external fields and/or exposing the system to patterned surfaces are
only a few examples (see, for instance,
\cite{Gong:17,Glotzer:07,Bianchi:17,Bianchi:17a,mikhael_2008,Chen:2011,Bianchi:13,Bianchi:14},
and references therein).  Our system marks the return to a simple,
classical case: in striking contrast to the aforementioned examples,
it represents with its elementary setup a surprisingly simple
alternative to study in a systematic manner complex self-assembly
scenarios by varying only two system parameters
Wigner bilayer systems can thus be viewed as encouraging setups to
study complex self-assembly scenarios of charged particles in a
systematic manner.

The paper is organized as follows. The subsequent Section is dedicated
to the specification of our model, the summary of the methods used,
and the tools that enabled us to identify the emerging
structures. This section contains also a brief, but comprehensive
summary of the ordered phases of the symmetric Wigner bilayer problem.
In Section III, we provide a general overview over the ordered ground
state configurations as they were identified in a representative range
of the $(\eta, A)$-plane with the analytical and the numerical
approaches, while Sections IV to VI contain detailed presentations and
discussions of these structures as they emerge at small, large, and
intermediate $\eta$-values, respectively. Section VII is dedicated to
the results obtained in MC simulations, carried out at small, finite
temperature. The main text is closed with concluding
remarks. Additional and more specific information are summarized in
Appendices. Preliminary accounts of part of this work have already
been published in \cite{Ant15}.




\section{Model and methods}
\label{sec:models_methods}

\subsection{Model} 
\label{subsec:model}

We consider two parallel plates (denoted by 1 and 2), which we assume
to be arranged perpendicular to the $z$-axis and separated by a
distance $d$.  The plates are of infinite extent in the $x$- and
$y$-directions, with surfaces $S_1=S_2=S$ tending towards infinity.
Both plates bear fixed, uniform surface charge densities $e \sigma_1$
and $e \sigma_2$, respectively, with $e$ the elementary charge.  The
electrostatic potential induced by the charged plates is given by
\begin{equation} \label{elpot}
\phi(z) = - 2\pi e (\sigma_1-\sigma_2) z  + {\rm const.} \qquad 0<z<d .
\end{equation}
The space between the plates is filled by $N (\to\infty)$ classical,
mobile particles of (negative) unit charge $-e$ which are
``counter-ions'' with respect to the charged plates.  The entire
system is assumed to be electro-neutral, i.e.  $(\sigma_1 + \sigma_2)
S = N $.  The particles are immersed in a solution of dielectric
constant $\varepsilon$ which, for convenience, we put equal to unity.
Also the walls have the same dielectric constant, $\varepsilon ' = 1$;
thus no image charges have to be considered.  The surface charge
densities on the plates and the particles interact via the
three-dimensional Coulomb potential $1/r$. Our task is to find the
(zero temperature) ground state of this system, having the lowest
energy.

Without loss of generality we assume $\sigma_1$ to be positive.
Further, we introduce the asymmetry parameter
\begin{equation}
A = \frac{\sigma_2}{\sigma_1} .
\end{equation} 
As a consequence of the exchange symmetry of the plates 1 and 2, we
can reduce the relevant range of $A$ to the interval $[-1, 1]$.
Excluding further the case $A \in [-1, 0)$, where all particles are
  trivially located on plate 1, it is eventually sufficient to focus
  our investigations to $A \in [0, 1]$.  For the symmetric case, i.e.,
  $A = 1$, the emerging ground state configurations have been fully
  identified by analytical approaches \cite{Samaj12_1,Samaj12_2} and
  simulation methods \cite{GoPe96,Weis:01,messina_2003,LoNe07,OgML09}.

As in the symmetric case, it is convenient to introduce the
dimensionless ``distance''
\begin{equation}
\eta = d \sqrt{\frac{\sigma_1 + \sigma_2}{2}} .
\end{equation}
Our system is entirely defined by $\eta$ and $A$. In a potential
experimental setup, it is natural to fix the asymmetry parameter $A$
and to change continuously the dimensionless distance $\eta$ from $0$
to $\infty$.

According to Earnshaw's theorem \cite{Earnshaw}, a classical system of
point charges under the action of direct (i.e., not image)
electrostatic forces alone cannot be in an equilibrium configuration;
thus the mobile particles are forced to be located on the plate
surfaces.  Let $N_1=n_1 S$ particles stick to plate 1 (and creating a
regular lattice structure $\alpha$), and $N_2=n_2 S$ particles stick
to plate 2 (creating a sublattice $\beta$).  In general $n_i \ne
\sigma_i$ ($i = 1, 2$); under these conditions each of the plates
carry a net charge (i.e. particle charges plus surface charge).  Since
the total number of particles $N=N_1+N_2$, the overall system
electro-neutrality requirement imposes that
\begin{equation} \label{neutrality}
\sigma_1 + \sigma_2 = n_1 + n_2 .
\end{equation}

Further we introduce the particle occupation parameter of the plates
as follows
\begin{equation} \label{occup}
x = \frac{N_2}{N} = \frac{n_2}{n_1+n_2} .
\end{equation}

In case each of the plates as a whole is neutral, i.e. $n_1=\sigma_1$
and $n_2=\sigma_2$, the occupation parameter becomes
\begin{equation} \label{x_neutr}
x_{\rm neutr} \equiv x^* = \frac{A}{1+A} .
\end{equation}
Figure \ref{fig:model} provides a sketch of the setup.

As the system is entirely defined by the parameters $\eta$ and $A$, we
can grasp the full information about the structures that the system
forms for given values of $\eta$ and $A$ by a systematic variation of
these two quantities.
The following limiting cases have been discussed in literature:
\begin{itemize}
\item[(i)] for $A=1$, we recover the symmetric case, which has been
  thoroughly discussed
  \cite{GoPe96,Weis:01,messina_2003,LoNe07,OgML09,Samaj12_1,Samaj12_2}.
  Each of the plates 1 and 2 as a whole (i.e., charge of the particles
  plus surface charge) is neutral.  In the one-dimensional diagram of
  states, which depends only on $\eta$, five ordered ground state
  configurations have been identified; they are termed \textrm{I},
  \textrm{I\kern -0.3ex I}, \textrm{I\kern -0.3ex I\kern -0.3ex I},
  \textrm{I\kern -0.3ex V}, and \textrm{V} and will play a key role in
  the diagram of states of the {\it asymmetric} Wigner bilayer
  problem, discussed in the following subsection;
\item[(ii)] for $\eta=0$ the system forms, irrespective of the
  $A$-value, a hexagonal (equilateral triangle) monolayer;
\item[(iii)] we encounter the same ordered ground state configuration
  on plate 1 for the limiting case $A=0$;
\item[(iv)] finally, for $\eta \to \infty$ the charges form on each of
  the layers two ideal hexagonal lattices, which are shifted with
  respect to each other.
\end{itemize}

While the analytic (Subsection \ref{subsec:methods_analytical}) and
the numerical approaches based on relatively small sets of particles
(Subsection \ref{subsec:methods_EA}) aim at a comprehensive
identification of the ordered ground state configurations, we have
performed complementary Monte Carlo simulation at a finite, but small
temperature (see Subsection \ref{subsec:methods_MC}).  These
investigations have been carried out with the intention to test the
numerical predictions for the structural features of the system on
much larger sets of particles and to investigate the thermal stability
of the predicted configurations.

\begin{figure}[htbp]
\begin{center}
\includegraphics[width=15cm]{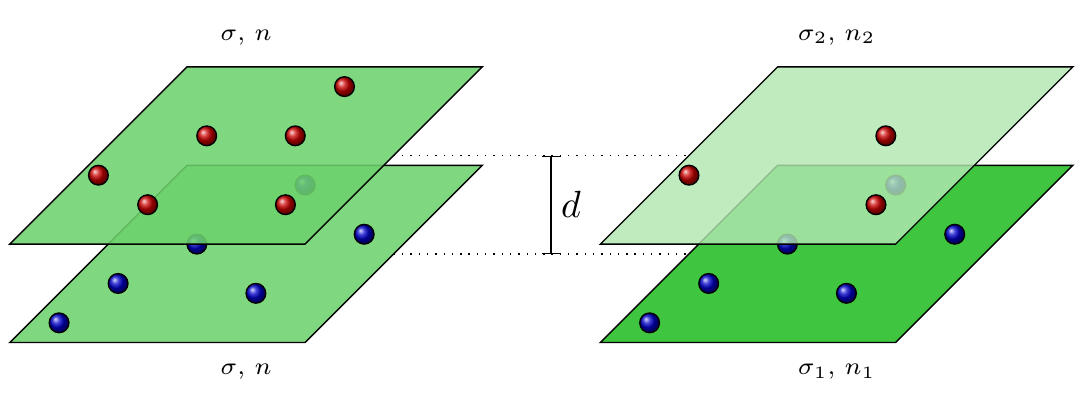}
\caption{(color online) Schematic views of the Wigner bilayer model.
  Red (blue) particles form (possibly ordered) lattices on the two
  parallel plates. Left panel: symmetric case, where both plates carry
  the same charge density (i.e., $\sigma_1 = \sigma_2$); they are
  occupied by the same number of particles and thus have both the same
  particle density (i.e., $n_1 = n_2$).  Right panel:
  asymmetric case, with $\sigma_1 > \sigma_2$ and in general, $n_1 \ne n_2$.}
\label{fig:model}
\end{center}
\end{figure}

\subsection{Symmetric case ($A = 1$): structures \textrm{I} through \textrm{V}}
\label{subsec:symmetric_case}

Before discussing the results obtained for the {\it asymmetric} Wigner
bilayers in Sections \ref{sec:results_vanishing} to
\ref{sec:intermediateeta}, we start by summarizing the results
obtained for the {\it symmetric} case
\cite{GoPe96,Weis:01,messina_2003,LoNe07,OgML09,Samaj12_1,Samaj12_2}.
Here, the availability of highly accurate data, accessible via pure
analytic calculations in Refs. \cite{Samaj12_1,Samaj12_2}, serve as a
stringent benchmark for our implementation of the numerical
Evolutionary Algorithm code.

\begin{figure}[htbp]
\begin{center}
\includegraphics[width=15cm]{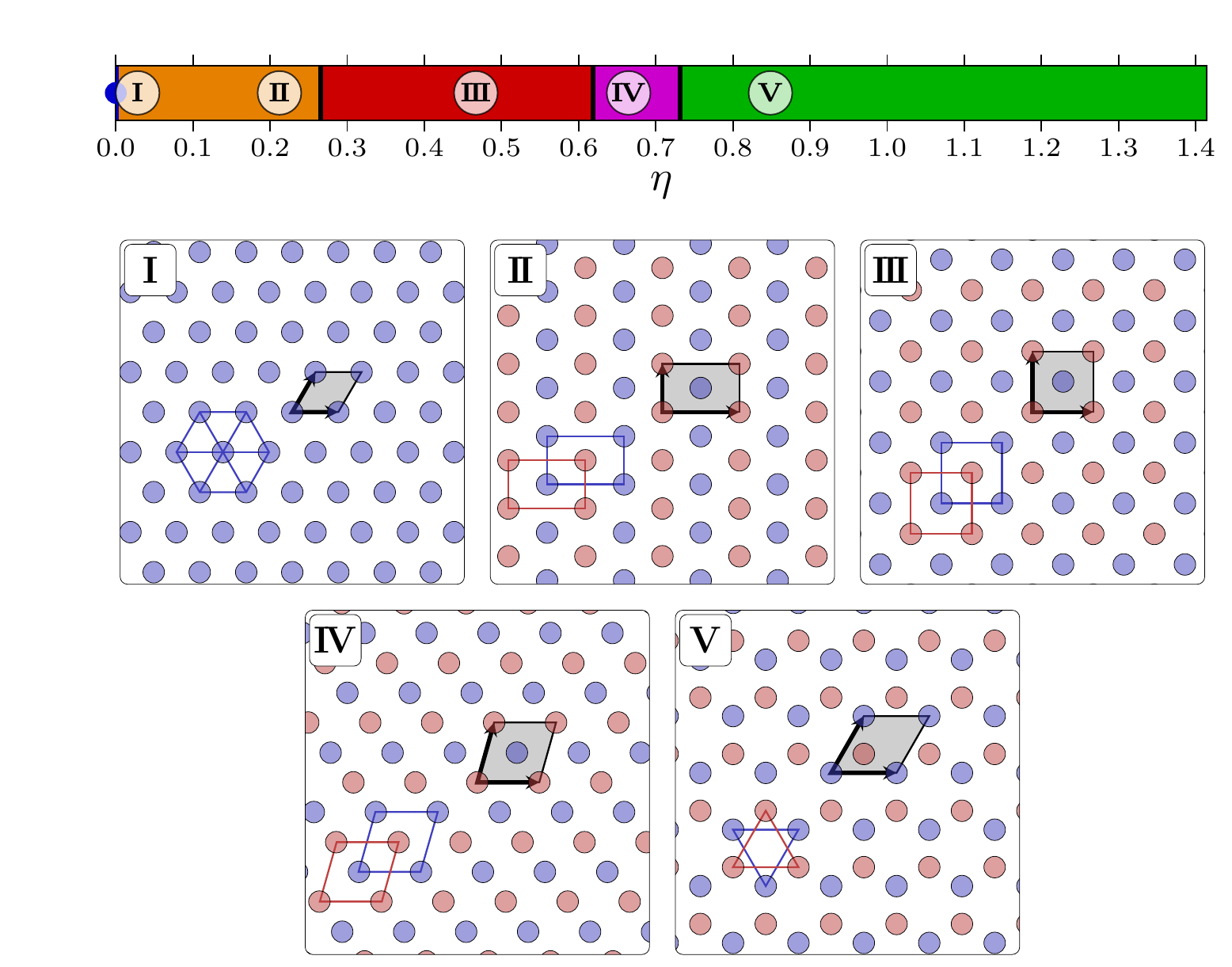}
\caption{(color online) {\it Symmetric} bilayer Wigner system ($A=1$).
  Top panel: diagram of states in terms of the emerging sequence of
  structures (\textrm{I} through \textrm{V}) and their respective
  regions of stability in terms of distance between the plates $\eta$
  (as labeled).  Structure \textrm{I} is only stable for $\eta=0$
  (indicated by a blue dot).  Bottom panels: representative snapshots
  of structures \textrm{I}, \textrm{I\kern -0.3ex I}, \textrm{I\kern
    -0.3ex I\kern -0.3ex I}, \textrm{I\kern -0.3ex V}, and \textrm{V}
  (as labeled).  Particles in layer 1 are colored blue, particles in
  layer 2 red.  The unit cell of the respective structure is indicated
  by the shaded area.  Blue and red lines highlight interesting
  structural features in layers 1 and 2, respectively.}
\label{fig:systems_wigner_symmetric}
\end{center}
\end{figure}

For $A=1$, five different structures have been predicted.  The top
panel of Figure \ref{fig:systems_wigner_symmetric} shows their
respective regions of stability.  For $\eta=0$, the hexagonal
monolayer (termed structure \textrm{I}) provides the lowest
energy. Phase \textrm{I} can also be viewed as the union of two
rectangular lattices where the aspect ratio of its edges, $\Delta$, is
given by $\Delta=\sqrt{3}$; the lattices are shifted with respect to
each other in both spatial directions by half of the respective side
lengths.

As soon as $\eta>0$, the Wigner monolayer is transformed to a
staggered rectangular bilayer, the so-called phase \textrm{I\kern
  -0.3ex I}, both rectangular sublattices having the same aspect ratio
$\Delta$. There were contentions that the value for the monolayer,
i.e., $\Delta=\sqrt{3}$, prevails in a small, but finite $\eta$-range
(see, e.g., Refs. \cite{GoPe96,Weis:01,messina_2003,OgML09}).  It was
shown in Refs. \cite{Samaj12_1,Samaj12_2} that as soon as $\eta$ is
nonzero, $\Delta<\sqrt{3}$, i.e., phase \textrm{I\kern -0.3ex I} takes
place (see corresponding panel of Figure
\ref{fig:systems_wigner_symmetric}).  This phase is stable in the
range $0< \eta \lesssim 0.263$ where $\Delta$ decreases continuously
to $1$ (corresponding to a square lattice) at $\eta \simeq 0.263$.  We
can specify structure \textrm{I\kern -0.3ex I} via the following set
of parameters: $x=1/2$, $\Psi_4^{(1,2)}=1$, and $0<\Psi_6^{(1,2)}<1$
(for the definition of the bond orientational order parameters
$\Psi_n^{(\alpha)}$ see Subsection \ref{subsec:methods_MC}).

At $\eta \simeq 0.263$, structure \textrm{I\kern -0.3ex I} transforms
via a second-order phase transition \cite{Samaj12_1,Samaj12_2} into
structure \textrm{I\kern -0.3ex I\kern -0.3ex I} (the staggered square
bilayer) which remains stable up to $\eta \simeq 0.621$ (see Figure
\ref{fig:systems_wigner_symmetric}).  Structure \textrm{I\kern -0.3ex
  I\kern -0.3ex I} can be considered a special case of the
neighbouring structures \textrm{I\kern -0.3ex I} and \textrm{I\kern
  -0.3ex V}; thus the transitions $\textrm{I\kern -0.3ex
  I}\to\textrm{I\kern -0.3ex I\kern -0.3ex I}$ and $\textrm{I\kern
  -0.3ex I\kern -0.3ex I}\to\textrm{I\kern -0.3ex V}$ are of second
order.  The critical exponents are of mean-field type
\cite{Samaj12_1,Samaj12_2}, in particular the index $\beta$, which is
related to the order parameter takes the mean-field classical value
$1/2$.  We can define structure \textrm{I\kern -0.3ex I\kern -0.3ex I}
using the set of parameters: $x=1/2$, $\Psi_4^{(1,2)}=1$ and
$\Psi_6^{(1,2)}=0$.

For $0.621 \lesssim \eta \lesssim 0.732$, we observe structure
\textrm{I\kern -0.3ex V} (see Figure
\ref{fig:systems_wigner_symmetric}), which is a staggered {\it
  rhombic} bilayer.  Particles in layer 2 are positioned above the
centers of the rhombs in layer 1, and {\it vice versa}.  The
deformation angle $\varphi$ of the rhombs decreases from $\pi/2$
(corresponding to a square and thus to structure \textrm{I\kern -0.3ex
  I\kern -0.3ex I}) at $\eta \simeq 0.621$ to a value of $\varphi
\simeq 0.386 \pi$ at $\eta \simeq 0.732$.  We can define structure
\textrm{I\kern -0.3ex V} using the following set of parameters:
$x=1/2$, $0<\Psi_4^{(1,2)}<1$ and $0<\Psi_6^{(1,2)}<1$.

Finally, for $0.732 \lesssim \eta$, we observe structure \textrm{V}
(see the corresponding panel of Figure
\ref{fig:systems_wigner_symmetric}) which is a staggered {\it
  hexagonal} bilayer.  Particles in layer 2 are positioned above the
centers of equilateral triangles in layer 1, and {\it vice versa}.
The transition between structures \textrm{I\kern -0.3ex V} and
\textrm{V} is of first-order as there is a jump in the deformation
angle $\varphi$; simultaneously, particles in each layer move from the
center of a rhomb to the projected center of a triangle of the other
layer.  Since particles form on both layers hexagonal lattices,
structure \textrm{V} covers also the asymptotic $\eta \to \infty$
case.  We can define structure \textrm{V} using the set of parameters:
$x=1/2$, $\Psi_4^{(1,2)}=0$ and $\Psi_6^{(1,2)}=1$.

\subsection{Analytical computations} 
\label{subsec:methods_analytical}

We establish in Appendix \ref{app:A} a connection between the
Coulombic energies of systems having the same point-charge
configuration on the plates, but otherwise arbitrary surface charges
$\sigma_1$ and $\sigma_2$. Of course, the electroneutrality constraint
should be enforced ($\sigma_1+\sigma_2=n_1+n_2$).  The resulting
expression, Eq. \eqref{crucial}, will prove useful in Subsection
\ref{subsec:methods_EA}. We also provide here relevant information on
the analytical method used to work out Coulombic energies.  It follows
the periodic lattice summation idea introduced for periodic structures
in Refs. \cite{Samaj12_1,Samaj12_2}. The starting point is the
$\Gamma$-identity for the $1/r$ potential
\begin{equation}
\frac{1}{\vert {\bf r}\vert} \equiv \frac{1}{\sqrt{r^2}} =
\frac{1}{\sqrt{\pi}} \int_0^{\infty} \frac{dt}{\sqrt{t}} e^{-r^2t}
\end{equation}
which enables one to transform a lattice Coulomb summation into an
integral over the products of two Jacobi theta functions with zero
argument, namely
\begin{equation} \label{theta}
\theta_2(q) = \sum_{j=-\infty}^{\infty} q^{(j-\frac{1}{2})^2} , \qquad
\theta_3(q) = \sum_{j=-\infty}^{\infty} q^{j^2} , \qquad
\theta_4(q) = \sum_{j=-\infty}^{\infty} (-1)^j q^{j^2} .
\end{equation}
The neutralizing background subtracts the $q\to 1$ singularities of
the product of theta functions.  Using a sequence of integral
transformations combined with the Poisson summation formula
\begin{equation} \label{PSF}
\sum_{j=-\infty}^{\infty} {\rm e}^{-(j+\phi)^2 t} =
\sqrt{\frac{\pi}{t}} \sum_{j=-\infty}^{\infty} {\rm e}^{2\pi{\rm
    i}j\phi} {\rm e}^{-(\pi j)^2/t}
\end{equation}
and specific properties of the Jacobi theta functions, the expression for 
the Coulomb lattice sum can be converted into a quickly converging series of 
special functions
\begin{equation} \label{specialf}
z_{\nu}(x,y) = \int_0^{1/\pi} \frac{{\rm d}t}{t^{\nu}} {\rm e}^{-xt} {\rm e}^{-y/t} ,
\qquad y>0 ,
\end{equation}
which are generalizations of the so-called Misra functions
\cite{Misra}.  In numerical calculations,
the truncation of the generalized Misra series at the
fourth term ensures an accuracy of the energy calculations for
approximately 17 significant decimal digits.

Near a critical point, the Misra functions can be expanded in powers
of the corresponding order parameter; in this way one derives an exact
Landau form of the ground state energy.  The critical point can thus
be specified up to an arbitrary accuracy as a nullity condition for a
coefficient and the critical exponents (usually of mean-field type)
can be determined. Thus, the above Jacobi-Misra reformulation is not
only useful for computing numerically energies, but also to obtain
explicit analytical results.

In real lattice structures with particles on both plates, there exist
vacancies due to a particle skip from one plate to the other.  They
cause local deformations of ideal structures which are negligible if
the plates are close to one another, but can be considerable at large
distances between the plates.  In the analytical approach, we ignore
these local deformations and consider instead of real structures their
idealized simplifications with a reasonable number of particles per
unit cell.  It is worthwhile to point out that this neglect leads to
only small differences in comparison with numerical approaches which
deal with realistic, deformed structures.

The analytical approach works well also in special regions of the
$(\eta,A)$-plane where the numerical methods fail.  A typical example
is the region of large distances $\eta$ where the interlayer energy is
too small to be detected numerically, while the analytical treatment
is able to predict the asymptotic form of the energy and the
asymptotic behavior of the occupation parameter.

\subsection{Evolutionary Algorithms (EAs)} 
\label{subsec:methods_EA}

To identify the ordered ground state configurations of
our system, we use an optimization tool based on ideas of Evolutionary
Algorithms (EAs) \cite{Gottwald:05}. EAs are heuristic approaches to
search for global minima in high dimensional spaces
\cite{Gol89} that are characterized by rugged energy landscapes.
We introduce a unit cell which creates (together with
its periodic images) a system of infinite extent. The periodic
boundary conditions are in compliance with the Ewald summation
technique (see Appendix \ref{appendix:ewald}). Inside this cell,
the particles are located in such a way as to minimize the energy of
the system, which is a lattice sum.

We initialize the algorithm by creating a set of random particle
arrangements. These configurations are graded by their fitness value,
a quantity that provides information on how suitable this
configuration is to solve the optimization problem. Since we are
interested in finding ground state structures, a high fitness value of
a particular configuration corresponds to a low value of the energy
per particle.  We then iteratively use existing configurations to
create new ones by applying alternatively one of two operations:
crossover and mutation. In the former one we first select two
configurations where this choice is biased by high fitness values of
the two configurations. Traits of both particle arrangements (such as
lattice vectors and/or particle positions) are then combined to form a
new configuration. The mutation operation, on the other hand,
introduces random changes to a randomly chosen configuration, such as
moving an arbitrarily chosen particle or distorting the lattice by
changing the underlying vectors. Typically 2000 iterations are
required for a particular state point until proper convergence towards
the minimum has been achieved.

Our implementation of EAs is memetic, i.e., we combine global and
local search techniques: every time a new configuration has been
created with one of the two above mentioned EA operations, we apply
the L-BFGS-B \cite{Byrd:95} algorithm which guides us to the nearest
local minimum. As all configurations obtained in this way are local
minima, our implementation is similar to basin-hopping techniques
\cite{Wales:97}.

So far, the method has been applied to a broad variety of systems
\cite{Fornleitner:08,Fornleitner:08a,Pauschenwein:08,Doppelbauer:10,Doppelbauer:12}
where it has been demonstrated that the concept is able to deal
successfully with strongly rugged energy surfaces in high dimensional
search spaces. The current application of EAs represents the so far
most challenging one, as competing structures are characterized by
extremely small energy differences.

We consider unit cells whose size ranges between one and 40 particles,
the latter value being imposed by computational limitations. In an
effort to find the optimized particle configuration we proceed as
follows:

\begin{itemize}
\item[(i)] We do not allow particles to move from one layer to the
  other and consider all possible values of $x ~ (\le 0.5)$ that are
  compatible with the number of particles per cell; according to our
  experience, this strategy improves the convergence speed when
  sampling the search space.
\item[(ii)] We then fix $A = 0$ and perform computations for 201
  evenly-spaced values of $\eta \in [0,\sqrt{2}]$; this range in $A$
  and $\eta$ covers the most essential features of our system. We thus
  obtain the optimized energy-values $E(\eta,A=0;x)$.
\item[(iii)] We then proceed to $A > 0$ and vary this quantity on a
  grid of 201 evenly-spaced values of $A \in [0,1]$. The optimized
  energy for these configurations $E(\eta,A;x)$ is then obtained by
  exploiting the $A$-dependence specified in Eq. (\ref{crucial}).  The
  same result is obtained by exploiting the $A$-dependence of the last
  two terms in Eq. (\ref{U_2D_e}) of Appendix \ref{appendix:ewald}.
\end{itemize}

For given distance $\eta$ and asymmetry parameter $A$, $E(\eta,A;x)$
is minimized over occupation ratio $x$.  For a closer investigation of
certain transitions between minima, we employ a related Energy
Minimization (EM) approach: here we construct starting configurations
suggested by the analytical approach (see Subsection
\ref{subsec:methods_analytical}) and then locally optimize the
particle positions using the L-BFGS-B \cite{Byrd:95} algorithm. This
strategy allows us to study specific problems on a considerably finer
grid in phase space and to increase, concomitantly, the size of the
unit cell to up to 101 particles.

\subsection{Bond orientational order parameters} 
\label{subsec:methods_BOOP}

The overall structure and the local particle arrangements realized on
each plates are quantified via different types of bond orientational
order parameters (BOOPs) \cite{Steinhardt:83,Mazars:08}.  Here, the
neighbors of a tagged particle (carrying index $i$) that populate the
same layer are identified via a Voronoi construction
\cite{Voronoi:Book}; the number of nearest neighbors of particle $i$
is denoted by $N_{i}$. Some examples for Voronoi constructions for
selected configurations obtained in MC simulations will be shown
later.

For the data originating from MC simulations, the average values of
BOOPs (i.e., averaged along the MC run) are defined by
\begin{equation} \label{BOOP}
\langle \Psi_n^{(\alpha)} \rangle = 
\frac{1}{N_{L}} \Big <
\mbox{\LARGE$\mid$} \sum_{i\in {L}}\frac{1}{N_{i}}
\sum_{j=1}^{N_{i}}W_{ij}\exp(in\theta_{ij}) \mbox{\LARGE$\mid$}
\Big>;
\end{equation}
the tagged particle (with index $i$) is taken from a layer (or from
layers) $L$ (as specified via the index $\alpha$ -- see below), which
hosts in total $N_{L}$ particles; $\theta_{ij}$ is the angle enclosed
by the projection of the interparticle vector ${\bf r}_{ij}$ onto one
of the planes and an arbitrary, but fixed direction, and $W_{ij}$ is a
weight introduced in \cite{Mickel:13} used to appreciate correctly the
length of the sides of the Voronoi cells of a given particle $i$; to
be more specific, $W_{ij}$ is computed via
\begin{equation} \label{W-BOOP}
W_{ij} = \frac{l_j}{\sum_{k=1}^{N_{i}} l_k}
\end{equation}
where $l_j$ the length of the side of the Voronoi cell that separates
particle $i$ from its neighbor $j$. The index $n$, appearing in the
definition of the $\langle \Psi_n^{(\alpha)} \rangle$ is an integer:
we have computed BOOPs for $n=4, 5, 6, 7, 8, 10, 12, 18$ and 24 both
in EA and MC calculations. Finally, the superscript $\alpha$ refers to
the four different methods of Voronoi construction that we have used
for calculating the BOOPs: for layer 1 ($\alpha = 1$), for layer 2
($\alpha = 2$), or for all particles after projecting them onto the
same plane ($\alpha = 3$); in addition, we have also calculated
modified BOOPs ($\alpha = 4$), which quantify the geometry of
``holes'', i.e., of particles in layer 2 and the surrounding particles
in layer 1.

The Voronoi constructions allows to estimate the (averaged)
distribution of the number of neighbors for particles in each layer
\cite{Leipold:15}; we denote the probability (as calculated from the
MC simulations) that a particle has $n$ neighbours in layer $\alpha$
by $p_{\alpha}(n)$.

\subsection{Monte Carlo simulations} 
\label{subsec:methods_MC}

In the calculations based on the analytical approach and on the EA,
the exploration of the diagram of states in the $(\eta,A)$-plane is
limited to a rather small number of particles within the primitive
cells (i.e., to $N \leq 40$). However, some of the EA based
calculations have revealed that crystal phases with a rather large
number of particles per primitive cell can exist. To provide an
estimate of the stability of the ordered structures predicted by the
EA investigations, we have performed Monte Carlo (MC) simulations at
finite, but small temperatures and for considerably larger systems
(typically $N \sim 4000$).  These simulations are carried out in the
canonical ensemble, assuming a variable shape of the simulation box
${\cal S}_0$ (but assuming a fixed surface area $S$). Trial moves for
the shape of the box in combination with the Ewald method
\cite{Mazars:11} are documented in Ref. \cite{Weis:01}; this method is
particularly well suited to study solid-solid and solid-liquid
transitions and has been successfully applied for the study of the
crystal phases of Coulomb \cite{Weis:01} and Yukawa bilayers
\cite{Mazars:08}.

For $\eta=0$, our system is equivalent to a one-component plasma
confined to a plane (OCP-2D); for this system the ground state is a
triangular lattice (corresponding to our structure \textrm{I}). The
only relevant thermodynamic variable that characterizes the OCP-2D
system is the coupling constant $\Gamma$, defined via $\Gamma
=e^2\sqrt{\pi(\sigma_1+\sigma_2)}/(k_{\rm B} T)$, $k_{\rm B}$ being
the Boltzmann constant. Melting of structure \textrm{I} of the OCP-2D
system occurs at $\Gamma \simeq 140$ \cite{Mazars:15}. In an effort to
remain as close to the ground state of the bilayer as possible, we
have chosen in all MC simulations of the present study the temperature
such that $1500 \lesssim \Gamma \lesssim 2200$.

We define a MC-cycle as $N$ trial moves of randomly chosen particles
and a trial change of the shape of the simulation box. A trial move of
a particle is realized either as spatial displacement within the layer
the particle belongs to (in 90 - 97 percent of the cases) or as a
trial move of this particle from one layer to the other (in the
remaining 3 - 10 percent of the cases). Equilibration is realized
during $0.3-1.6 ~ 10^6$ MC-cycles; subsequently ensemble averages are
taken over $0.3-1.0 ~ 10^6$ MC-cycles \cite{rque20}.

In a first set of simulations we have used as initial configurations
those particle arrangements that have either been identified in
preceding EA runs, or ordered structures found for the symmetric
bilayer ($A=1$), or random particle configurations. However, since for
the first case the number of particles per primitive cell, $N$, can
differ substantially between two neighboring state points, it is
difficult to observe transitions between two ordered structures in MC
simulations when some fixed value of $N$ is assumed {\it a priori}.
To overcome this problem, we have considered specific sets of systems
for which the ordered structures are throughout compatible with the
number of particles used in the MC simulations: to this end we have
performed simulations for states that populate domains in the
($\eta,A$)-plane where the value of $x = N_2/N$ is essentially
constant. In Figure \ref{fig:runs-MC}, we highlight a few of these
domains as they are predicted via the EA-based approach.  They are
characterized by a fixed rational value of $x$, the largest of these
regions are found to be those characterized by $x = $ 3/7, 1/3, 1/4
and 1/5.

Since the ordered structures that populate the ($x=1/2$)-domain are
identical to those that have been identified for the symmetric bilayer
(cf. discussion in Subsection \ref{subsec:symmetric_case}), we have
focused in our MC simulations on domains specified by $x < 1/2$; to be
more specific, we discuss in Section \ref{sec:results_finite} and in
the Appendix \ref{appendix:MC} results obtained for four selected
$x$-values. In an effort to explore these regions systematically, we
have defined for each of them in an empiric manner simple polynomial
curves, $A_{x = {\rm const.}}(\eta)$, which define within numerical
accuracy pathways through these domains; the expressions for these
polynomials are collected for the different domains in Appendix
\ref{appendix:MC}.

The state points that have been investigated with MC simulations along
these curves are marked by symbols in Figure \ref{fig:runs-MC}. For
each of these four pathways an (ordered) initial configuration has
been chosen according to the predictions of the EA approach for this
specific state point (highlighted by a red triangle in Figure
\ref{fig:runs-MC}). This particular configuration then served as a
starting configuration for all the other states located along the
corresponding line of constant $x$.

\begin{figure}[htbp]
\begin{center}
\includegraphics[width=8.2cm]{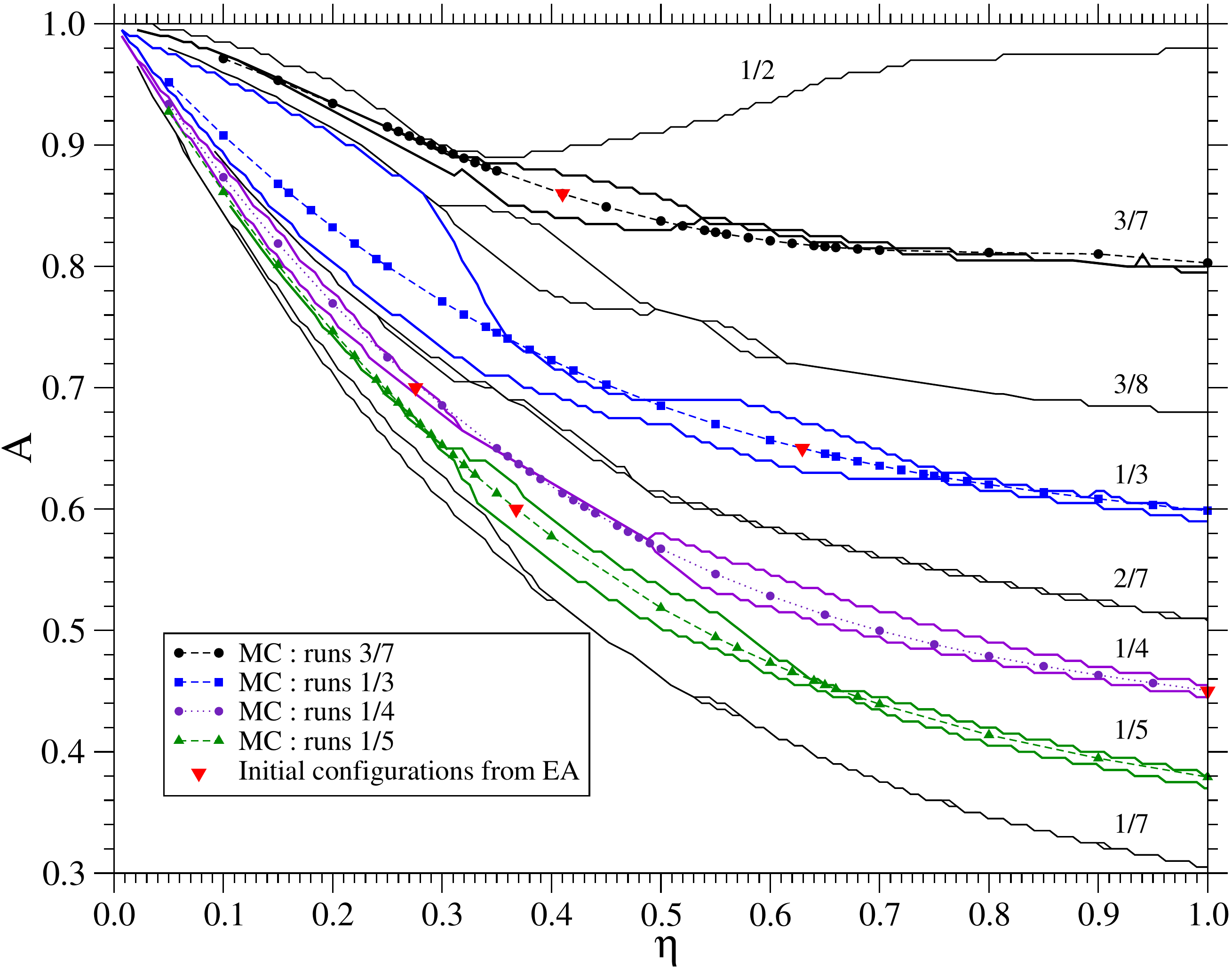}
\caption{(color online) Domains in the $(\eta,A)$-plane where --
  according to the EA predictions -- the value of $x = N_2/N$ is
  constant; regions for eight selected $x$-values are shown (as
  labeled). The dotted lines are simple polynomial fits $A = A(\eta)$
  which pass within numerical accuracy through the respective regions
  of constant $x$.  The red triangles represent on each of these
  curves those states which were used as initial configurations
  (predicted by EA calculations) of the subsequent MC runs of all the
  other states, located along these lines (marked by the
  colored dots).}
\label{fig:runs-MC}
\end{center}
\end{figure}

Additional structural information can be extracted from MC simulations
via the intra- and inter-layer pair correlation functions,
respectively defined via
\begin{eqnarray}
\label{g_of_r}
g_{\alpha}(s) & = & \frac{1}{4\pi} \frac{1}{s\sigma_\alpha (N_{\alpha}-1)}
{\Big <} \sum_{i\in L_{\alpha}} \sum_{\substack{j\in L_{\alpha} \\ j\neq i}}
\delta(s-\mid {\bf s}_{ij}\mid ) 
{\Big >} ~~~~~ \alpha = 1, 2 \\ \nonumber
g_{3}(s) & = & \frac{1}{2\pi} \frac{1}{s\sigma_1\sigma_2 S}
{\Big <} \sum_{i\in L_{1}} \sum_{j\in L_{2}} \delta(s-\mid {\bf s}_{ij}\mid)
{\Big >} .
\end{eqnarray}
Here ${\bf s}_{ij}$ represents the vector between particles $i$ and
$j$ and $N_\alpha$ ($\alpha = 1, 2$) is the number of particles in
layer $\alpha$. In an effort to capture the long-range orientational
order, we have also computed the bond orientational correlation
function for each layer $\alpha$ via
\begin{equation}
\label{Boop-gr}
G_{n,\alpha}(s) = \frac{1}{g_{\alpha}(s)} 
\Big< \Psi^{(\alpha)}_{n}(0) \Psi^{(\alpha)}_{n}(s)\Big>
~~~~~ \alpha = 1, 2, 3 ~~~~~ n~ {\rm integer.}
\end{equation}
If $\langle \Psi_n^{(\alpha)} \rangle \neq 0$, a long-range
orientational order can be identified via the bond orientational
correlation functions $G_{n,\alpha}(s)$, which then fulfill the
relation
\begin{equation}
\label{Boop-gr-lim}
\displaystyle \lim_{s \rightarrow \infty} G_{n,\alpha}(s) = 
\langle | \Psi_n^{(\alpha)} |^2 \rangle.
\end{equation}




\section{Structural informations and taxonomy} 
\label{sec:results_vanishing}

Structural informations are compiled in the different diagrams of
state of Figures \ref{fig:diagram_of_states_rgb} to
\ref{fig:diagram_of_states_x}.  Covering a representative range of the
$(\eta,A)$-plane, these figures highlight on one hand those regions
where the analytical approach predicts the stability of the emerging
structures; these areas are specified by the respective labels and are
delimited by solid curves.  On the other hand, these figures provide
on a pixel-based presentation information about the results obtained
via the EA approach; each of the $\sim$ 35000 pixels contain via a
color- or a shade-code the structural information for the respective
state point: these encoding schemes were either based on the BOOPs
(Fig. \ref{fig:diagram_of_states_rgb}), the number of particles per
unit cell (Fig. \ref{fig:diagram_of_states_n}), or the occupation
fraction $x$ (Fig. \ref{fig:diagram_of_states_x}).  In particular the
BOOPs (in combination with $x$) played a central and indispensable
role in identifying the ordered ground state configurations (see
below).  Panels of Fig. \ref{fig:diagram_of_states_rgb} are
constructed by assigning to each pixel
a color depending on the values of the BOOPs (see the caption).

\begin{figure}[htbp]
\begin{center}
\includegraphics[width=6cm]{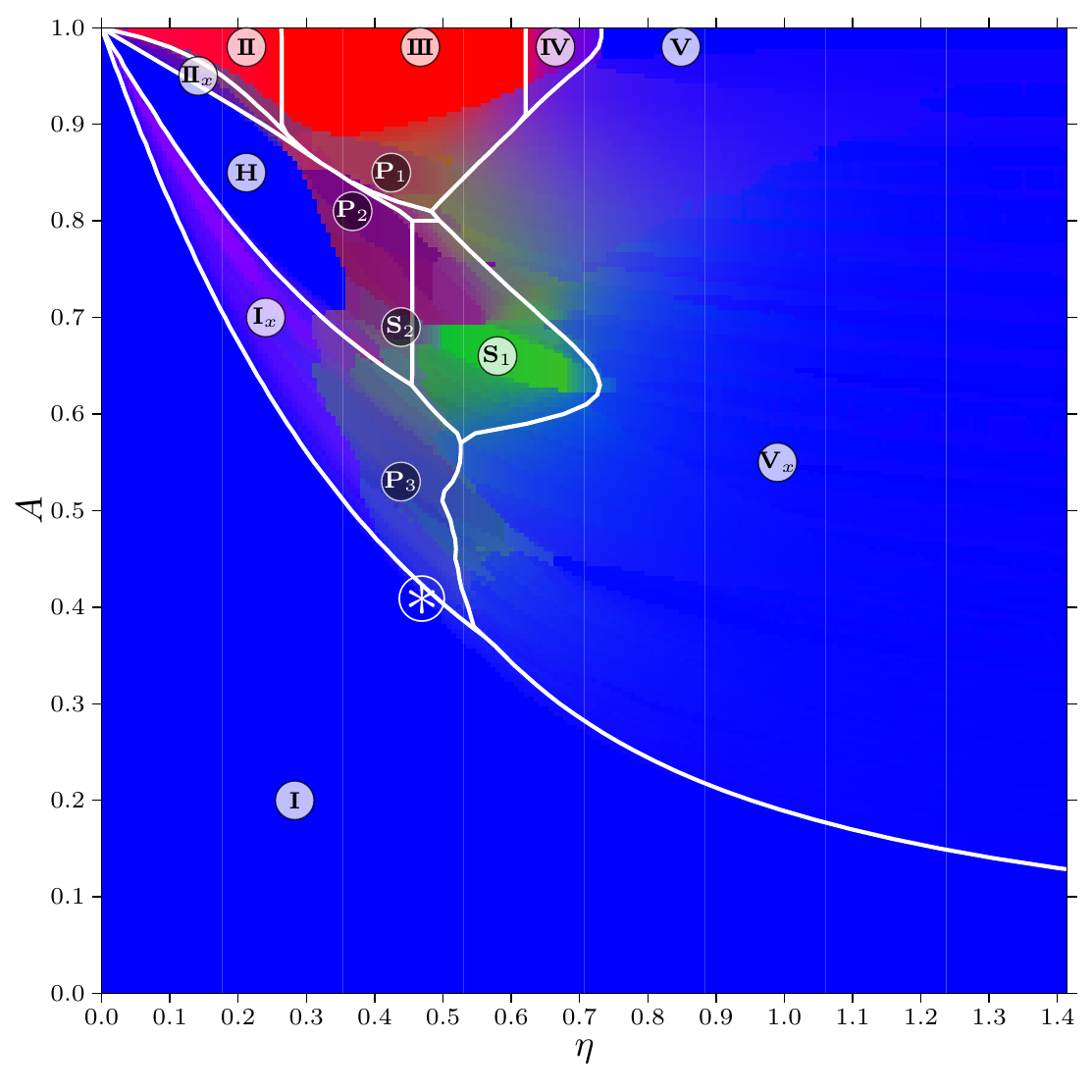}
\hspace{0.1cm}
\includegraphics[width=6cm]{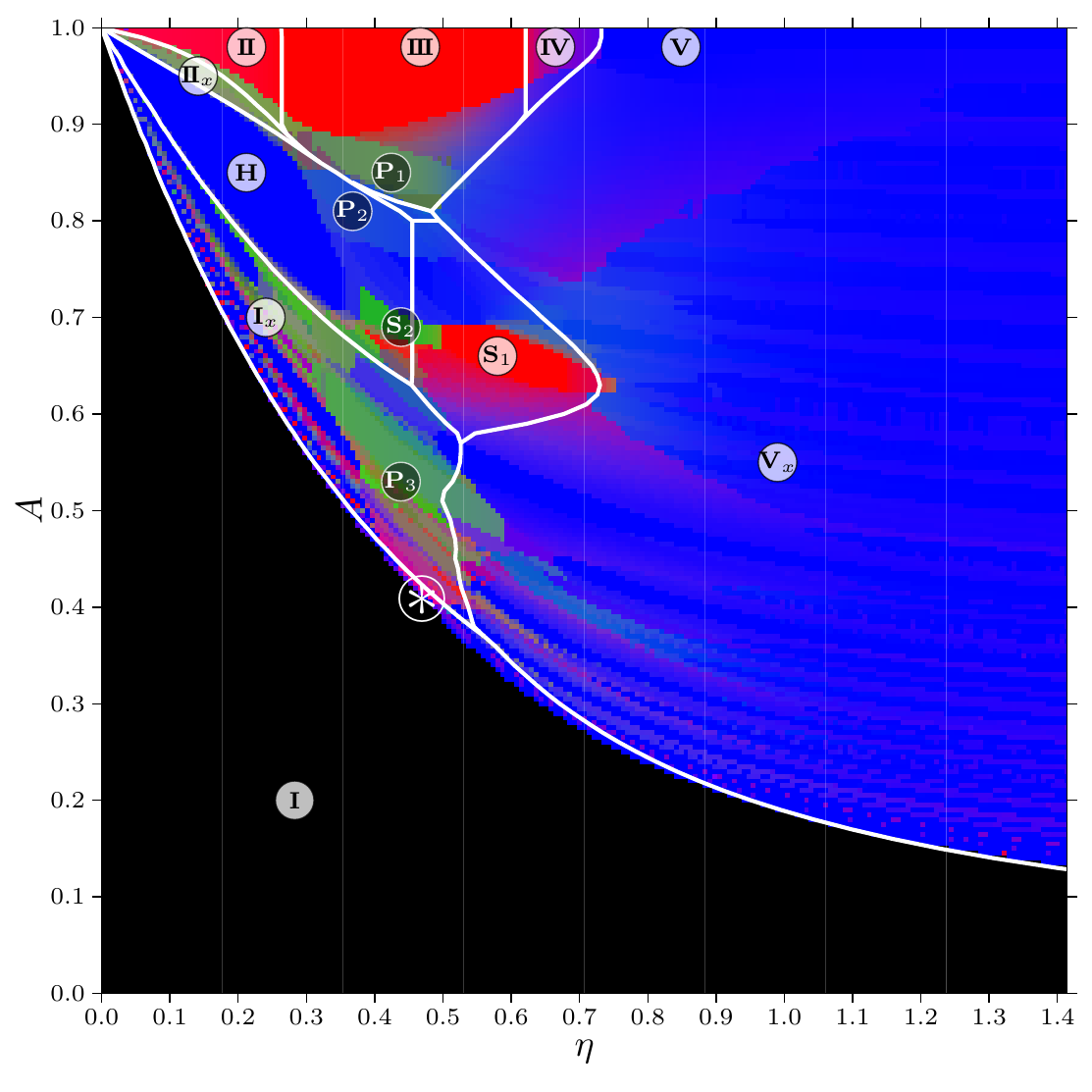}
\vspace{0.2cm}
\includegraphics[width=6cm]{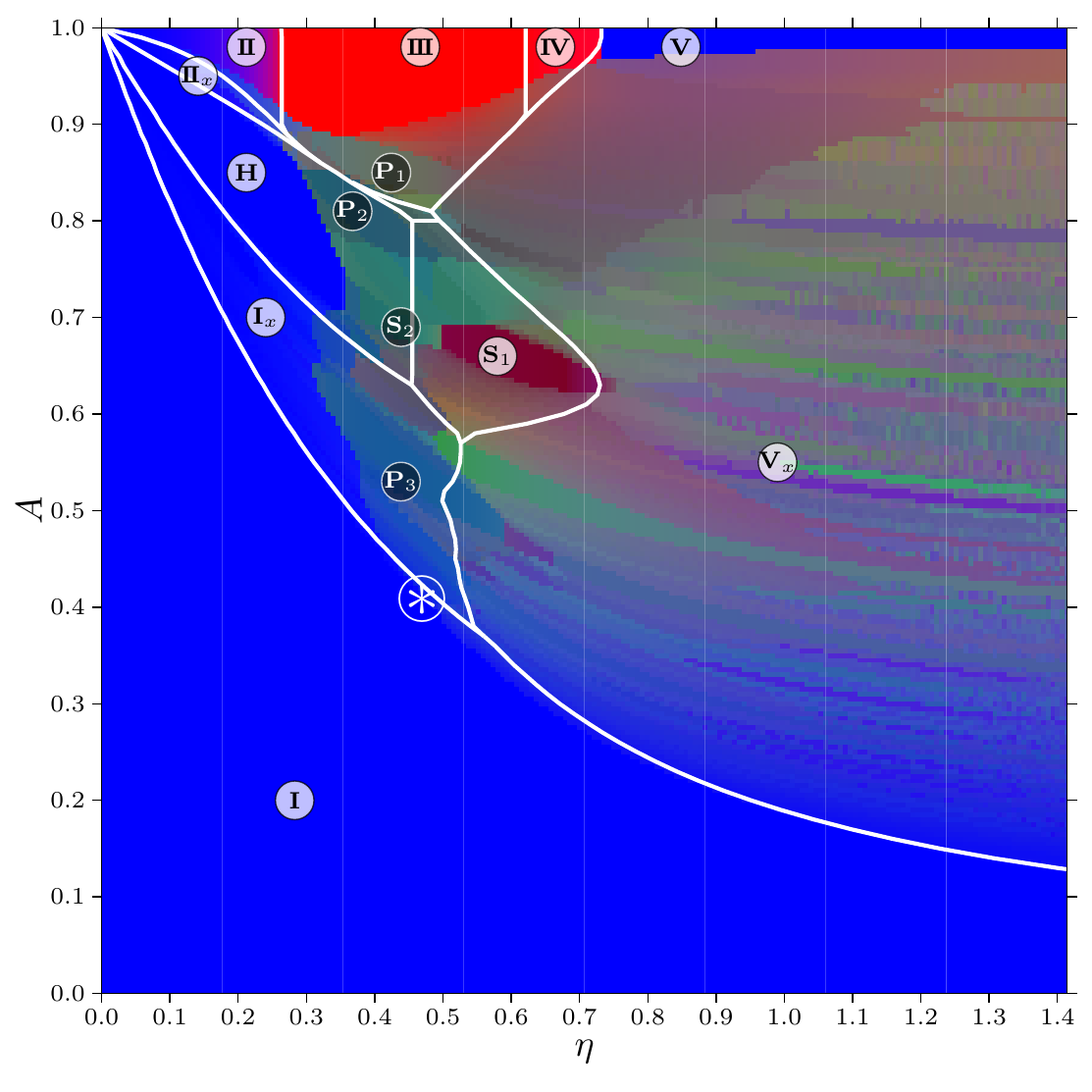}
\hspace{0.1cm}
\includegraphics[width=6cm]{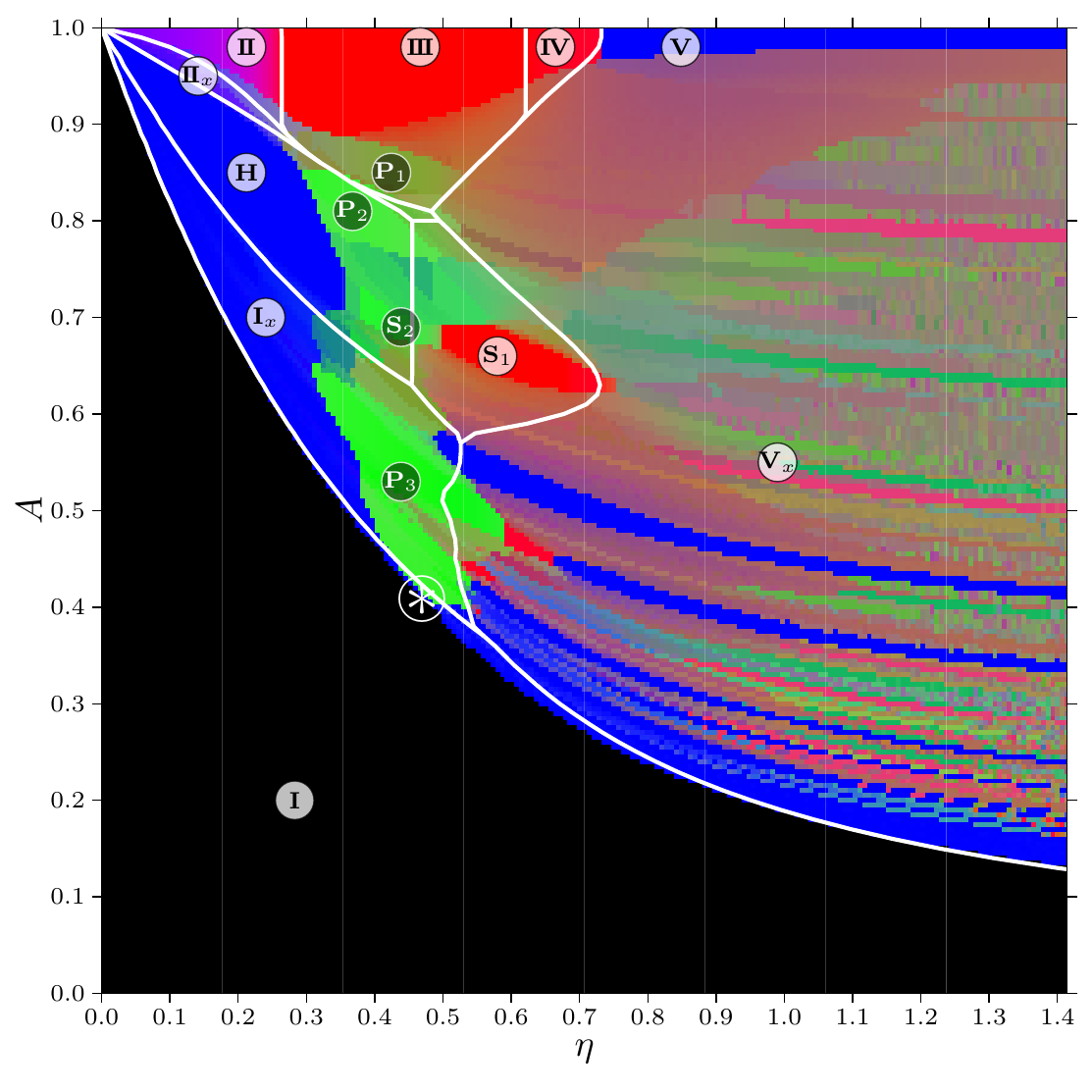}
\caption{(color online) Diagrams of states in the $(\eta,A)$-plane in
  terms of the BOOPs. Each pixel (corresponding to a state point) is
  assigned a color, which is based on the respective values of the
  specific parameters of the respective ground state configuration:
  within this $\Psi$RGB-color scheme, the value of $\langle
  \Psi_4^{(\alpha)} \rangle$ sets the red-component, the value of
  $\langle \Psi_5^{(\alpha)} \rangle$ sets the green-component, and
  the value of $\langle \Psi_6^{(\alpha)} \rangle$ sets the
  blue-component. Top left panel: order parameters $\langle \Psi^{(1)}
  \rangle$ for layer 1; top right panel: order parameters $\langle
  \Psi^{(2)} \rangle$ for layer 2. Bottom left panel: $\langle
  \Psi^{(3)} \rangle$ involving both layers; bottom right panel:
  $\langle \Psi^{(4)} \rangle$ based on the geometry of ``holes''; for
  the respective definitions of the BOOPs see Subsection
  \ref{subsec:methods_BOOP}. White lines mark regions where the
  analytical approach predicts the stability of the respective
  structure (as labeled). Particle arrangements marked by dark labels
  are too complicated to be amenable to the analytical approach. The
  white star marks a bi-critical point (see text).}
\label{fig:diagram_of_states_rgb}
\end{center}
\end{figure}

In our investigations, the numerical and the analytical approaches are
complementary in the following sense: (i) the EA-based optimization
methods suggested particle arrangements that have been further
analyzed with the analytical approach; (ii) results based on the
latter method represented a stringent test for the data obtained via
the EA route.  The EA-based part of the studies has been carried out
for approximately 35000 state points: for each of them the number of
particles per unit cell was systematically increased from simple
lattices to cells with up to 40 basis particles. As a consequence the
numerical resolution in $x$ in the EA approach is limited: in
particular, the largest value for $x < 1/2$ that can be obtained is $x
= 19/39 = 0.487$. Thus it cannot be excluded that significantly larger
unit cells could allow for a more complicated two-dimensional particle
arrangement which might be energetically more favourable.  The
analytical framework uses the simplifying assumption that the
competing structures on both plates are undistorted (i.e., ideal).
The colored region, in contrast, covers data obtained via the
numerical approach which is able to grasp appropriately the emerging
minute deviations of the particle configurations from ideal lattices.
The mentioned limitations of the analytical approach explain small
discrepancies between the limiting white curves and the border of the
colored region.

When identifying ordered structures, we first classify particle
arrangements by the respective value of $x$.  Then, further refinement
is achieved by a classification scheme, involving one or more BOOPs
$\langle \Psi_n^{(\alpha)} \rangle$.  The relevant criteria for
identifying structures in the EA approach are summarized in Table
\ref{tab:table_order}.
While the detailed discussion of the emerging structures is postponed
to the following sections, a few general remarks are in order:
\begin{itemize}
\item the relatively large regions of uniform and pure colors (i.e.,
  red, green, or blue) occurring in the panels for the BOOPs $\langle
  \Psi^{(1)}_n \rangle$ and $\langle \Psi^{(2)}_n \rangle$ in Fig.
  \ref{fig:diagram_of_states_rgb} for most of the state points
  investigated indicate that the particles form simple, ordered
  structures with four-, five-, or six-fold symmetry in the respective
  layers;
\item the degree of structural commensurability of the two sublattices
  in the two layers is reflected by the respective colors encoded in
  the values of $\langle \Psi^{(3)}_n \rangle$ and $\langle
  \Psi^{(4)}_n \rangle$: the effort of the system to guarantee a high
  degree of structural commensurability leads to pure colors of the
  respective state points; this is for instance the case along the
  stripe-shaped regions in the domain where the structure
  $\textrm{V}_x$ is stable: within each of these stripes the value $x$
  is essentially constant;
\item related observations can also be made for the shade-coded plot
  of $N$, the number of particles per unit cell (Fig.
  \ref{fig:diagram_of_states_n}). The white/bright regions
  characterize state points with a simple, ordered structure (i.e.,
  with only a few particles per unit cell) and a high degree of
  commensurability between the two sub-structures. This also holds for
  the stripe-shaped regions (along which $x$ is essentially constant)
  located within the domain where structure $\textrm{V}_x$ is stable.
  In contrast, large $N$ values (i.e. dark regions in Fig.
  \ref{fig:diagram_of_states_n}) indicate the occurrence of complex,
  incommensurate structures.
\end{itemize}

\begin{table}[h]
\begin{center}
\caption{Classification scheme used to identify the observed ordered
  structures (first and last columns) in the asymmetric Wigner bilayer
  system, based on their respective values of $x$ (second column) and
  of the BOOPs (third column): the first criterion is the value of
  $x$; then, further refinement is achieved by using BOOPs or linear
  combinations thereof. Note that the threshold values for the BOOPs
  (specified in the third column) are to some extent arbitrary. For
  convenience, we have dropped the symbols that indicate the averaged
  values of the BOOPs (i.e., $\Psi^{(\alpha)}_n$ stands for $\langle
  \Psi^{(\alpha)}_n \rangle$). The occupation index $x$ is defined in
  Eq. (\ref{occup}).}
\label{tab:table_order}

\begin{tabular}{| l | l | l | l |}
\hline
\textrm{I} & $x=0$ & & hexagonal monolayer \\
\hline
\textrm{I\kern -0.3ex I} & $x=1/2$ & $\Psi_4^{(1,2)}=1$, $0<\Psi_6^{(1,2)}<1$ & rectangular bilayer \\
\textrm{I\kern -0.3ex I\kern -0.3ex I} & $x=1/2$ & $\Psi_4^{(1,2)}=1$, $\Psi_6^{(1,2)}=0$ & square bilayer \\
\textrm{I\kern -0.3ex V} & $x=1/2$ & $0<\Psi_4^{(1,2)}<1$, $0<\Psi_6^{(1,2)}<1$ & rhombic bilayer \\
\textrm{V} & $x=1/2$ & $\Psi_4^{(1,2)}=0$, $\Psi_6^{(1,2)}=1$ & hexagonal bilayer \\
\hline
$\textrm{I}_x$ & $0<x<1/3$ & $0.9 < \Psi_6^{(3)}$ & \\
\textrm{H} & $x=1/3$ & $0.9 < \Psi_6^{(3)}$ & honeycomb (layer 2) \\
$\textrm{I\kern -0.3ex I}_x$ & $1/3<x<1/2$ & $0.9 < \Psi_6^{(3)}$ & \\
$\textrm{V}_x$ & $0<x<x^*$ & $0.9 < (1-x) \Psi_6^{(1)}+x \Psi_6^{(2)}$ & hexagonal bilayer \\
$\textrm{DV}_x$ & $2/5\leq x <1/2$ & $0.5\leq \Psi_6^{(1,2)}$, $\Psi_4^{(1)}\sim 0.4$, $\Psi_5^{(2)}\sim 0.3$ & distorted hexagons \\
$\textrm{S}_1$ & $x=1/3$ & $0.9 < \Psi_5^{(1)}$, $0.9 < \Psi_4^{(2)}$ & snub square (layer 1) \\			
\hline
$\textrm{S}_2$ & $x=1/3$ & $0.45 < \Psi_5^{(2)}$ & snub square (layer 2) \\
\textrm{P}-type & $1/3<x<1/2$ & $0.45 < \Psi_5^{(2)}$ & pentagonal in layer 2\\
			                         & or $0<x<1/3$ & or $0.9 < \Psi_5^{(4)}$ & pentagonal holes \\
\hline
\end{tabular}
\end{center}
\end{table}

\begin{figure}[htbp]
\begin{center}
\includegraphics[width=8cm]{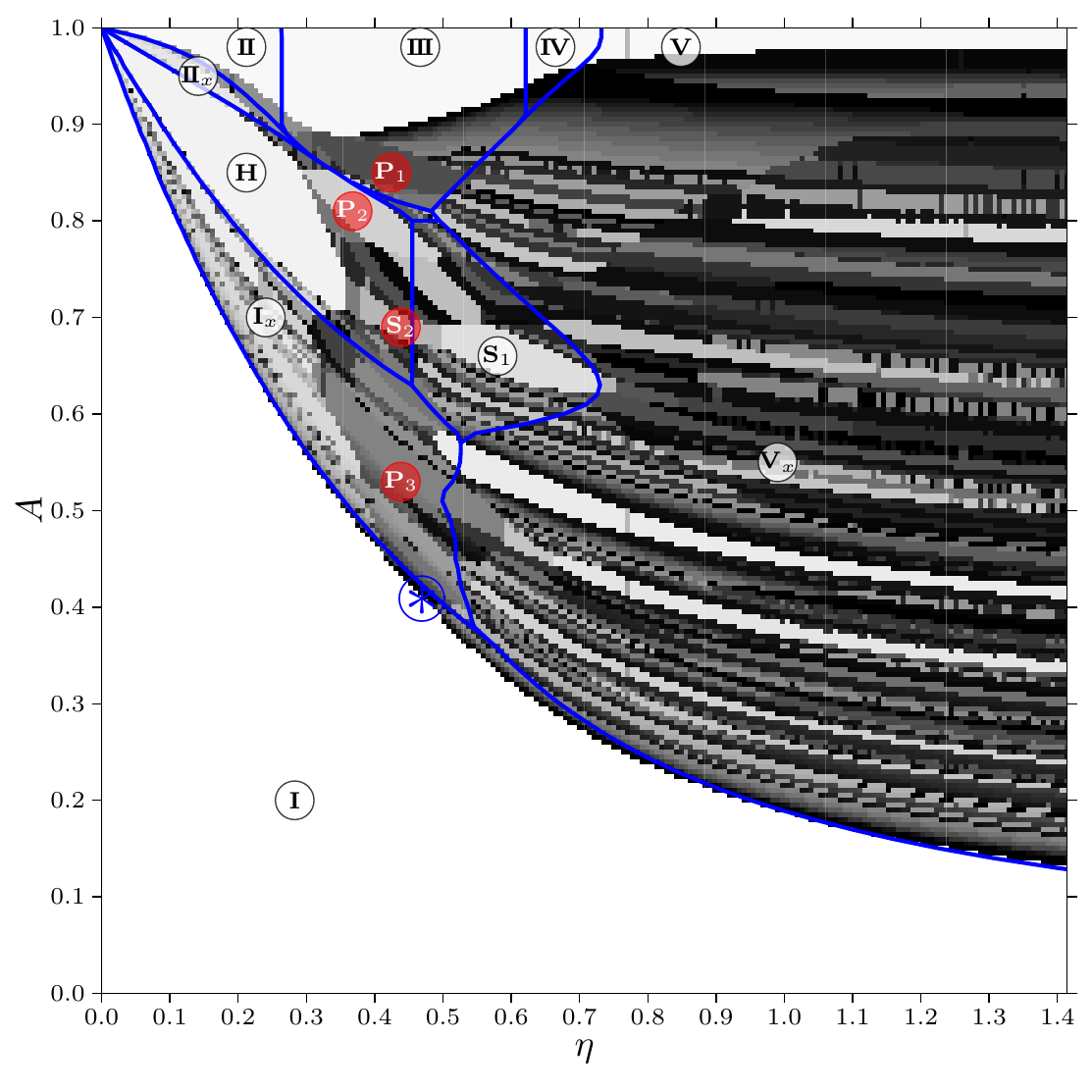}
\caption{(color online) Diagram of states in terms of numbers of
  particles per unit cell ($N$), as obtained via the EA-based
  approach. The following gray-scale encoding was used: $N=2$
  corresponds to white, $N=40$ to black. Simple structures that are
  easily tractable with the analytical approach (specified by bright
  labels, with their respective regions of stability delimited by blue
  lines) appear thus as bright regions. Structure $\textrm{V}_x$
  represents an exception to this rule, since for this case the
  structures of the two layers are not necessarily strongly
  correlated. Structures with red-colored labels are too complicated
  to be amenable to the analytical approach. The blue star marks a
  bi-critical point (see text).}
\label{fig:diagram_of_states_n}
\end{center}
\end{figure}

\begin{figure}[htbp]
\begin{center}
\includegraphics[width=8cm]{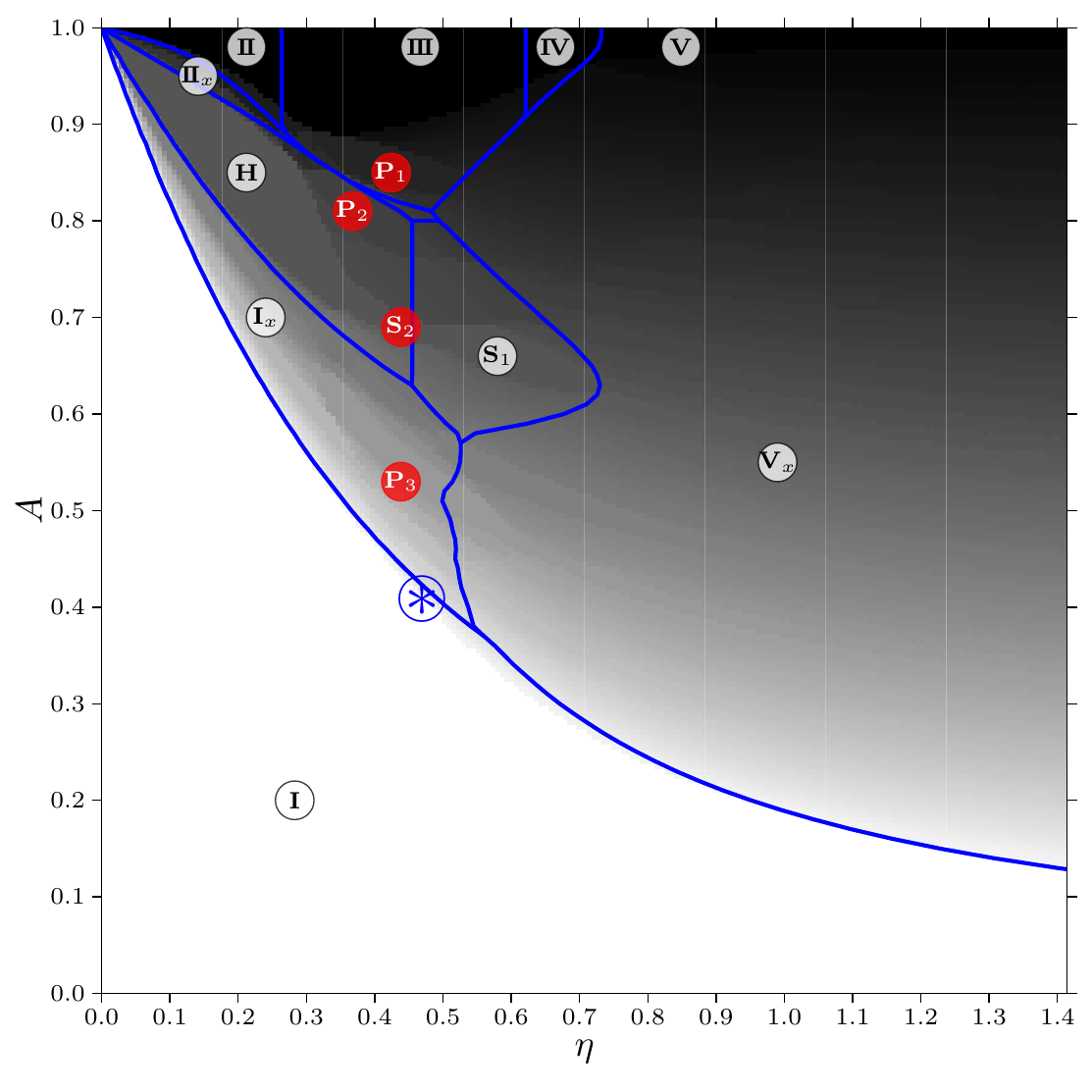}
\caption{(color online) Same as Fig. \ref{fig:diagram_of_states_n},
  now in terms of the order parameter $x = n_2/n$.  The following
  gray-scale encoding was used: $x = 0$ corresponds to white, $x =
  1/2$ to black.  Note that the value $x = 1/3$ is of particular
  relevance (see regions surrounding the labels of structures
  \textrm{H}, $\textrm{S}_2$, and $\textrm{S}_1$).
  }
\label{fig:diagram_of_states_x}
\end{center}
\end{figure}

\section{Structures emerging at small $\eta$: 
\textrm{I}, $\textrm{I}_x$, \textrm{H}, and $\textrm{I\kern -0.3ex I}_x$}
\label{sec:smalleta}

\subsection{Phase \textrm{I}} 
\label{app:I_1}

When the two plates are at contact ($\eta=0$), the lowest energy of
the system corresponds to the hexagonal Wigner monolayer (structure
I). Each of the triangles is shared by three particles and each
particle is surrounded by six triangles; hence, there are just $6
\times 1/3 = 2$ triangles per particle. Therefore the lattice spacing
$a$ is imposed by the requirement of electro-neutrality as
$ \sqrt{3}\, a^2 (\sigma_1+\sigma_2) = 2$ .
The hexagonal lattice can be considered as the union of two
rectangular lattices with the aspect ratio $\Delta =
a_2/a_1=\sqrt{3}$, shifted with respect to each other in both spatial
directions by half of the respective side lengths.  Since for
$\eta=0$, the monolayer is neutral by definition, we find -- using the
formalism developed in Appendix \ref{app:A} -- for the energy
$E_{\textrm{I}}(\eta=0)=E_{pp}^{\rm neutr}$, where
\begin{equation}
\frac{E_{pp}^{\rm neutr}}{N} = \sum_{(j,k)\ne (0,0)} \frac{e^2}{2 a\sqrt{j^2+3 k^2}}
+ \sum_{j,k} \frac{e^2}{2 a\sqrt{(j+1/2)^2+3 (k+1/2)^2}} - \mbox{backgr.}
\end{equation}
Here, the lattice Coulomb summations extend over all integers;
infinite constants in the summations are regularized by the
neutralizing background.  We define the Madelung structural constant
$c$ of the hexagonal lattice in the following way
\begin{equation}
c \equiv \frac{E_{\textrm{I}}(\eta=0)}{N e^2\sqrt{\sigma_1+\sigma_2}} .
\end{equation}

Using the technique put forward in Refs. \cite{Samaj12_1,Samaj12_2},
the lattice Coulomb summations can be transformed into integrals over
the Jacobi theta functions with zero argument (\ref{theta}).  In terms
of the function
\begin{equation} \label{cdelta}
c(\Delta) \equiv \frac{1}{2^{3/2}\sqrt{\pi}} \int_0^{\infty} 
\frac{{\rm d}t}{\sqrt{t}} \left\{ \left[ \theta_3({\rm e}^{-\Delta t}) 
\theta_3({\rm e}^{-t/\Delta}) - 1 - \frac{\pi}{t} \right] 
+ \left[ \theta_2({\rm e}^{-\Delta t}) \theta_2({\rm e}^{-t/\Delta}) 
- \frac{\pi}{t} \right] \right\} , 
\end{equation}
the Madelung constant is given by $c=c(\sqrt{3})$.  The neutralizing
background subtracts the ($t\to 0$)-singularities $\pi/t$ of the
products of two $\theta_3$- and two $\theta_2$-functions.  Based on
results of Refs. \cite{Samaj12_1,Samaj12_2}, the expression for $c$
can be transformed into a quickly converging series of the generalized
Misra functions (\ref{specialf}) and we obtain the well-known value $c
= - 1.960515789\ldots$ .

For $A<1$ and at sufficiently small distances $\eta$ between the
plates, all particles forming the hexagonal Wigner crystal will remain
at their positions on plate 1; such a monolayer phase will also be
coined as phase \textrm{I}. Since $x = 0$ in phase \textrm{I}, the
corresponding energy is given -- according to the ``neutralization''
analysis of Appendix \ref{app:A} -- by expression (\ref{resequation})
as follows
\begin{equation} \label{energyI}
\frac{E_{\textrm{I}}(\eta,A)}{N e^2\sqrt{\sigma_1+\sigma_2}} = c +
2^{3/2} \pi \eta \left( \frac{A}{1+A} \right)^2 .
\end{equation}

Whether or not phase I is stable can be tested qualitatively by moving
one of the particles perpendicularly from plate 1 at $z=0$ to plate 2
at $z=d$.  This move is accompanied by the increase of the potential
energy of the particle by
\begin{equation} \label{poten}
\Delta E_{\rm pot} = -e \left[ \phi(d) - \phi(0) \right] 
= 2\pi e^2 (\sigma_1-\sigma_2) d .
\end{equation}
Simultaneously, since the distance of the reference particle to all
other particles is increased, its interaction energy is decreased by
$\Delta E_{\rm int}\sim - e^2 C d^2$ ($C > 0$ being a structure
constant of the hexagonal Wigner lattice) due to the symmetry $[z \to
  -z]$ of the interaction potential.  As soon as $A<1$, the total
energy change of this operation
\begin{equation} \label{balance}
\Delta E = \Delta E_{\rm pot} + \Delta E_{\rm int} \sim e^2 [2\pi
  (\sigma_1-\sigma_2) d - C d^2]
\end{equation}
is dominated by the linear potential term for small $d$. $\Delta E$ is
therefore positive and the particle prefers to remain in its lattice
position within phase I.  Since we proceed here by necessary condition
for stability, this provides a hint that phase I is always stable at
sufficiently small distances.

The way of how the monolayer phase \textrm{I} transforms into another
bilayer phase at a specific distance $d_c$ (or, equivalently,
$\eta_c$) depends on the value of the asymmetry parameter $A$; these
values form in the diagram of states the line $\eta_c(A)$, or,
equivalently, $A_c(\eta)$. Two scenarios will be discussed in the
following: one valid for $A$ close to 1 where $d_c$ is small and the
transition is due to the perpendicular moves of particles from plate 1
to plate 2 and the other for small $A$, where $d_c$ is somewhat
larger. In the latter case, the moves of the particles from plate 1 in
Structure I to plate 2 to form the ground state are in a direction
that is no longer perpendicular to the plates.


\subsection{Phase $\textrm{I}_x$} 
\label{app:I_11}

Starting from the monolayer, keeping $A$ fixed to a value close to
unity and increasing $\eta$, more and more charges will shift their
location to layer 2: they leave distorted hexagonal holes in layer 1
and form, in turn, a new, ordered particle arrangement in layer
2. This is the origin of the so-called family of structures
\textrm{I}$_x$.
To be more specific, phase \textrm{I}$_x$ can be defined as a bilayer
structure where the projections of the particles of both layers onto
one plane form a hexagonal phase (which can be slightly distorted).
The parameter $x = N_2/N$ specifies the number of particles that have
been shifted in a perpendicular direction from the hexagonal monolayer
on plate 1 to plate 2 (see snapshots in
Fig. \ref{fig:snapshots_Ix_H_IIx}).

The essentially unrestricted search of the EA-based optimization
algorithm provides evidence that upon increasing distance $\eta$ at a
fixed large $A$, structure \textrm{I}$_x$ transforms first into
structure \textrm{H} and then into phase \textrm{I\kern -0.3ex I}$_x$
(to be discussed in detail in Subsection \ref{subsec:phaseII}). Both
of these phases are characterized by the feature that the projected
particle positions of both layers form an almost perfect (i.e.,
possibly slightly distorted) hexagonal lattice; we can characterize
this family of structures via the criterion $0.9 < \Psi_6^{(3)}$ (see
Table \ref{tab:table_order}). The difference between these three
structures can be quantified via the occupation parameter $x$; the
respective ranges of stability are displayed in
Fig. \ref{fig:diagram_of_states_rgb}:

\begin{itemize}
\item[$\bullet$] structure \textrm{I}$_x$ (with a representative
  snapshot in the left panel of Fig.  \ref{fig:snapshots_Ix_H_IIx} for
  $x = 1/4$) has $0<x<1/3$;
\item[$\bullet$] structure \textrm{H} (central panel of Fig.
  \ref{fig:snapshots_Ix_H_IIx}) is characterized by $x = 1/3$ and can
  be considered as a special case of both neighbouring structures,
  i.e., of \textrm{I}$_x$ and \textrm{I\kern -0.3ex I}$_x$; structure
  \textrm{H} consists of a honeycomb lattice in layer 1 and a
  hexagonal lattice in layer 2 where particles of the latter are
  located above the centers of the hexagonal rings in layer 1;
\item[$\bullet$] eventually, structure \textrm{I\kern -0.3ex I}$_x$
  (see right panel of Fig. \ref{fig:snapshots_Ix_H_IIx}), having $1/3
  < x < 1/2$.
\end{itemize}


\begin{figure}[htbp]
\begin{center}
\includegraphics[width=15cm]{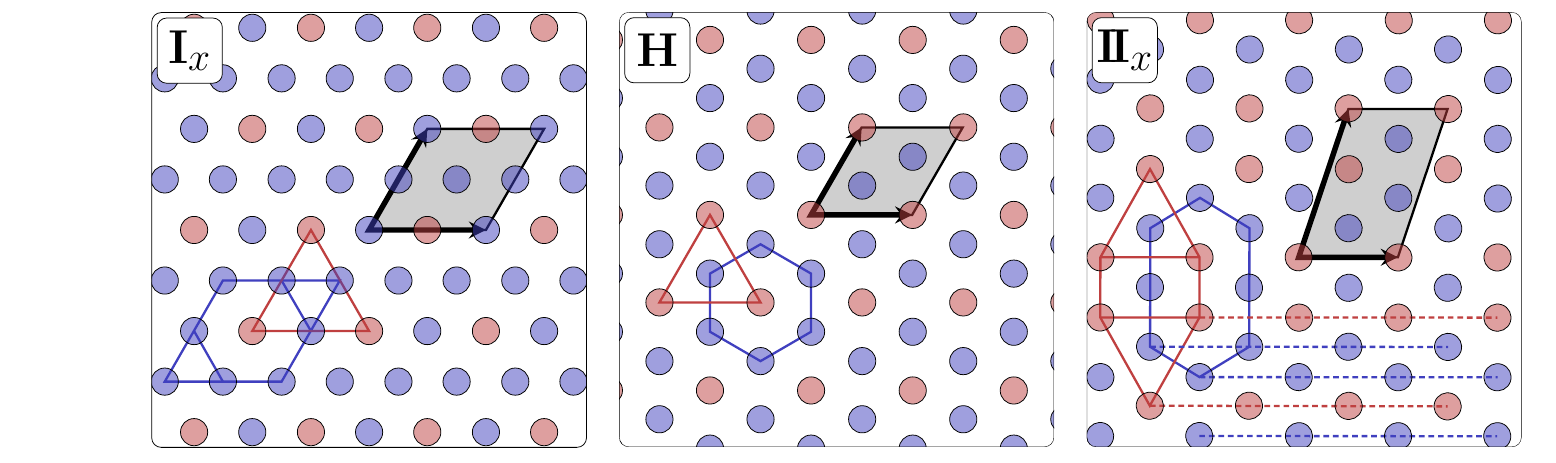} \\
\caption{(color online) Representative snapshots of structures
  $\textrm{I}_x$, \textrm{H}, and $\textrm{I\kern -0.3ex I}_x$ (see
  text).  Particles in layer 1 are colored blue, particles in layer 2
  red. The unit cell of the respective structure is indicated by the
  shaded area. Blue and red lines highlight interesting structural
  features in layers 1 and 2, respectively. For the dotted, colored
  lines see text. Left panel: structure $\textrm{I}_x$ emerging for
  $\eta=0.184$ and $A=0.775$, with $x=1/4$.  Center panel: a special
  case of structure $\textrm{I}_x$, \textrm{H}, for $\eta=0.198$ and
  $A=0.85$, with $x=1/3$.  Right panel: structure $\textrm{I\kern
    -0.3ex I}_x$ for $\eta=0.148$ and $A=0.95$, with $x=2/5$.}
\label{fig:snapshots_Ix_H_IIx}
\end{center}
\end{figure}

Within the analytic approach it is not possible to fully capture the
features of all the emerging phases, as $x$ is essentially
continuous. With a reasonable amount of computational effort the
analytic route is able to grasp those \textrm{I}$_x$ phases, where the
two sublattices (with lattice spacings $a$ and $b$, respectively) are
commensurate hexagonal layers. These lattices form a limited subset of
the whole structural family \textrm{I}$_x$, where the corresponding
values of $x$ are restricted to a subset of $\mathbb{Q}$, as detailed in
the following.  To specify the possible values of $b/a$ (with $b >
a$), which guarantee commensurability of the two sublattices on plates
1 and 2, we notice that joining two arbitrary vertices of lattice
$\alpha$ implies a side of the hexagonal lattice $\beta$ whose all
points also belong to $\alpha$.  The primitive vectors of the
hexagonal lattice $\alpha$ are
\begin{equation}
{\bf a}_1 = a(1,0), \qquad {\bf a}_2 = \frac{a}{2} (1,\sqrt{3}) . 
\end{equation}
Choosing the lattice vector of sublattice $\beta$ as ${\bf b} = j {\bf
  a_1} + k {\bf a}_2$ with $(j,k)$ two arbitrary positive
integers such that $j+k \neq 0,1$ [i.e., $(j,k)=(0,2)$, $(1,1)$,
  $(0,3)$, $(1,2)$, $(0,4)$, $(1,3)$, $(2,2)$, etc.] we find that $b^2
  = a^2(j^2+jk+k^2)$.  Since $S/N_2 = \sqrt{3} b^2/2$, the possible
  values of $x$ are constrained to
\begin{equation} \label{discretevalues}
x \equiv \frac{N_2}{N} = \frac{a^2}{b^2} = \frac{1}{j^2+jk+k^2} , \qquad
x\in \left\{ \frac{1}{3},\frac{1}{4}, \frac{1}{7}, \frac{1}{9}, \frac{1}{12},
\frac{1}{13}, \frac{1}{16}, \ldots \right\} .  
\end{equation} 
The admissible discrete values of the occupation parameter $x$ become
essentially dense when $x\to 0$ and we can take $x$ as a
quasi-continuous variable in that limit.

Among the structures $\textrm{I}_x$ the one with the largest
occupation parameter, namely $x=1/3$, is pictured in the center panel
of Fig. \ref{fig:snapshots_Ix_H_IIx}; it is the aforementioned
structure \textrm{H}.  Structure \textrm{H} has a special property:
due to a high degree of symmetry of the internal architecture, no
local distortions of the two sublattices on plates 1 and 2 can be
observed.  Therefore analytical results match perfectly the numerical
data of EA-based method.  This particularly stable internal
architecture guarantees a relatively large region of the parameter
space $(\eta,A)$ over which this phase represents the energetically
most favorable candidate.

\subsection{Transition $\textrm{I}\to\textrm{I}_x$} 
\label{subsec:transition_I_Ix}

Whether the system remains in its monolayer configuration \textrm{I}
or populates the second layer (leading thus to structure
$\textrm{I}_x$) is of course the result of an energetic competition,
to which the analytical approach has -- despite the above mentioned
limitations -- essentially full access.  Let a reference particle 1 be
located on sublattice $\alpha$ of plate 1.  The occurring energy
change of phase $\textrm{I}_x$ with respect to phase \textrm{I} is
given by
\begin{eqnarray} 
\frac{E_{\textrm{I}_x}(\eta,A;x)-E_{\rm I}(\eta,A)}{e^2 N_2} & = & 
2\pi (\sigma_1-\sigma_2) d + \sum_{j\in \alpha\atop j\ne 1} 
\left( \frac{1}{\sqrt{R_{1j}^2+d^2}} - \frac{1}{R_{1j}} \right) \nonumber \\
& & - \sum_{j\in \beta} \left( \frac{1}{\sqrt{R_{1j}^2+d^2}} - \frac{1}{R_{1j}} 
\right) . \label{energychange}
\end{eqnarray}
The first term on the rhs of this equation corresponds to the increase
of the potential energy by taking $N_2$ particles from plate 1 to 2.
The second term is the change in the interaction energy of a particle
transferred from plate 1 to 2, with respect to particles remaining in
sublattice $\alpha$.  The particles located in sublattice $\beta$
should {\it not} be included in that sum as the mutual interaction
energy of particles in sublattice $\beta$ is unchanged by their
simultaneous transfer to plate 2, so the third term in the above
relation is simply the compensation sum.

Using methods outlined in Refs. \cite{Samaj12_1,Samaj12_2}, we obtain
the following integral representation of the energy change, specified
in Eq.  (\ref{energychange}):
\begin{eqnarray}
\frac{E_{\textrm{I}_x}(\eta,A;x)-E_{\textrm{I}}(\eta,A)}{
e^2 N \sqrt{\sigma_1+\sigma_2}} = x \Bigg\{ 2^{3/2}\pi \frac{1-A}{1+A} \eta 
\phantom{aaaaaaaaaaaaaaaaaaaaaaaaaaaaaaaaaaa} 
\nonumber \\ - \frac{1}{\sqrt{2\pi}} 
\int_0^{\infty} \frac{{\rm d}t}{\sqrt{t}} \left( 1 - {\rm e}^{-\eta^2 t}\right)
\left[ \theta_3({\rm e}^{-\sqrt{3}t}) \theta_3({\rm e}^{-t/\sqrt{3}}) - 1
+ \theta_2({\rm e}^{-\sqrt{3}t}) \theta_2({\rm e}^{-t/\sqrt{3}}) \right]
 \nonumber \\   + \frac{\sqrt{x}}{\sqrt{2\pi}} 
\int_0^{\infty} \frac{{\rm d}t}{\sqrt{t}} \left( 1 - {\rm e}^{-\eta^2 x t}\right)
\left[ \theta_3({\rm e}^{-\sqrt{3}t}) \theta_3({\rm e}^{-t/\sqrt{3}}) - 1
+ \theta_2({\rm e}^{-\sqrt{3}t}) \theta_2({\rm e}^{-t/\sqrt{3}}) \right] \Bigg\} .
\label{xdepend}
\end{eqnarray}
The integrals over the Jacobi theta functions are expressible via the
$K$-function defined in Eq. (\ref{K}) of Appendix \ref{app:B} as
follows:
\begin{equation}
\frac{E_{\textrm{I}_x}(\eta;x)-E_{\textrm{I}}(\eta)}{ e^2 N
  \sqrt{\sigma_1+\sigma_2}} = \frac{x}{\sqrt{2}} \left[ - 8 \pi
  \frac{A}{1+A} \eta + 4\pi \eta x - K(\sqrt{3},\eta) + \sqrt{x}
  K(\sqrt{3},\sqrt{x}\eta) \right] .
\end{equation}
Compared to the expression (\ref{energyI}) for the energy of phase
\textrm{I}, the energy of phase $\textrm{I}_x$ is now given by
\begin{equation} \label{energyIx}
\frac{E_{\textrm{I}_x}(\eta,A;x)}{e^2 N \sqrt{\sigma_1+\sigma_2}} = 
2^{3/2} \pi \eta \left( x - \frac{A}{1+A} \right)^2 + c +
\frac{x}{\sqrt{2}} \left[ - K(\sqrt{3},\eta) + 
\sqrt{x} K(\sqrt{3},\sqrt{x}\eta) \right] .
\end{equation}
Using the series representation of $K(\sqrt{3},\eta)$ presented in
Appendix \ref{app:B}, this expression becomes suitable for numerical
calculations.

The transition from phase I (with $x=0$) to phase $\textrm{I}_x$ (with
$x > 0$) is continuous, i.e. of second-order (as discussed in the
following).  In an effort to find a formal anallogy of our system of
classical particles at zero temperature with a statistical model at
finite temperature, we keep in mind that the role of the inverse
temperature is played in our case by the dimensionless distance
between the plates $\eta$, while the role of the free energy is played
by the energy given in Eq. (\ref{xdepend}), or equivalently in Eq.
(\ref{energyIx}).  The order parameter, which increases from 0 just at
the critical point continuously to finite values, is the occupation
number $x$.

For small $x$, the expression for the energy (\ref{xdepend}) can be
expanded in powers of $x$ as follows
\begin{equation} \label{xinvolved}
\frac{E_{\textrm{I}_x}(\eta,A;x)-E_{\textrm{I}}(\eta,A)}{
e^2 N \sqrt{\sigma_1+\sigma_2}}
\simeq f(\eta) x + \frac{2^{3/2}\pi}{\lambda} \eta^2 x^{5/2} + O(x^{7/2}) ,
\end{equation} 
where
\begin{eqnarray}  \label{f_eta}
f(\eta) & = & 2^{3/2}\pi \frac{1-A}{1+A} \eta - \frac{1}{\sqrt{2\pi}} 
\int_0^{\infty} \frac{{\rm d}t}{\sqrt{t}} \left( 1 - {\rm e}^{-\eta^2 t}\right)
\left[ \theta_3({\rm e}^{-\sqrt{3}t}) \theta_3({\rm e}^{-t/\sqrt{3}}) - 1
\right. \nonumber \\ & & \left. + \theta_2({\rm e}^{-\sqrt{3}t}) 
\theta_2({\rm e}^{-t/\sqrt{3}}) \right] \label{functionf} ;
\end{eqnarray}
the constant $\lambda$ is defined in Eqs. (\ref{asymptexp}) and
(\ref{latticesum}).  Note that the expansion of the energy in the
order parameter $x$, given in Eq.  (\ref{xinvolved}) is {\it not}
analytic due to the long-range Coulomb interaction of the charged
particles. This feature is in striking contrast to the standard
mean-field, Landau-type theory of phase transitions where the
thermodynamic potential (in our case the energy), assumed to be a
smooth function of the order parameter, is expanded in {\it integer}
powers of the order parameter, reflecting the symmetry of the system.
Our energy change (\ref{xinvolved}) does not show the symmetry
invariance with respect to a transformation of $x$, which explains the
occurrence of rational powers in the order parameter $x$; we emphasize
that our expansion (\ref{xinvolved}) starts with $x$ as the leading
term, which is in contrast to the well-known Landau expansions,
starting -- in the absence of an external field -- with a term
proportional to $x^2$.

The free variable $x$ has to be chosen in such a way that it provides
the minimal value of the energy.  The extremum condition for
$E_{\textrm{I}_x}(\eta;x)$, i.e., $\partial_x E_{\textrm{I}_x}(\eta;x)
= 0$, when applied to relation (\ref{xinvolved}) takes the form
\begin{equation} \label{extremum}
0 \simeq f(\eta) + \frac{5 \sqrt{2}\pi}{\lambda} \eta^2 x^{3/2} + O(x^{5/2}) .
\end{equation}
For a given value of $A$ the critical point $\eta_c$ is identified by
the condition $f(\eta_c)=0$, i.e.,
\begin{equation} \label{border}
4\pi \frac{1-A}{1+A} \eta_c = 
\frac{1}{\sqrt{\pi}}\int_0^{\infty} \frac{{\rm d}t}{\sqrt{t}} 
\left( 1 - {\rm e}^{-\eta_c^2 t}\right)
\left[ \theta_3({\rm e}^{-\sqrt{3}t}) \theta_3({\rm e}^{-t/\sqrt{3}}) - 1 + 
\theta_2({\rm e}^{-\sqrt{3}t}) \theta_2({\rm e}^{-t/\sqrt{3}}) \right] .
\end{equation}

As can be seen in Fig. \ref{fig:diagram_of_states_rgb}, this analytic
estimate of critical points $\eta_c = \eta_c(A)$ (i.e., the white line
that separates phases $\textrm{I}$ and $\textrm{I}_x$ and ending at
the bi-critical point with the latter one marked by the star)
coincides well with the EA results. In the limit $\eta\to 0$ (or,
equivalently $A\to 1$), expression (\ref{border}) reduces to the exact
asymptotic relation
\begin{equation} \label{asymptexp}
\eta_c(A) \mathop{\sim}_{A\to 1} \lambda \frac{1-A}{1+A} , \qquad 
\lambda = \frac{4\pi}{\frac{1}{\sqrt{\pi}}\int_0^{\infty} {\rm d}t\, \sqrt{t} 
\left[\theta_3({\rm e}^{-\sqrt{3}t}) \theta_3({\rm e}^{-t/\sqrt{3}}) - 1 + 
\theta_2({\rm e}^{-\sqrt{3}t}) \theta_2({\rm e}^{-t/\sqrt{3}}) \right]} .
\end{equation}
Using the general theory of lattice sums \cite{Zucker_74,Zucker_75} it
can be shown that
\begin{eqnarray} 
\frac{1}{\sqrt{\pi}} \int_0^{\infty} {\rm d}t\, \sqrt{t} 
\left[ \theta_3({\rm e}^{-\sqrt{3}t}) \theta_3({\rm e}^{-t/\sqrt{3}}) - 1 + 
\theta_2({\rm e}^{-\sqrt{3}t}) \theta_2({\rm e}^{-t/\sqrt{3}}) \right] \nonumber \\
= 3^{1/4}\zeta\left(\frac{3}{2}\right)
\left[ \zeta\left(\frac{3}{2},\frac{1}{3}\right) - 
\zeta\left(\frac{3}{2},\frac{2}{3}\right) \right] , \label{latticesum}
\end{eqnarray}
where $\zeta(z,q) = \sum_{j=0}^{\infty}1/(q+j)^z$ is the generalized
Riemann zeta function and $\zeta(z) \equiv \zeta(z,1)$. The prefactor
$\lambda \simeq 0.999215$ in Eq. (\ref{asymptexp}) is thus very close,
but not equal, to 1.

The function $f(\eta)$ given in Eq.  (\ref{functionf}) is dominated
for small $\eta$ by the linear term, so that $f(\eta)>0$ for
$\eta<\eta_c$, while $f(\eta)<0$ for $\eta>\eta_c$; thus we can write
in the neighborhood of the critical point $\eta_c$ that $f(\eta)\sim g
(\eta_c-\eta)$ with a positive prefactor $g>0$.  The consequent
extremum condition reads as (cf. Eq. (\ref{f_eta}))
\begin{equation} \label{extremum2}
g(\eta-\eta_c) = \frac{5 \sqrt{2}\pi}{\lambda} \eta_c^2 x^{3/2}(\eta) .
\end{equation}

\begin{itemize}
\item In the region $\eta\ge \eta_c$, the extremum condition
  (\ref{extremum2}) has only one real solution, namely
\begin{equation} \label{eq:xthreshold}
x(\eta) \simeq \left( \frac{\lambda g}{5\sqrt{2}\pi\eta_c^2}
\right)^{2/3} (\eta-\eta_c)^{\beta} , \qquad {\rm with} ~~~
\beta=\frac{2}{3} .
\end{equation}
Here, we use the standard notation for the critical index $\beta$
describing the non-analytic growth of the order parameter.  The
corresponding energy change of phase $\textrm{I}_x$ with respect to
phase \textrm{I}, i.e.,
\begin{equation}
\label{e_diff_Ix}
\frac{E_{\textrm{I}_x}(\eta;x(\eta))-E_{\textrm{I}}(\eta)}{
e^2 N \sqrt{\sigma_1+\sigma_2}}
\simeq - 3 \left( \frac{g^5 \lambda^2}{2\pi^2 5^5 \eta_c^4} \right)^{1/3}
(\eta-\eta_c)^{2-\alpha} , \qquad {\rm with} ~~~ \alpha=\frac{1}{3} 
\end{equation} 
is negative; hence the extremum is indeed a minimum as it should be.
Here, we use the standard notation for the critical index $\alpha$,
defined by the relation for the ``heat capacity''
\begin{equation} \label{heatcapacity}
\frac{{\rm d}^2E_{\textrm{I}}(x(\eta),\eta)}{{\rm d}\eta^2} \propto
\frac{1}{(\eta-\eta_c)^{\alpha}} .
\end{equation}
Note that the energy of phase I, as given in Eq.  (\ref{energyI}), is
linear in $\eta$ and therefore does not contribute to
Eq. (\ref{heatcapacity}).

\item In the region $\eta<\eta_c$, the extremum condition
  (\ref{extremum2}) has no real solution for $x$.  Since the energy
  (\ref{xinvolved}) is a monotonously increasing function of $x$ in
  that region, the accepted ``physical'' value $x=0$ corresponds to a
  threshold for non-negative real $x$-values, i.e. to phase I.  Since
  the energy of phase I is linear in $\eta$, its second derivative
  with respect to $\eta$ vanishes and the critical index $\alpha'$ has
  no meaning.
\end{itemize}

\begin{figure}[htbp]
\begin{center}
\includegraphics[width=8cm]{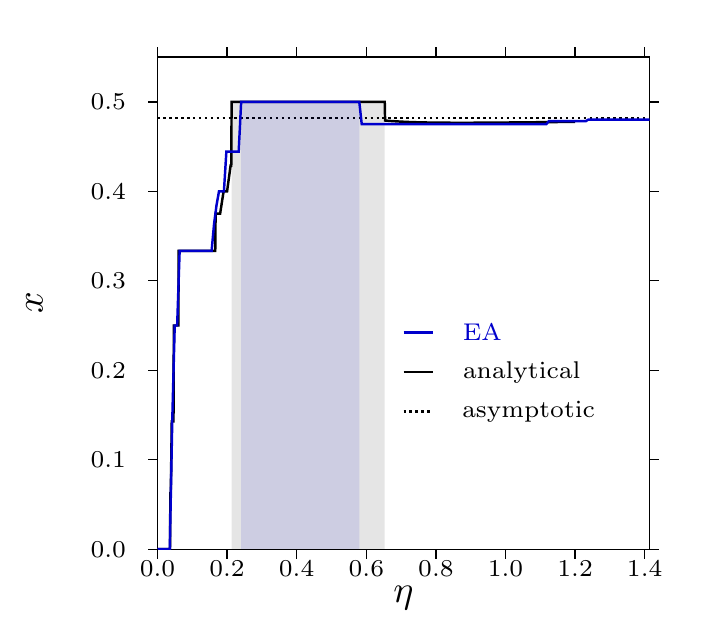}
\hspace{0.0cm}
\includegraphics[width=8cm]{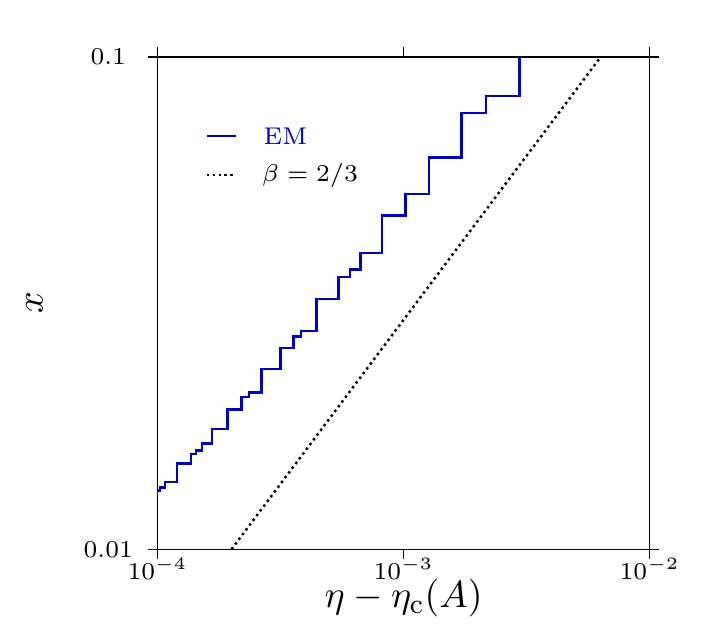}
\caption{(color online) $x(\eta)$ for $A=0.93$.  Left panel:
  occupation index $x$ vs dimensionless distance $\eta$. We identify
  successively structures \textrm{I}, $\textrm{I}_x$, \textrm{H},
  $\textrm{I\kern -0.3ex I}_x$, $\textrm{P}_1$, \textrm{I\kern -0.3ex
    I}, \textrm{I\kern -0.3ex I\kern -0.3ex I}, a range of
  unclassified phases, and finally structures $\textrm{V}_x$ (see also
  Fig. \ref{fig:diagram_of_states_rgb}).  EA results (blue) and
  analytical results (black) are shown.  In the region of stability of
  structure $\textrm{I}_x$ (i.e., for $x \le 1/3$), the discrete
  $x$-values as predicted by the analytic approach (and assuming an
  idealized version of phase \textrm{I}$_x$ -- see related $x$-values
  specified in Eq. (\ref{discretevalues})) characterize the
  staircase-like form of the curve.  The regions of constant $x$ are
  very thin for this value of $A$, close to unity. In addition,
  $x$-values that are not compatible with Eq. (\ref{discretevalues})
  could be identified within the EA approach.  Shaded areas indicate
  overcharging, i.e., where $x>x^*$, see Eq. (\ref{x_neutr}). The
  value of $x^*$ is indicated by a dashed line; see the discussion in
  Subsec. \ref{subsec:II_III_IV_overcharging} for overcharging.  Right
  panel: double-logarithmic plot of $x$ versus $(\eta-\eta_{\rm c})$,
  as obtained in the EM calculations.  The dotted black line is a
  guide to the eye, indicating the critical exponent $\beta=2/3$, see
  Eq. (\ref{eq:xthreshold}).}
\label{fig:systems_wigner_1x}
\end{center}
\end{figure}

A more thorough discussion of critical features is available in
Appendix \ref{app:crit}. We recall that the above analytical treatment
is rigorous only in the asymptotic limit $A\to 1$ (i.e., when
$\eta_c\to 0$), due to absence of local deformations of the structures
on the plates.  For other values of the asymmetry parameter $A$, the
values of the critical indices have to be checked numerically along
the whole critical line, separating phases \textrm{I} and
$\textrm{I}_x$. An example is given in
Fig. \ref{fig:systems_wigner_1x}: for $A=0.93$, the left panel of this
figure shows the $x(\eta)$-curves as calculated analytically and by
using the EA approach.  One observes that $x$ grows quickly with
$\eta$ for $\eta > \eta_c$, the curve being characterized by very thin
plateaus at the anticipated discrete values -- see Eq.
(\ref{discretevalues}).  According to Eq.  (\ref{eq:xthreshold}), the
analytical approach predicts that the transition
$\textrm{I}\to\textrm{I}_x$ is of second-order with a critical
exponent $\beta=2/3$ for the order parameter $x$ along the whole
critical line $\eta_c = \eta_c(A)$ that separates phases \textrm{I}
and \textrm{I}$_x$.  Our numerical EA and EM data corroborate this
prediction.  For the particular value of $A=0.93$, the plot of $x$
vs. $\eta$ close to $\eta_c$ is presented in a
double-logarithmic representation in the right panel of Fig.
\ref{fig:systems_wigner_1x}.  Although even small inaccuracies in the
determination of $\eta_{\rm c}$ can severely change the slope of this
curve, the shape of $x(\eta)$ does seem compatible with the analytical
prediction (the dotted black line).  Analogous results were obtained
for other values of $A$ when the transition $\textrm{I} \to
\textrm{I}_x$ takes place.
			
\subsection{Phase $\textrm{I\kern -0.3ex I}_x$}
\label{subsec:phaseII}

In phase \textrm{I\kern -0.3ex I}, with its structure shown in the
right panel of Fig. \ref{fig:systems_wigner_symmetric}, parallel rows
of blue (to be indexed `b') and red (to be indexed `r') particles
appear in an alternating sequence on plates 1 and 2, respectively,
connected in Fig. \ref{fig:snapshots_Ix_H_IIx} by dotted horizontal
lines.  We can formally assign to this particular periodic repetition
of rows the symbol [br], thus $x=1/2$.

The entire family of \textrm{I\kern -0.3ex I}$_x$ structures can be
constructed by combining the two building elements [br] and [bbr];
phases \textrm{I\kern -0.3ex I}$_x$ can be characterized by $x$-values
in the range $x\in[1/3,1/2]$. Examples for structures \textrm{I\kern
  -0.3ex I}$_x$ are given in Fig.  \ref{fig:snapshots_Ix_H_IIx}: (i)
the previously discussed phase \textrm{H} (being an intermediate
structure between phases \textrm{I}$_x$ and \textrm{I\kern -0.3ex
  I}$_x$) is specified by the sequence of rows [bbr] and $x = 1/3$,
thus \textrm{I\kern -0.3ex I}$_{x = 1/3}$ = \textrm{H} (see central
panel of Fig.  \ref{fig:snapshots_Ix_H_IIx}). Phase $\textrm{I\kern
  -0.3ex I}_x$ with $x=2/5$, shown in the right panel of Fig.
\ref{fig:snapshots_Ix_H_IIx}, is formally represented by the
periodically repeated sequence [br][bbr].  From a more global
perspective, the family of structures $\textrm{I\kern -0.3ex I}_x$
represents the transition phase from structure \textrm{H} to structure
\textrm{I\kern -0.3ex I} and eventually to phase \textrm{I\kern -0.3ex
  I\kern -0.3ex I}.

From an alternative point of view, the lattices on layer 2 of the
family of structures \textrm{I\kern -0.3ex I}$_x$ can be viewed as a
sequence of (slightly distorted) triangular and rectangular rows. Lines
that connect particles of layer 1 (2), respectively, (as shown as an
example in the right panel of Fig.  \ref{fig:snapshots_Ix_H_IIx}), can
generate rows of triangles and rectangles via the following simple
rules: (i) a blue line followed by a red line produces a row of
rectangles, while (ii) two blue lines followed by a red line lead to a
row of equilateral triangles. With these two building entities at
hand, $x$ can be varied continuously between the values $1/3$ and
$1/2$, i.e., a range of $x$-values characteristic for the structures
\textrm{I\kern -0.3ex I}$_x$. It should be emphasized that this
decomposition into rectangles and triangles represents an idealized
view of structures \textrm{I\kern -0.3ex I}$_x$ as they are identified
via the numerical tools. These combinations of structural units lead
in layer 1 to rings which can be quite elongated or can have more
complicated shapes, while the lattice in layer 2 consists of slightly
distorted rectangles and triangles.  We note that similar, alternating
sequences of triangles and rectangles have been identified in
colloidal structures as precursors of quasi-crystalline structures
\cite{mikhael_2008}.

Within the analytic approach the series representations of the
energies of the phases $\textrm{I\kern -0.3ex I}_x$ can be derived in
an analogous way as for phase \textrm{I\kern -0.3ex I}, using,
however, a more general application of the Poisson summation formula
(\ref{PSF}). As an example, we outline in the following how to obtain
the series representation of the energy of phase \textrm{H}
(corresponding to a [bbr] sequence of rows) with $x=1/3$.  Denoting by
$\Delta$ and $a \Delta$ the lattice spacings of the rectangular
structure, we have
\begin{equation}
a^2 \Delta  = \frac{2}{\sigma_1+\sigma_2} = \frac{2}{n_1+n_2} .
\end{equation}
The total energy per particle of this phase can be written as 
\begin{equation}
\frac{E_{\textrm{I\kern -0.3ex I}_x}(\eta;x=1/3)}{N e^2 \sqrt{\sigma_1 + \sigma_2}} = 
2^{3/2} \pi
\eta \left( \frac{1}{3} - \frac{A}{1+A} \right)^2 + \frac{1}{3}
\frac{\sqrt{\Delta}}{2^{3/2}} \left( 2 E_{\rm b} + E_{\rm r} \right) ,
\end{equation}
where
\begin{eqnarray}
E_{\rm b} & = & \sum_{j,k\atop (j,k)\ne (0,0)} \frac{1}{\sqrt{(3j)^2+\Delta^2 k^2}}
+ \sum_{j,k} \frac{1}{\sqrt{(3j+1/2)^2+\Delta^2 (k+1/2)^2}} \nonumber \\
& & + \sum_{j,k} \frac{1}{\sqrt{(3j+1)^2+\Delta^2 k^2+(d/a)^2}} 
+ \sum_{j,k} \frac{1}{\sqrt{(3j+3/2)^2+\Delta^2 (k+1/2)^2}} \nonumber \\
& & + \sum_{j,k} \frac{1}{\sqrt{(3j+2)^2+\Delta^2 k^2}}
+ \sum_{j,k} \frac{1}{\sqrt{(3j+5/2)^2+\Delta^2 (k+1/2)^2+(d/a)^2}} \nonumber \\
& & - \mbox{backgr.}
\end{eqnarray}
is the (dimensionless) energy counted from the point of view of {\it
  blue} (index 'b') particles on plate 1 and
\begin{eqnarray}
E_{\rm r} & = & \sum_{j,k\atop (j,k)\ne (0,0)} \frac{1}{\sqrt{(3j)^2+\Delta^2 k^2}}
+ \sum_{j,k} \frac{1}{\sqrt{(3j+1/2)^2+\Delta^2 (k+1/2)^2+(d/a)^2}} \nonumber \\
& & + \sum_{j,k} \frac{1}{\sqrt{(3j+1)^2+\Delta^2 k^2+(d/a)^2}}
+ \sum_{j,k} \frac{1}{\sqrt{(3j+3/2)^2+\Delta^2 (k+1/2)^2}} \nonumber \\
& & + \sum_{j,k} \frac{1}{\sqrt{(3j+2)^2+\Delta^2 k^2+(d/a)^2}}
+ \sum_{j,k} \frac{1}{\sqrt{(3j+5/2)^2+\Delta^2 (k+1/2)^2+(d/a)^2}} \nonumber \\
& & - \mbox{backgr.}
\end{eqnarray}
is the energy with respect to {\it red} (index 'r') particles on plate
2.  After a series of transformations akin to those presented in Refs.
\cite{Samaj12_1,Samaj12_2}, the total energy is expressible in terms
of the $K$-function (\ref{K}) and the function $c(\Delta)$, specified
in Eq.  (\ref{cdelta}), as follows
\begin{equation} \label{energyIIx}
\frac{E_{\textrm{I\kern -0.3ex I}_x}(\eta;x=1/3)}{N} 
= 2^{3/2} \pi \eta \left( \frac{1}{3}
- \frac{A}{1+A} \right)^2 + c(\Delta) + \frac{1}{3 \sqrt{2}}
\left[ - K(\Delta,\eta) + \frac{1}{\sqrt{3}} 
K(3/\Delta,\eta/\sqrt{3}) \right] .
\end{equation}

For the $\Delta$-value of the hexagonal phase, i.e.,
$\Delta=\sqrt{3}$, phase $\textrm{I\kern -0.3ex I}_{x=1/3}$ becomes
identical to phase $\textrm{H}$ pictured in the central panel of
Fig. \ref{fig:snapshots_Ix_H_IIx}: actually, assuming
$\Delta=\sqrt{3}$ in Eq. (\ref{energyIIx}) and recalling that the
Madelung constant of the hexagonal lattice is given by
$c=c(\sqrt{3})$, one indeed recovers the energy
$E_{\textrm{I}_{x=1/3}}$ specified in Eq. (\ref{energyIx}), i.e., the
energy of the phase \textrm{H}; these considerations provide a check
for the internal consistency of the formalism.

Since the projected positions of the particles of both layers form an
almost perfect hexagonal lattice both in phases $\textrm{I}_x$ and
$\textrm{I\kern -0.3ex I}_x$, these structures can be characterized by
the criterion $0.9 < \Psi_6^{(3)}$ (see Table \ref{tab:table_order}).
Structures $\textrm{I}_x$ and $\textrm{I\kern -0.3ex I}_x$ are
complementary from the point of view of the occupation parameter $x$
which is constrained to $0 < x < 1/3$ for phases $\textrm{I}_x$ and to
$1/3<x<1/2$ for phases $\textrm{I\kern -0.3ex I}_x$.  Finally, due to
the complexity of the emerging structures within the family of phases
the $\textrm{I\kern -0.3ex I}_x$, the last phase that could be taken
into account within the analytical treatment corresponds to the
sequence [br][bbr][bbr]; it is characterized by $x=3/8$.

\section{Structures emerging for large $\eta$: $\textrm{V}_x$}
\label{sec:largeeta}

If the asymmetry parameter $A$ is small, the prefactor in
Eq. (\ref{poten}) is large and the transition from phase \textrm{I} to
another competitive phase occurs at larger $\eta_c$; to be more
specific, this particular scenario occurs for $A \lesssim 0.408$, or
equivalently for $\eta \gtrsim 0.450$.  At large distances between the
plates, a particle moving from plate 1 to plate 2 can ``loose'' the
information about its Wigner monolayer position in phase \textrm{I}
and can create, together with all the other displaced particles, a
completely new, energetically favorable structure.  Since local
deformations of the sublattices $\alpha$ and $\beta$ are now
substantial for large $\eta_c$ the analytical approach is no longer
trustworthy, as it cannot grasp the distortions; this refers
especially to the identification of the order of phase transition
which should rather be investigated by numerical tools.

\subsection{Phase $\textrm{V}_x$}

Starting again off from the monolayer structure \textrm{I}, an
alternative strategy to populate layer 2 is realized below $A \simeq
0.408$.
The emerging family of structures is termed $\textrm{V}_x$, as it
maintains many of the characteristic features of structure \textrm{V}
which is the lowest-energy phase for symmetrically charged plates at
sufficiently large distances $\eta$ (see Subsection
\ref{subsec:symmetric_case}). We mention that one way to describe the
emergence of this structure was discussed in Ref. \cite{Ant14} via the
so-called ``in-branch'', i.e., by approaching one single charge from
infinite distance to a perfectly ordered hexagonal monolayer of
charges.

In its idealized version (amenable to the analytical treatment), phase
$\textrm{V}_x$ consists of two hexagonal structures, sublattice
$\alpha$ (spacing $a$) at plate 1 with $N_1=(1-x)N$ particles and
sublattice $\beta$ (spacing $b$) at plate 2 with $N_2= x N$ particles;
the sublattices are shifted with respect to one another in such a way
that all particles of $\beta$, when projected to plate 1, are located
in the center of some of the triangles of sublattice $\alpha$.  The
spacings of the sublattices are related by the equality
\begin{equation}
\frac{S}{N} = \frac{\sqrt{3}}{2} a^2 (1-x) = \frac{\sqrt{3}}{2} b^2 x .
\end{equation}  
Similarly to the case of phase $\textrm{I}_x$, joining arbitrary two
vertices of sublattice $\alpha$ implies a ``commensurate'' spacing of
sublattice $\beta$, so the sublattice spacings $a$ and $b$ are constrained by
\begin{equation} 
\frac{a^2}{b^2} = \frac{1}{j^2+jk+k^2} =\frac{n_2}{n_1} .
\end{equation} 
Now all possible integer values are allowed for $(j,k)$, except for
$(0,0)$; thus we obtain
\begin{equation} \label{eq:discretevaluesV}
x = \frac{n_2}{n_1+n_2} = \frac{1}{1+ j^2+jk+k^2}, \qquad
x\in \left\{ \frac{1}{2},\frac{1}{4},\frac{1}{5},\frac{1}{8},\frac{1}{10} 
\ldots \right\} .  
\end{equation} 

In general, both commensurate and incommensurate versions of structure
$\textrm{V}_x$ exist, as can be seen from the examples given in
Fig. \ref{fig:snapshots_Vx}.  As the two substructures $\alpha$ and
$\beta$ are characterized by two (perfect or non-ideal) hexagonal
lattices located at each of the plates, we can identify phase
\textrm{V}$_x$ via the BOOPs $0.9 < (1-x) \Psi_6^{(1)}+x \Psi_6^{(2)}$
(see also Table \ref{tab:table_order}).

The panels of Fig. \ref{fig:snapshots_Vx} show representative
snapshots of structure $\textrm{V}_x$ for three selected state
points. These panels confirm that in many -- but not all --
realizations of this phase, particles in layer 2 are positioned above
the centers of triangles in layer 1; further, slight distortions are
commonly encountered for intermediate values of $\eta$. For large
$\eta$-values the two layers form essentially uncorrelated hexagonal
lattices.
	
\begin{figure}[htbp]
\begin{center}
\includegraphics[width=15cm]{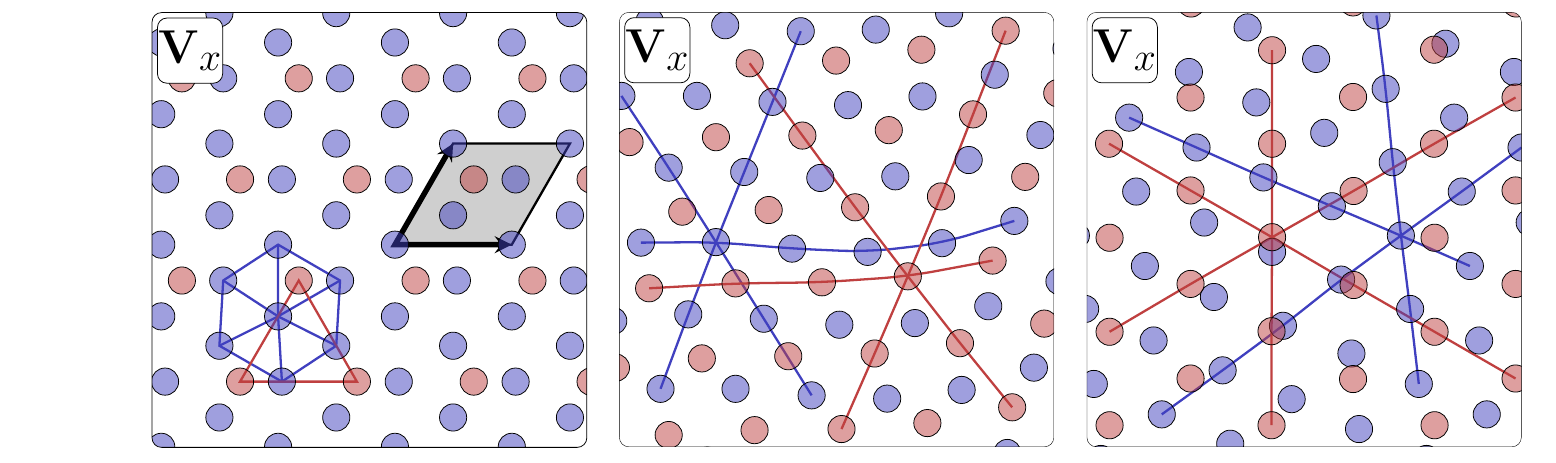} \\
\caption{(color online) Representative snapshots of structure
  $\textrm{V}_x$ (see text), with same graphical convention as in
  Fig. \ref{fig:snapshots_Ix_H_IIx}.
  structural features in layers 1 and 2,respectively.  Left panel:
  structure $\textrm{V}_x$ for $\eta=0.707$ and $A=0.5$, with $x=1/4$;
  this $x$-value allows for perfect commensurability of the two
  sublattices.  Center panel: structure $\textrm{V}_x$ for $\eta=0.7$
  and $A=0.9$, with $x=13/28 \simeq 0.464$.  Note the wave-like
  modulation of the hexagonal sublattices (as emphasized by the red
  and the blue lines). Right panel: structure $\textrm{V}_x$ for
  $\eta=1.061$ and $A=0.7$, with $x=12/31 \simeq 0.387$.  For large
  values of $\eta$, there is essentially no correlation between the
  two layers, leading to a Moir\'e-type pattern \cite{moire}.}
\label{fig:snapshots_Vx}
\end{center}
\end{figure}

Within the idealized assumption of the analytic approach, each
particle of sublattice $\beta$, when projected onto plate 1, is
located in the center of a triangle of sublattice $\alpha$ and
therefore sees the same relative array of lattice-$\alpha$ sites,
i.e., particles of $\beta$ have topologically equivalent positions.
Note that this is no longer true for $\alpha$-particles which group
into more topologically non-equivalent sets.  When calculating the
interaction energy between particles on sublattice $\alpha$ and
particles on sublattice $\beta$, it is advantageous to evaluate the
{\em full} interaction energy of one $\beta$-particle with all
$\alpha$-particles and then simply multiply the result by $N_2$.
Using the summation techniques developed in Ref. \cite{Samaj12_2} and
the formula (\ref{resequation}),
we obtain the energy of phase $\textrm{V}_x$ in the form
\begin{equation} \label{phaseVx}
\frac{E_{\textrm{V}_x}(\eta,A;x)}{N e^2 \sqrt{\sigma_1+\sigma_2}} = 
2^{3/2} \pi \eta \left( x - \frac{A}{1+A} \right)^2 
+ c \left[ (1-x)^{3/2} + x^{3/2} \right] + J(x,\eta) ,
\end{equation}
where
\begin{eqnarray}    \label{J_x_eta}
J(x,\eta) & = & x\sqrt{1-x} \frac{1}{2^{3/2}\sqrt{\pi}}
\int_0^{\infty} \frac{{\rm d}t}{\sqrt{t}} 
\left[ - {\rm e}^{-t\eta^2(1-x)} + \sqrt{3} {\rm e}^{-3 t\eta^2(1-x)} \right]
\nonumber \\ & & \times
\left\{ \left[ \theta_3({\rm e}^{-\sqrt{3}t}) \theta_3({\rm e}^{-t/\sqrt{3}}) 
- 1 - \frac{\pi}{t} \right] + \left[ \theta_2({\rm e}^{-\sqrt{3}t}) 
\theta_2({\rm e}^{-t/\sqrt{3}}) -\frac{\pi}{t} \right] \right\} . \label{integral}
\end{eqnarray}

The first term on the rhs of Eq. (\ref{phaseVx}) is the excess energy
due to the non-neutrality of each of the plates, the second term
corresponds to the neutralized intra-layer sums of hexagonal
structures within plate 1 and within plate 2; finally, the integral
$J(x,\eta)$ describes the interlayer interaction between
electro-neutral plates 1 and 2.  For the special case $x = 0$,
Eq. (\ref{phaseVx}) in combination with relation (\ref{J_x_eta}),
reduces to the energy of phase \textrm{I}, specified in
Eq. (\ref{energyI}). The series representation of the energy
difference $[E_{\textrm{V}_x}(\eta,A;x)-E_{\textrm{I}}(\eta,A)]$,
suitable for numerical calculations, is presented in
Eq. (\ref{energydif}) of Appendix \ref{app:B}.

Strictly speaking, the energy formula (\ref{phaseVx}) was derived for
an idealized phase $\textrm{V}_x$ with commensurate discrete values of
$x$ given by Eq. (\ref{eq:discretevaluesV}). However, we ignore
henceforward the discreteness of $x$ and apply the formula given in
Eq. (\ref{phaseVx}) also to {\it continuous} values of $x$.

\subsection{Transition \textrm{I} $\to$ \textrm{V}$_x$}

The analytical approach, based on the comparison of the energies
(\ref{energyI}) and (\ref{phaseVx}) of the competing phases \textrm{I}
and \textrm{V}$_x$, predicts a transition line \textrm{I} $\to$
$\textrm{V}_x$ which is in a very good agreement with the EA estimate,
as shown in Fig.  \ref{fig:diagram_of_states_rgb}. However,
within the analytic approach, the phase transitions are found to be
{\it discontinuous} (i.e., of first order), accompanied by a
discontinuous change of $x$ from 0 to a small, non-vanishing value at
the transition point; this scenario differs from the previously
discussed second-order transitions between phases \textrm{I} and
$\textrm{I}_x$. In contrast, our numerical results (based on the EA)
provide evidence that the transition between phases \textrm{I} and
$\textrm{V}_x$ is {\it continuous} as well, with constant values of
critical exponents along the whole critical line $\eta_c(A)$.  These
values coincide with those obtained for the transition \textrm{I}
$\to$ $\textrm{I}_x$ (see Subsection
\ref{subsec:transition_I_Ix}). Hence we conclude that the neglect of
local lattice distortions within the analytical treatment is a
non-adequate simplification of the problem.

For the special value $A=0.38$, the left panel of
Fig. \ref{fig:systems_wigner_5x} shows the analytical and EA estimates
of the curves $x(\eta)$, along which we identify only structures
\textrm{I} and $\textrm{V}_x$.  While we know that some preferred,
discrete values of $x$ exist for structure $\textrm{V}_x$ [see
  Eq. (\ref{eq:discretevaluesV})], the $x(\eta)$-curve turns out to be
basically continuous and smooth.  This is due to the fact that at
these transitions $\eta$ assumes rather large values, leading to a
relatively weak effective interaction between the two layers such that
the commensurability of the lattice spacings is no longer crucial.
Interestingly, for this value of $A$, $x(\eta)$ appears to converge
only very slowly towards the asymptotic value $x^*$ -- see
Eq. (\ref{x_neutr}).  As mentioned before, deformations in the
sublattices are rather pronounced which manifests itself in the
visible difference in $\eta_{\rm c}$ between the analytic and the
numerical approaches.  The right panel of
Fig. \ref{fig:systems_wigner_5x} emphasizes the bahaviour close to
$\eta_{\rm c}(A)$. A critical exponent $\beta=2/3$ ensues, as for the
transition \textrm{I} $\to$ $\textrm{I}_x$, addressed in
Fig. \ref{fig:systems_wigner_1x}.

\begin{figure}[htbp]
\begin{center}
\includegraphics[width=8cm]{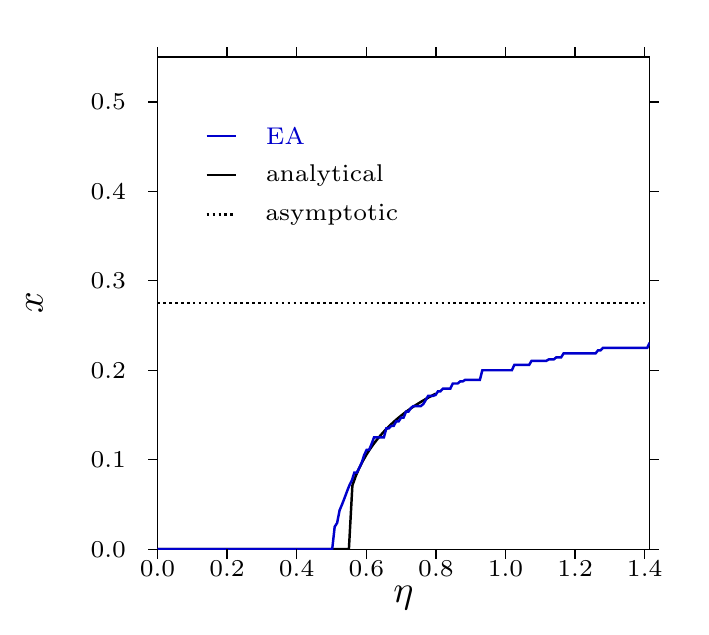}
\hspace{0.0cm}
\includegraphics[width=8cm]{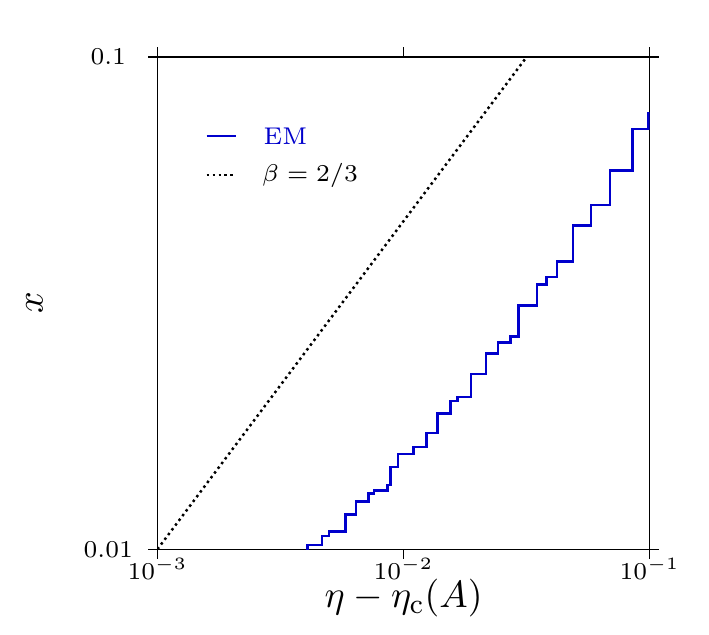}
\caption{(color online) Plot of $x(\eta)$ for $A=0.38$.  Left panel:
  $x(\eta)$; we identify structures \textrm{I} $(x=0)$ and
  $\textrm{V}_x$ $(x>0)$.  EA results are in blue, analytical results
  in black.  As $\eta_c$ is rather large, the two layers are
  correlated only weakly, leading to a smooth curve $x(\eta)$.  The
  asymptotic value $x^*$ (see Eq. (\ref{x_neutr})) is indicated by a
  dotted line.  Note the visible difference in $\eta_{\rm c}$ between
  the two approaches; this difference is due to deformations in the
  two sublattices, which are rather pronounced in structure
  $\textrm{V}_x$ close to the bi-critical point (see also panels of
  Fig. \ref{fig:diagram_of_states_rgb}).  Right panel:
  double-logarithmic plot of $x$ versus $(\eta-\eta_{\rm c}(A))$.  The
  dotted black line is a guide to the eye, indicating a critical
  exponent $\beta=2/3$.}
\label{fig:systems_wigner_5x}
\end{center}
\end{figure}

In Fig. \ref{fig:diagram_of_states_rgb}, we show the regions where the
monolayer structure \textrm{I} competes with the bilayer structures
$\textrm{I}_x$ and $\textrm{V}_x$.  The bi-critical point (with index
'bi'), where these three stability regions meet, was calculated within
the EM approach, with the result $\eta_{\rm bi} \simeq 0.470$, $A_{\rm
  bi} \simeq 0.4085$, and within the EA approach (using smaller cell
sizes than in the EM approach, i.e, up to $N=40$ particles per cell),
leading to $\eta_{\rm bi} \simeq 0.477$, $A_{\rm bi} \simeq 0.4075$;
this point is shown in the panels of Fig.
\ref{fig:diagram_of_states_rgb} by the white circled asterisk.  Close
to the bi-critical point, deformations in structures $\textrm{I}_x$
and $\textrm{V}_x$ are most pronounced: (i) compared to the
$\eta$-values where structure $\textrm{I}_x$ is stable, $\eta_{\rm
  bi}$ represents now a rather large value, causing the holes that are
left by those particles that moved to layer 2 to contract
significantly; (ii) in contrast, for $\eta$-values where structure
\textrm{V}$_x$ is stable, $\eta_{\rm bi}$ can be considered to be
small and the triangles in layer 1 surrounding particles in layer 2
are distorted significantly.  Since in either of the two cases the
respective distortions are neglected within the analytical approach,
the boundaries separating structure \textrm{I} and structures
$\textrm{I}_x$ and $\textrm{V}_x$ (as predicted by the numerical
approach), respectively, differ noticeably in the neighborhood of the
bi-critical point (see panels of
Fig. \ref{fig:diagram_of_states_rgb}).  In contrast, agreement of the
analytical and numerical approaches is found to be excellent both for
small and large $\eta$-values, where lattice distortions are small.

\subsection{Large-distance behavior of phase $\textrm{V}_x$}
\label{subsection:large-distance}

Numerical approaches have serious convergence problems when dealing
with two plates that are separated by large distances due to the fact
that the effective interaction energy of the plates is small.  On the
other hand, an analytical treatment of the large-distance
characteristics of phase $\textrm{V}_x$ is relatively simple; in fact
it becomes exact at asymptotically large values of $\eta$.

A saddle-point calculation presented in Appendix \ref{app:C} shows
that the integral $J(x,\eta)$ (\ref{integral}), which describes the
interaction energy between plates 1 and 2 (each of them begin neutral
as a whole) in Eq. (\ref{phaseVx}), behaves at large $\eta$ as follows
\begin{equation} \label{exponential}
J(x,\eta) \mathop{\sim}_{\eta\to\infty} - \frac{3^{5/4}}{\sqrt{2}} x \sqrt{1-x}
\exp\left( - \frac{4\pi\sqrt{1-x}}{3^{1/4}} \eta \right) . 
\end{equation}
In the symmetric case $x=1/2$, this result has already been obtained
in Ref. \cite{Samaj12_2}.  We emphasize that the exponential decay of
the interaction between two plates is not related to the hexagonal
structures on the plates, but holds more generally for any pair of
plates, each of which is as a whole electro-neutral.  We can therefore
neglect in the large-$\eta$ limit the inter-layer integral $J(x,\eta)$
in Eq. (\ref{phaseVx}) and consider only intra-layer interactions
(from which algebraic decay ensues, as will become clear below):
\begin{equation}
\frac{E_{\textrm{V}_x}(\eta;x)}{N e^2 \sqrt{\sigma_1+\sigma_2}} \simeq
2^{3/2} \pi \eta \left( x - \frac{A}{1+A} \right)^2 
+ c \left[ (1-x)^{3/2} + x^{3/2} \right] \qquad \mbox{for~large $\eta$.}
\end{equation}
One recognizes in this expression the same structure as invoked in
Ref. \cite{Messina01}. In the case of interest here, each plate is as
a whole not neutral ($x\neq x^*$). In the following we derive the
optimal occupation index $x$.  The energy minimization condition
\begin{equation}
\frac{\partial E_{\textrm{V}_x}(\eta;x)}{\partial x} = 0 = 2^{5/2}
\pi\eta \left( x - \frac{A}{1+A} \right) + \frac{3}{2} c \left(
\sqrt{x} - \sqrt{1-x} \right)
\end{equation}
implies the asymptotic behaviour for $\eta\to\infty$
\begin{equation} \label{eq:xlarged}
x \mathop{\sim}_{\eta\to\infty} x^* - \frac{3(-c)}{2^{7/2}\pi}
\frac{1-\sqrt{A}}{\sqrt{1+A}} \frac{1}{\eta} . 
\end{equation}

This relation proves that, as soon as $A\ne 1$, the plates (each as a
whole) remain charged up to infinite distance.  Since the Madelung
constant $c$ is a negative number, $x$ approaches to its asymptotic
``neutral'' value $x^*$ from below.  Note that the case $A=1$ is
specific in the sense that we always have $x=1/2$, irrespective of the
value of $\eta$.  For this case the energy of structure \textrm{V}
behaves as
\begin{equation}
\frac{E_{\textrm{V}}(\eta,A)}{N e^2 \sqrt{\sigma_1+\sigma_2}} 
\mathop{\sim}_{\eta\to\infty} 
c \left[ \left( \frac{1}{1+A} \right)^{3/2} + \left( \frac{A}{1+A} \right)^{3/2}
\right] - \frac{9 c^2}{2^{11/2}\pi} 
\frac{(1-\sqrt{A})^2}{1+A} \frac{1}{\eta} . 
\end{equation}
Finally one obtains
\begin{equation} \label{eq:EVlarged}
\frac{E_{\textrm{V}}(\eta,A)-E_{\textrm{V}}(\eta\to\infty,A)}{
N e^2 \sqrt{\sigma_1+\sigma_2}} \mathop{\sim}_{\eta\to\infty} 
- \frac{9 c^2}{2^{11/2}\pi} \frac{(1-\sqrt{A})^2}{1+A} \frac{1}{\eta} ,
\end{equation}
i.e., at large distances also the ground state energy approaches its
asymptotic value from below and the two plates attract each other.  As
one can see from Eq. (\ref{eq:EVlarged}), the asymptotic approach of
the lhs of Eq. (\ref{eq:EVlarged}) is rather slow (i.e., as $1/\eta$),
due to the non-neutrality of the plates, except for the symmetric
plates $A=1$ when the prefactor vanishes and one recovers an
exponential decay with distance \cite{GoPe96,Samaj12_1}.

For completeness, we also write the inter-plate pressure following
from the energy difference specified in relation (\ref{eq:EVlarged}),
now in terms of the unscaled distance $d$:
\begin{equation} \label{eq:Plarged}
P = -(\sigma_1+\sigma_2) \frac{\partial}{\partial d} \frac{E_{\textrm{V}}}{N} \sim 
- (\sigma_1+\sigma_2) e^2 \, \frac{9 c^2}{2^5\pi}
\frac{(1-\sqrt{A})^2}{1+A}  \frac{1}{d^2}.
\end{equation}
%
It should be emphasized that this equation holds except for $A=0$:
indeed, when plate 2 is neutral ($\sigma_2=0$), phase \textrm{I} is
stable for any interplate separation $\eta$ (see the discussion of
limiting cases discussed in Subsection \ref{subsec:model}) and $P=0$.
In other words, we face two non-commuting limits:
\begin{equation}
\lim_{d\to\infty}\lim_{A \to 0} \, d^2 P = 0 , \qquad
\lim_{A \to 0}\lim_{d\to\infty} \, d^2 P \neq 0.
\end{equation}

%
%

Further we learn from the panels of
Fig. \ref{fig:diagram_of_states_rgb} that for special $x$-values the
respective structures are able to extend over larger $\eta$-ranges,
leading to the characteristic stripe pattern in the $\Psi$RGB color
schemes shown in of Fig. \ref{fig:diagram_of_states_rgb}.  There is,
however, a {\it representative} region in the $(\eta, A)$-plane where
a different mechanism appears to be at work: for $A \gtrsim 0.7$, the
$\Psi$RGB-color schemes provide evidence of regions where the colors
change smoothly.  A closer look at the corresponding snapshots reveals
a wave-like modulation of the hexagonal sub-lattices in the two layers
of the respective \textrm{V}$_x$ structures, allowing for an optimized
correlation between the lattices in the two layers without
significantly decreasing the hexagonal order of either of the layers
or preventing a favorable value of $x$ (see center panel of Figure
\ref{fig:snapshots_Vx}).  In contrast, a different mechanism is at
work for structures $\textrm{V}_x$ for $A \lesssim 0.7$ and large
$\eta$-values, where the two sublattices are essentially uncorrelated
hexagonal particle arrangements, creating thereby a Moir\'e-type
pattern \cite{moire} (see right panel of Fig. \ref{fig:snapshots_Vx}).

\section{Structures emerging at intermediate $\eta$: \textrm{I\kern -0.3ex I}, 
\textrm{I\kern -0.3ex I\kern -0.3ex I}, \textrm{I\kern -0.3 ex V}; 
$\textrm{S}_{\bm 1}$, $\textrm{S}_{\bm 2}$; $\textrm{P}_{\bm 1}$, 
$\textrm{P}_{\bm 2}$ and $\textrm{P}_{\bm 3}$}
\label{sec:intermediateeta}

\subsection{Structures \textrm{I\kern -0.3ex I}, 
\textrm{I\kern -0.3ex I\kern -0.3ex I}, and \textrm{I\kern -0.3 ex V};
overcharging}
\label{subsec:II_III_IV_overcharging}

We now return to the structures \textrm{I\kern -0.3ex I},
\textrm{I\kern -0.3ex I\kern -0.3ex I}, and \textrm{I\kern -0.3ex V},
identified as the ground state structures in the {\it symmetric}
setup, and investigate their role in the diagram of states as charge
asymmetry is introduced. Surprisingly we have found that these
structures do display a significant role for $A$-values down to $\sim
0.9$ (see the orange, red, and cyan regions in the panels of Fig.
\ref{fig:diagram_of_states_rgb}).

In the asymmetric case, the analytical results show that the two
layers remain charged up to arbitrarily large plate separations
$\eta$.  In an overwhelming portion of the $(\eta, A)$-plane layer 1
carries more point charges than required to compensate for the
neutralizing background, leading therefore to a negative net charge on
this layer (corresponding to $x < x^*$); consequently layer 2 is
``underpopulated'' by particles and it carries a positive net charge,
the so-called {\it undercharging} scenario. Here, at a fixed value
$A$, $x(\eta)$ tends monotonically increasing towards the limiting
value $x^*(A) ( < 1/2)$ -- see Fig. \ref{fig:systems_wigner_5x}.

However, since the three symmetric structures \textrm{I\kern -0.3ex
  I}, \textrm{I\kern -0.3ex I\kern -0.3ex I} and \textrm{I\kern -0.3ex
  V} are characterized by $x = 1/2 > x^*(A)$, their appearance for $A
\gtrsim 0.9$ implies that now layer 2 has to carry a negative net
charge, the so-called {\it overcharging} scenario. As can be seen in
Fig. \ref{fig:systems_wigner_1x}, $x(\eta)$ is now for a fixed value
of $A$ a non-monotonous function, which exceeds in the $\eta$-range
where the overcharging scenario takes place the threshold value
$x^*(A)$ and then tends towards this limiting value ``from below''.

For the {\it symmetric} case (i.e., $A=1$), a rigorous analysis of
second-order transitions \textrm{I\kern -0.3ex I} $\to$ \textrm{I\kern
  -0.3ex I\kern -0.3ex I} and \textrm{I\kern -0.3ex I\kern -0.3ex I}
$\to$ \textrm{I\kern -0.3 ex V} has been presented
Refs. \cite{Samaj12_1,Samaj12_2}.  These transitions belong to the
Landau family with the mean-field value of the critical index
$\beta=1/2$.  The analysis can be readily extended to the {\it
  asymmetric} case (i.e., $A < 1$) and leads to the same result,
namely $\beta=1/2$.  Consequently and rather noteworthy, fixing the
asymmetry parameter $A$ at an arbitrary value within the interval
$[0.9,1]$ and changing continuously the distance $\eta$ from 0 to
$\infty$, at least two different kinds of second-order phase
transitions take place: the first one (\textrm{I} $\to$
$\textrm{I}_x$) is characterized by the non-classical critical index
$\beta=2/3$, while the other ones (i.e., \textrm{I\kern -0.3ex I}
$\to$ \textrm{I\kern -0.3ex I\kern -0.3ex I} and \textrm{I\kern -0.3ex
  I\kern -0.3ex I} $\to$ \textrm{I\kern -0.3 ex V)} are of mean-field
type with $\beta=1/2$.

\subsection{Snub square structures $\textrm{S}_1$ and $\textrm{S}_2$}

In addition to the honeycomb phase \textrm{H} (see
Sec. \ref{sec:smalleta}), we have identified within the EA approach
two further structures that are characterized by $x=1/3$; these phases
occupy a substantial region for intermediate $\eta$-values in
Fig. \ref{fig:diagram_of_states_rgb}.  Due to the specific features of
their lattices that are reminiscent of the ideal snub square lattice
(see, for instance, \cite{Ant11} and references therein), we denote
them as snub square structures, $\textrm{S}_1$ and $\textrm{S}_2$.
Representative snapshots are shown in two panels of Figure
\ref{fig:snapshots_H_S2_S1}.  Structure $\textrm{S}_1$ is essentially
amenable to an analytical calculation, unlike $\textrm{S}_2$ that is
too complex.  The two structures are characterized by the following
geometric features:

\begin{itemize}
\itemsep=0pt
\item[$\bullet$] Structure $\textrm{S}_1$ consists of a slightly
  distorted snub square particle arrangement in layer 1, built up by
  squares and equilateral triangles.  Particles in layer 2 are
  positioned above the centers of the squares in layer 1, thus forming
  a square lattice (see right panel of Fig.
  \ref{fig:snapshots_H_S2_S1}).  This structure can be characterized
  by $x=1/3$, $0.9 < \Psi_5^{(1)}$ and $0.9 < \Psi_4^{(2)}$;
\item[$\bullet$] Structure $\textrm{S}_2$ consists of a strongly
  distorted snub square particle arrangement in layer 2 and pentagonal
  structural units in layer 1 (see center panel of Fig.
  \ref{fig:snapshots_H_S2_S1}).  The structure can be characterized by
  $x=1/3$ and $0.45 < \Psi_5^{(2)}$.
\end{itemize}

\begin{figure}[htbp]
\begin{center}
\includegraphics[width=15cm]{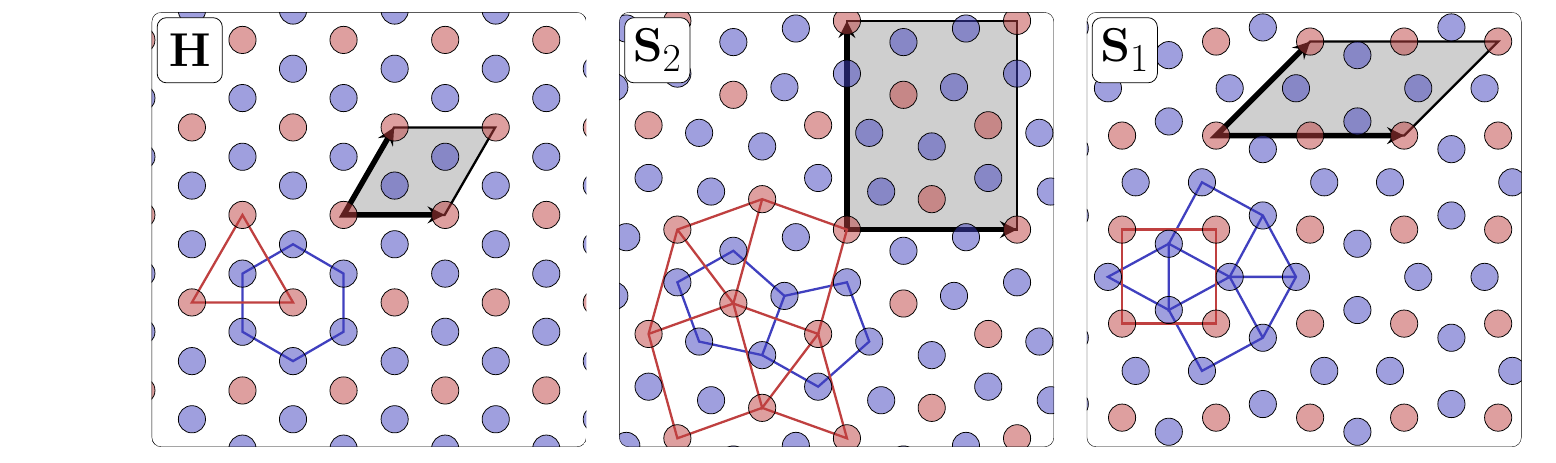} \\
\caption{(color online) Representative snapshots of structures
  \textrm{H}, $\textrm{S}_2$, and $\textrm{S}_1$ (see text), with the
  same graphical conventions as earlier.
Left panel: structure \textrm{H} for $\eta=0.198$ and $A=0.85$.
Center panel: structure $\textrm{S}_2$ for $\eta=0.417$ and $A=0.7$.
Right panel: structure $\textrm{S}_1$ for $\eta=0.622$ and $A=0.675$.
In all cases $x = 1/3$.}
\label{fig:snapshots_H_S2_S1}
\end{center}
\end{figure}

The reason why snub square lattices lead to significant values of the
five-fold BOOPs $\Psi_5^{(1)}$ and $\Psi_5^{(2)}$ is related to the
angles that are required for the formation of such a lattice
(considering, in particular, its idealized version).  As in all of the
Archimedean tilings \cite{Gruenbaum87,Ant11}, each vertex of the snub
square tiling (represented by a particle) has exactly the same
geometrical surrounding.  Since the particular sequence of polygons
which characterize the vertices of the (ideal) snub square tiling is
specified by [$\triangle - \Box - \triangle - \Box - \triangle$], the
required bond angles are given by (see dotted black lines in Fig.
\ref{fig:snub_square_schematic}):
\begin{equation}
\phi_{\rm snub-square} = \left\{ 
\begin{array}{l} 0 , \cr \displaystyle\pm \pi/3 
, \cr \pm \displaystyle 5 \pi/6 \simeq \pm 0.833 \pi .
\end{array} \right. 
\end{equation}
The values of these angles turn out be very close to the bond-angles
of a perfect {\it pentagonal} surrounding of the tagged particle (see
dotted green lines in Fig. \ref{fig:snub_square_schematic}), namely
\begin{equation}
\phi_{\rm pentagonal} = \left\{ 
\begin{array}{l} 0 , \cr \pm \displaystyle 2 \pi/5 , \cr 
\pm \displaystyle 4 \pi/5 = \pm 0.800 \pi . 
\end{array} \right.
\end{equation}

\begin{figure}[htbp]
\begin{center}
\includegraphics[width=5cm]{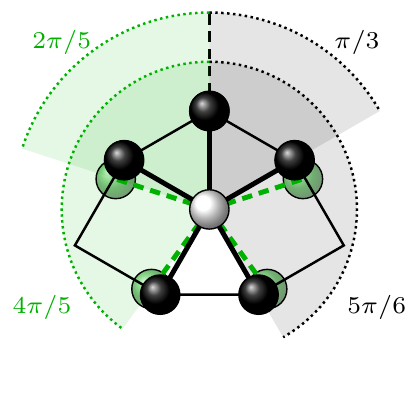}
\caption{(color online) Schematic view of the immediate neighborhood
  of a tagged particle (in white) of the ideal snub square tiling,
  formed by the neighboring black particles.  Proceeding clockwise
  from the top, the vertices are formed by the following sequence of
  surrounding polygons: $\triangle - \Box - \triangle - \Box -
  \triangle$.  The respective values of the bond angles (indicated by
  the grey shaded sectors, limited by the dotted black lines) are
  close to the ones of a perfect pentagonal arrangement (indicated by
  the green particles, the green shaded sectors, limited by the dotted
  green lines).}
\label{fig:snub_square_schematic}
\end{center}
\end{figure}
			
In the analytical approach, structure $\textrm{S}_1$ is assumed to
have a perfect snub square lattice in layer 1 and a perfect square
arrangement in layer 2 (see Fig. \ref{fig:Snub}).  Even though the
ensuing number of particles per cell (namely $N = 6$) is relatively
small for the numerical calculations, this value hits the limit for
the analytical approach. Here, the snub square phase is constructed by
projecting red particles on plate 2 (which there form a square lattice
of side $a$) onto plate 1, which is occupied by blue particles.  The
resulting unit cell of spatial extent $(2a) \times (2a)$ contains for
this structure eight blue and four red particles, so that indeed
$x=4/(4+8)=1/3$.

The relative positions of particles in layer 1 with respect to the
square lattice (defined by particles in layer 2) is quantified via the
parameter $\varepsilon$ (see Fig. \ref{fig:Snub}).  The value of
$\varepsilon$ is fixed by the requirement that the distance between
particles 1 and 2 is equal to the distance between particles 1 and 3
(see Fig. \ref{fig:Snub}), i.e.,
\begin{equation}
(a-2\varepsilon)^2 = \left( \frac{a}{2}+\varepsilon \right)^2
+ \left( \frac{a}{2}-\varepsilon \right)^2 .
\end{equation} 
This equation implies that
\begin{equation}
\varepsilon = a \left( 1 - \frac{\sqrt{3}}{2} \right) .
\end{equation}
Eventually, the value of square lattice spacing $a$ follows from the
electro-neutrality condition:
\begin{equation}
a = \sqrt{\frac{3}{\sigma_1+\sigma_2}} = \sqrt{\frac{3}{n_1+n_2}} .
\end{equation}

The positions of the particles on plate 2 on the square lattice can be
simply enumerated as the multiples of $a$ in terms of integers $j$ and
$k$, i.e., $(ja, ka)$. On the other hand, the positions of the
particles on plate 1 can be generated from eight basic positions in an
elementary cell of spatial extent $(2a)\times (2a)$ (see
Fig. \ref{fig:Snub}): (i) $(a/2,\varepsilon)$, (ii)
$(a/2,a-\varepsilon)$, (iii) $(a+\varepsilon,a/2)$, (iv)
$(2a-\varepsilon,a/2)$, (v) $(\varepsilon,-a/2)$, (vi)
$(a-\varepsilon,-a/2)$, (vii) $(3a/2,-\varepsilon)$, and (viii)
$(3a/2,-a+\varepsilon)$; all other positions of particles in layer 1
are obtained by shifting these positions in both spatial directions,
i.e., by adding $(2aj, 2ak)$ with $j$ and $k$ being integers.

\begin{figure}[ht]
\begin{center}
\includegraphics[width=0.2\textwidth,clip]{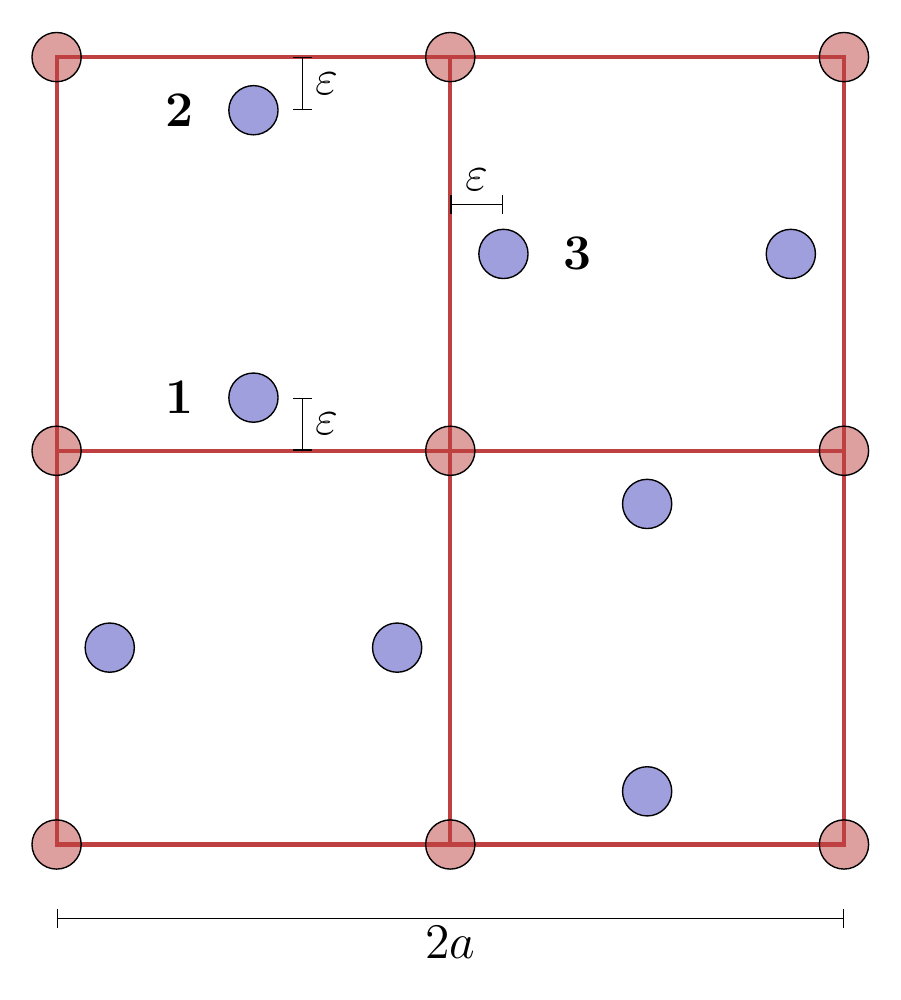}
\caption{Schematic view of an extended cell for the
  individual sublayers of spatial extent $(2a \times 2a)$ for the
  ideal snub square phase. Particles in layer 1 are colored blue,
  particles in layer 2 are colored red. For the definition of
  $\varepsilon$ and the particle labels 1, 2, and 3 see text.}
\label{fig:Snub}
\end{center}
\end{figure}

Using translation and reflection symmetries for this particular
lattice in the lattice sums, it can be shown that the total energy is
given by
\begin{eqnarray}
\frac{E_{\rm S_1}(\eta)}{N e^2\sqrt{\sigma_1+\sigma_2}} & = & 
\frac{1}{3^{3/2}} \Bigg\{ \sum_{j,k\atop (j,k)\ne (0,0)} \frac{1}{\sqrt{j^2+k^2}}
+ \frac{1}{2} \sum_{j,k} \frac{1}{\sqrt{(j+1/2)^2+(k+1/2)^2}} \nonumber \\
& & + \frac{1}{2} \sum_{j,k} \frac{1}{\sqrt{j^2+[k+(1+\sqrt{3})/2]^2}} 
+ \frac{1}{2} \sum_{j,k} \frac{1}{\sqrt{(j+1/2)^2+(k+\sqrt{3}/2)^2}}
\nonumber \\ & &
+ \sum_{j,k} \frac{1}{\sqrt{[j-(\sqrt{3}+1)/4]^2+[k+(\sqrt{3}-1)/4]^2}}
\nonumber \\ & &
+ \sum_{j,k} \frac{1}{\sqrt{[j-(\sqrt{3}-1)/4]^2+[k+(\sqrt{3}+1)/4]^2}}
\nonumber \\ & &
+ \sum_{j,k} \frac{1}{\sqrt{(j+1/2)^2+(k+\sqrt{3}/2)^2+2\eta^2/3}} \Bigg\} 
- \mbox{backgr.} \label{energysnub1}
\end{eqnarray}

Poisson summation formula, Eq. (\ref{PSF}) allows to express the
lattice summations as quickly convergent series of generalized Misra
functions, Eq.  (\ref{specialf}).  For an example of an explicit
series representation, we choose the last term in
Eq. (\ref{energysnub1}) which is the only term in this relation that
depends on the plate distance $\eta$:
\begin{eqnarray}
& & \sum_{j,k} \frac{1}{\sqrt{(j+1/2)^2+(k+\sqrt{3}/2)^2+2\eta^2/3}} - 
\mbox{backgr.} \phantom{aaaaaaaaaaaaaaaaaaaaaa} \nonumber \\ 
& = & \frac{1}{\sqrt{\pi}} \int_0^{\infty} \frac{{\rm d}t}{\sqrt{t}}
{\rm e}^{-2\eta^2 t/3} \left[ \theta_2({\rm e}^{-t}) 
\sum_k {\rm e}^{-(k+\sqrt{3}/2)^2 t} - \frac{\pi}{t} \right] \nonumber \\
& = & \frac{1}{\sqrt{\pi}} \Bigg\{
2 \sum_{j=1}^{\infty} [\cos(\pi\sqrt{3}j)+(-1)^j] z_{3/2}[2(\pi\eta)^2/3,j^2]
+ 4 \sum_{j,k=1}^{\infty} (-1)^j \cos(\pi\sqrt{3}k)  
z_{3/2}[2(\pi\eta)^2/3,j^2+k^2] \nonumber \\ 
& + & 2 \sum_{j=1}^{\infty} \sum_{k=-\infty}^{\infty} 
z_{3/2}[0,2\eta^2/3+(j-1/2)^2+(k+\sqrt{3}/2)^2] - \pi z_{1/2}(0,2\eta^2/3)
\Bigg\} .
\end{eqnarray}

In doing so, the energy of the (ideal) snub square phase can be
calculated within the analytic approach rather easily and in an
efficient manner.

The situation is considerably more complicated for structure
$\textrm{S}_2$: here a simplified, yet faithful approximation of this
structure requires $N = 12$ particles per unit cell, making thus in
practice an analytical treatment of this particular phase is currently
out of reach.

\subsection{Pentagonal structures $\textrm{P}_1$, $\textrm{P}_2$
and $\textrm{P}_3$}
\label{subsec:pentagonal}

The ordered structures that populate those regions of the $(\eta,
A)$-plane that have not been discussed so far are characterized by
complicated geometries and symmetry features which prevent them from
being amenable to the analytical framework. As $\eta$ and $A$ are
varied, these structures change their structural features
continuously, rendering a classification of distinct structures
difficult. Here we could rely on the support provided by the combined
analysis of different sets of BOOPs.

\begin{figure}[htbp]
\begin{center}
\includegraphics[width=15cm]{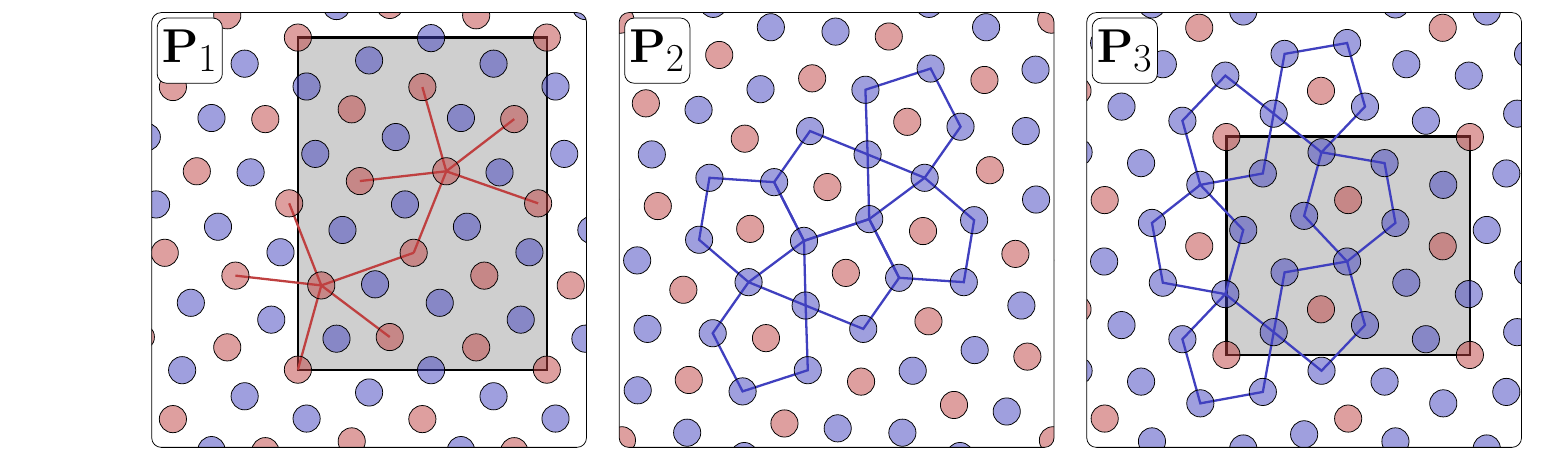}
\caption{(color online) Representative snapshots of structures with
  pentagonal features, labeled as structures $\textrm{P}_1$,
  $\textrm{P}_2$, and $\textrm{P}_3$ (see text).  They appear as
  contiguously colored regions in the $\Psi$RGB-phase diagrams (see
  panels of Fig. \ref{fig:diagram_of_states_rgb}).
  The unit cell is again indicated by the shaded area.
  Left panel: structure $\textrm{P}_1$ for $\eta=0.381$ and $A=0.85$,
  with $x = 12/28 \simeq 0.429$.  Center panel: structure
  $\textrm{P}_2$ for $\eta=0.346$ and $A=0.825$, with $x=3/8=0.375$.
  Right panel: structure $\textrm{P}_3$ for $\eta=0.410$ and $A=0.6$,
  with $x=4/18 \simeq 0.222$.}
\label{fig:snapshots_pent}
\end{center}
\end{figure}

We have put particular focus on structures with pentagonal features
due to their potential importance with respect to formation of
quasicrystals.  In the diagram of states (cf. panels of Fig.
\ref{fig:diagram_of_states_rgb}) we have highlighted three regions of
stability of structures that are characterized by a local pentagonal
order: in particular, they are characterized by a greenish color in
the RGB presentations of $\Psi^{(4)}$, indicating thus the dominance
of a local pentagonal order.  The corresponding representative
snapshots are shown in the panels of Fig. \ref{fig:snapshots_pent}:
Structures $\textrm{P}_1$ and $\textrm{P}_2$ appear for $x > 1/3$,
where they compete with structures \textrm{I\kern -0.3ex I\kern -0.3ex
  I} and \textrm{H}.  Structure $\textrm{P}_3$ is characterized by
$x<1/3$ and competes strongly with structure $\textrm{I}_x$:

\begin{itemize}
\item Structure $\textrm{P}_1$ is characterized by a complicated
  particle arrangement and exhibits a pronounced five-fold symmetry in
  layer 2 (see left panel of Fig.  \ref{fig:snapshots_pent}).  We can
  characterize phase $\textrm{P}_1$ via $1/3<x<1/2$ and $0.45 <
  \Psi_5^{(2)}$.
\item The characteristic feature of structure $\textrm{P}_2$ is the
  large number of pentagonal holes in layer 1: particles of layer 2
  occupy positions above the centers of the pentagons in layer 1 (see
  center panel of Fig. \ref{fig:snapshots_pent}).  These particles of
  layer 2 form a rather well-defined hexagonal lattice.  We can
  characterize phase $\textrm{P}_2$ via $1/3<x<1/2$ and $0.9 <
  \Psi_5^{(4)}$.
\item In a similar manner, structure $\textrm{P}_3$ consists of a
  large number of pentagonal holes in layer 1, albeit with a much
  lower density (see right panel of Fig.  \ref{fig:snapshots_pent}).
  The EA results imply that the region of stability of structure
  $\textrm{P}_3$ also reaches the bi-critical point (marked by a
  circled asterisk in the panels of
  Fig. \ref{fig:diagram_of_states_rgb}).  This might be an indication
  that structure $\textrm{P}_3$ represents a transitory phase between
  structures $\textrm{I}_x$ and $\textrm{V}_x$.  However, even our
  very fine resolution in the parameters $\eta$ and $A$ within the EA
  approach is not of sufficient quality for a closer study of this
  phenomenon; in addition, the complicated geometry of structure
  $\textrm{P}_3$ precludes a more accurate analytical investigation.
  We have characterized phase $\textrm{P}_3$ via $0<x<1/3$ and $0.45 <
  \Psi_5^{(4)}$.
\end{itemize}



\section{Results at finite temperature: Monte Carlo simulations}
\label{sec:results_finite}

\begin{figure}
\includegraphics[width=5.5in]{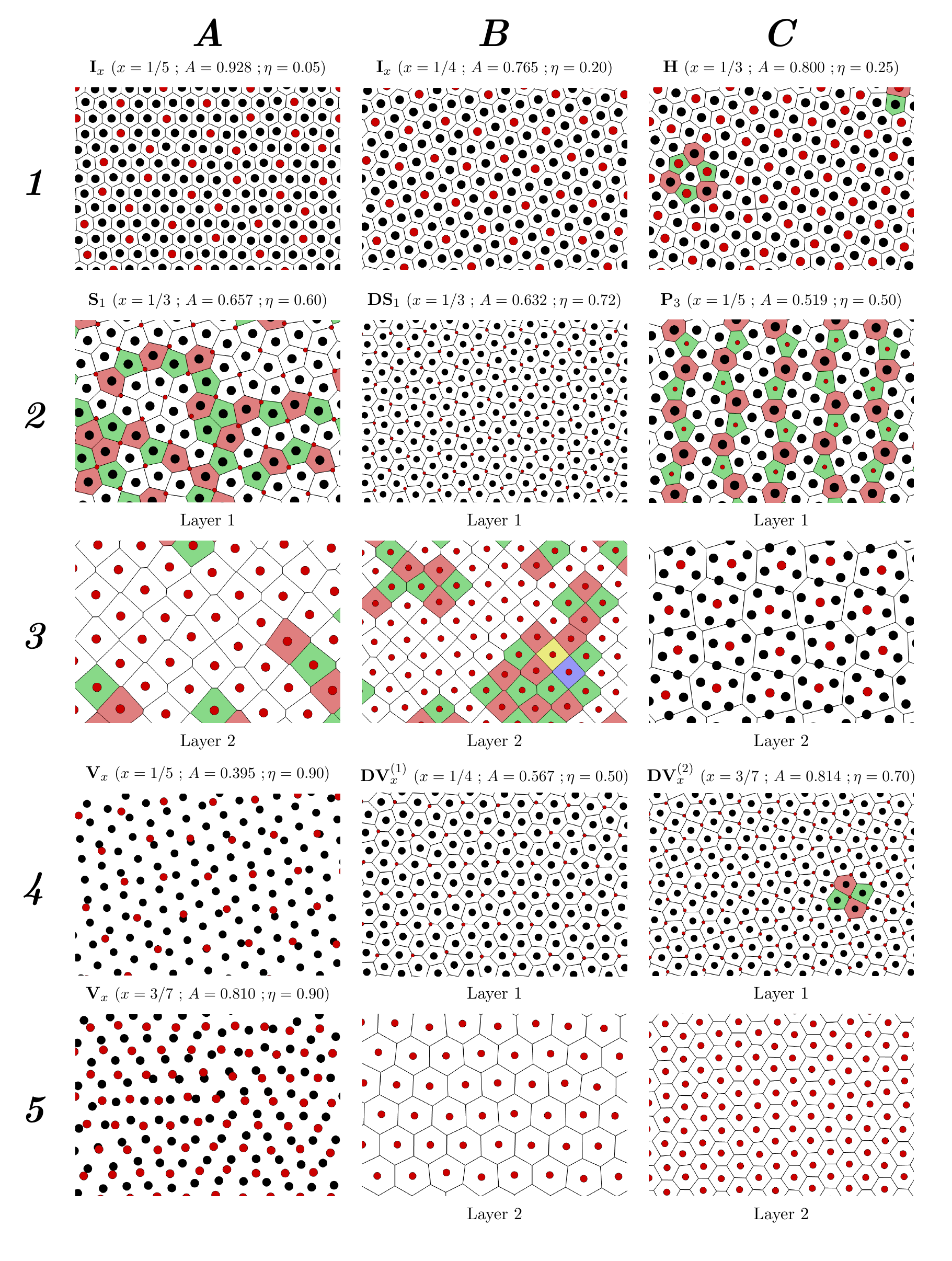}
\caption{(color online) Snapshots of the principal ordered states
  obtained in MC simulations along the four domains of constant
  $x$-values (specified in the text); the respective values of $x$,
  $A$, and $\eta$ are indicated for each of the snapshots
  (panels). Particles in layer 1 are indicated in black, while
  particles in layer 2 are colored in red. For each of the snapshots
  the respective Voronoi construction is indicated.  The color code
  for the Voronoi cells is the following (color and number of edges):
  yellow (four), green (five), white (six), red (seven), and blue
  (eight). For phases \textrm{I}$_x$ and \textrm{H}, the Voronoi
  constructions have been done with all particles projected onto one
  plane. Panels are addressed in the text by specifying the column (A
  to C) and the row (1 to 5).}
\label{fig:selected_snapshots}
\end{figure}

This section focuses on our comprehensive study of the thermal
stability of those ground state configurations that were predicted for
vanishing temperature. These investigations, performed via MC
simulations at small, but finite temperatures had to be restricted --
as a consequence of the high computational costs -- to selected state
points and relevant regions of the diagram of states.  Relevant
pathways in the $(\eta, A)$-plane were selected, which are
characterized -- according to the EA predictions -- by constant values
of $x$ \cite{polynomials}. From the seven respective domains
highlighted in Fig. \ref{fig:runs-MC}, we have chosen for the
subsequent discussion the ones for $x = $ 3/7, 1/3, 1/4, and 1/5. As
can be seen, these pathways cross the regions of stability of several
phases. On each of these pathways, the red triangle in
Fig. \ref{fig:runs-MC} represents the state point which was used as
initial configuration, taken from the EA calculation. The subsequent
MC runs are then carried out along the lines indicated, for state
points located along these pathways (marked by colored symbols).
Fig. \ref{fig:selected_snapshots} displays enlarged snapshots of
particle configurations as obtained along these pathways; they will be
addressed in the following discussions. Additional structures are
shown in Appendix \ref{appendix:MC}.

The conclusions drawn from our extensive MC simulation are the
following:
\begin{itemize}
\item structural data extracted agree remarkably well with the
  corresponding predictions of the EA based investigations; this holds
  in particular for those state points in the $(\eta, A)$-plane, which
  are located well within regions of stability of the different
  structures. This excellent agreement is observed even on a
  quantitative level, confirmed both by the values of the respective
  BOOPs as well as by the values of $x$, i.e. the quantity which
  specifies the particle population on the two planes;
\item in the case that these pathways cross limits of stability of
  competing structures, MC results are able to reflect these
  competitions faithfully: by a careful structural analysis based on
  Voronoi constructions, features of the involved competing structures
  could be identified.
\end{itemize}

Our MC based observations are detailed in the following, focusing on
the above mentioned four domains in the $(\eta, A)$-plane,
characterized by a constant $x$-value.

\subsection{The domain $x=3/7$}
\label{subsec:x=3/7}

\begin{figure}[htbp]
\begin{center}
\includegraphics[width=8.cm]{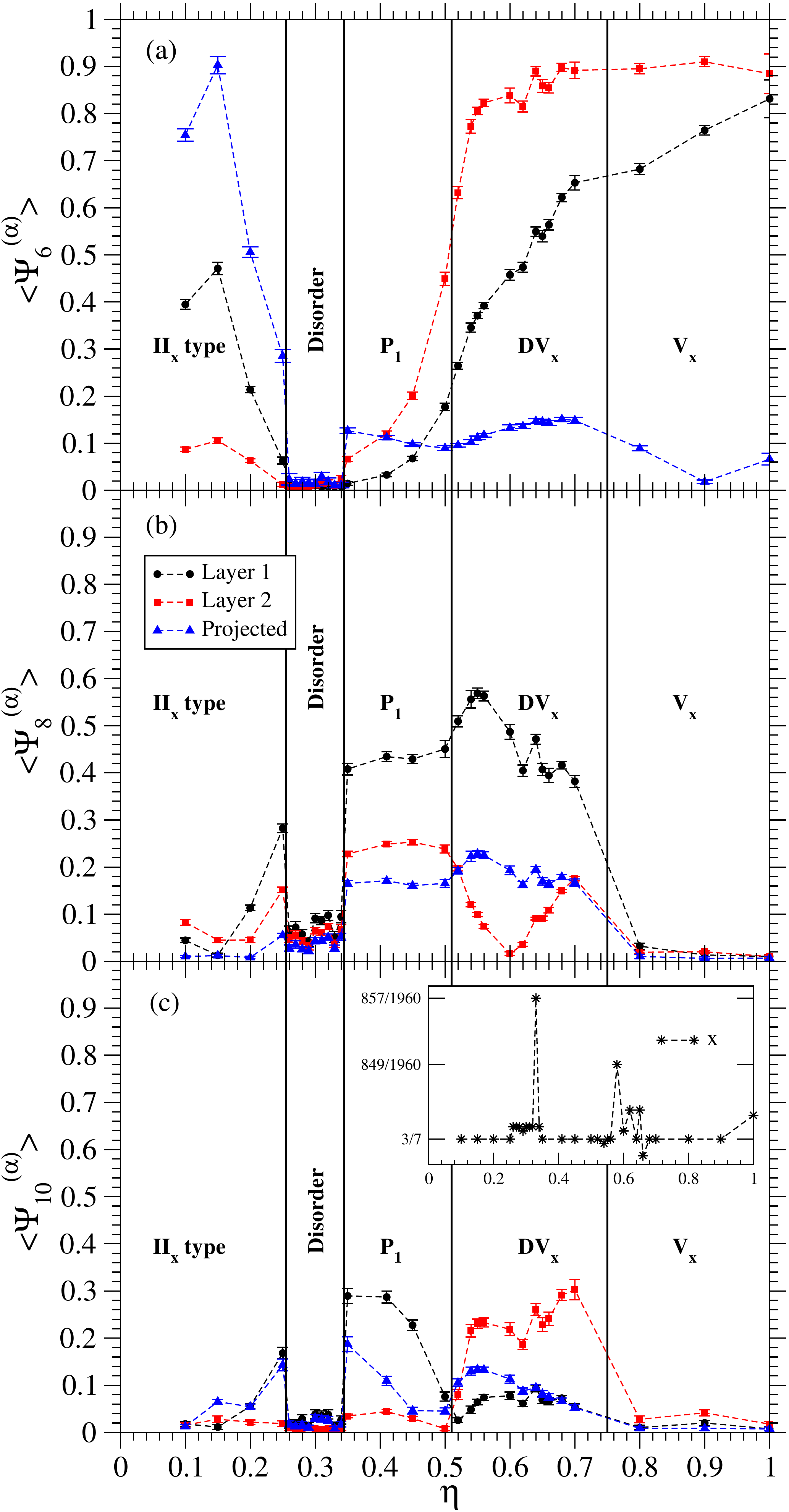}
\caption{(color online) Bond orientational order parameters computed
  in MC simulations for runs within the domain $x = 3/7$ as functions
  of the dimensionless distance $\eta$: (a) $\langle \Psi_6^{(\alpha)}
  \rangle$, (b) $\langle \Psi_8^{(\alpha)} \rangle$, and (c) $\langle
  \Psi_{10}^{(\alpha)} \rangle$; results obtained for the different
  layers ($\alpha$ =1, 2, or 3) are colored according to the
  labels. In the inset of panel (c), the value of $x=N_2/N$ is shown.
  Occurring ordered (and disordered) structures are labeled.}
\label{fig:Boop-3d7}
\end{center}
\end{figure}

\begin{figure}[htbp]
\begin{center}
\includegraphics[width=7.5cm]{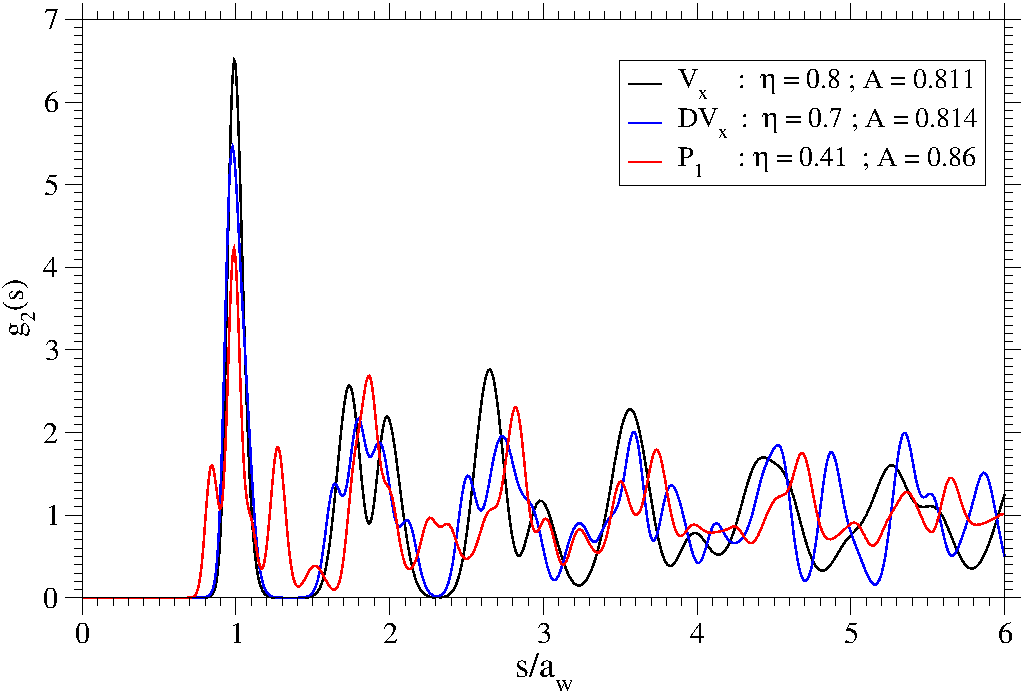}
\caption{(color online) Center-to-center pair correlation functions as
  obtained from MC simulations along the domain $x =3/7$. $g_{2}(s)$
  for the following three phases: \textrm{V}$_x$, \textrm{DV}$_x$, and
  \textrm{P}$_1$ (for state points as labeled in the inset). The
  length used to rescale distances is $a_{\rm w}= (2/\sqrt{3}
  n_2)^{1/2}$, $n_2$ being the density of the layer.
  }
\label{fig:gr-3d7}
\end{center}
\end{figure}

\begin{figure}[htbp]
\begin{center}
\includegraphics[width=7.3cm]{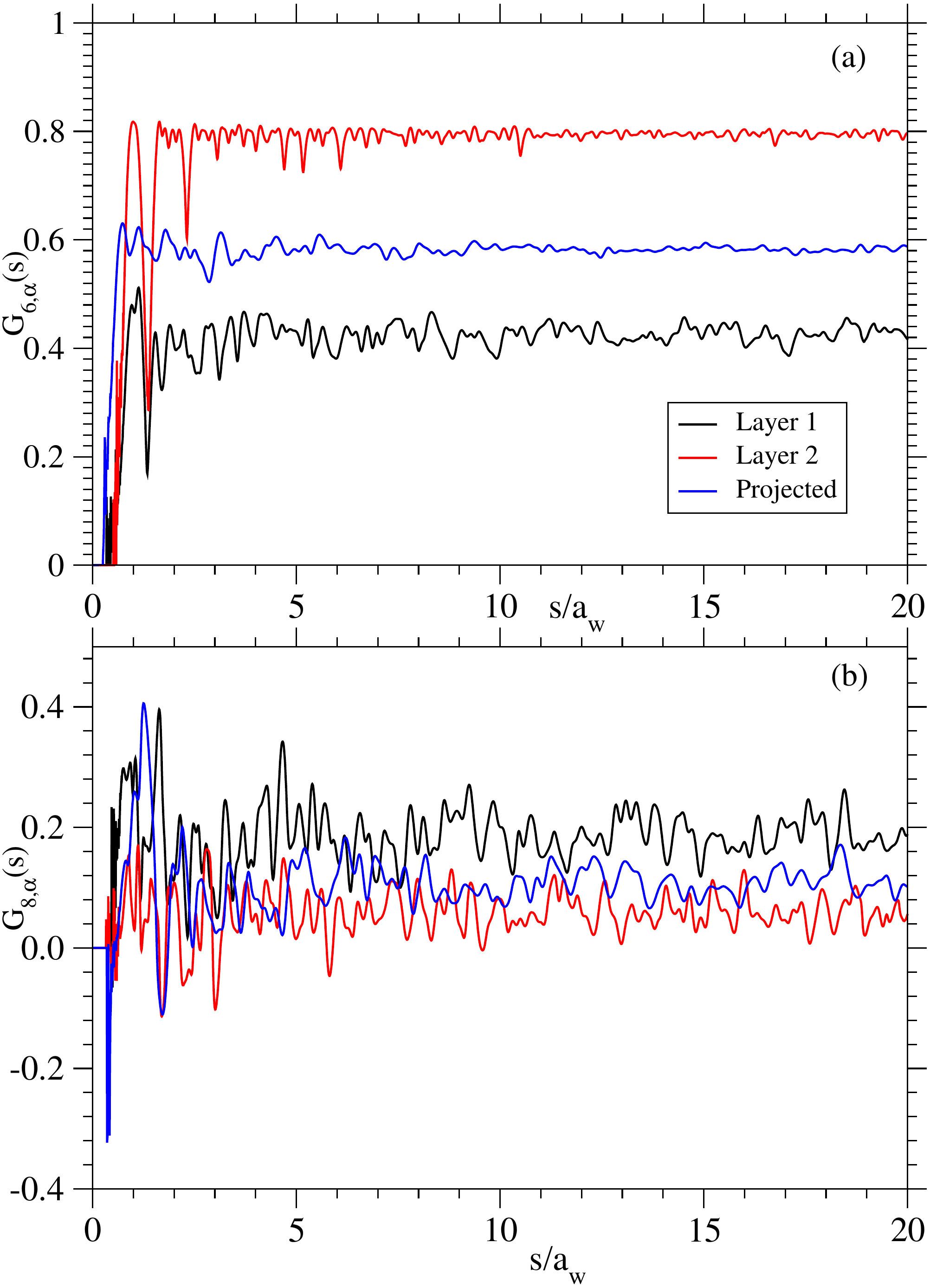}
\caption{(color online) Bond orientational correlation functions
  $G_{n, \alpha}(s)$ as defined in Eq. (\ref{Boop-gr}) with $n$ and
  $\alpha$ as labeled as obtained in our simulations along the domain
  $x =3/7$. (a) $G_{6, \alpha}(s)$ for the \textrm{DV}$_x$ phase
  ($\eta=0.7$, $A=0.814$) for layer 1, layer 2, and for all particles
  projected in the same plane (as labeled). (b) $G_{8, \alpha}(s)$ for
  the \textrm{P}$_1$ phase ($\eta = 0.41$, $A=0.86$) for layer 1,
  layer 2, and for all particles projected in the same plane (as
  labeled). The lengths used to rescale distances are $a_{i; {\rm w}}=
    (2/\sqrt{3} n_i)^{1/2}$, $n_i$ being the density of the respective
    layer.}
\label{fig:G-3d7}
\end{center}
\end{figure}

The initial configuration for MC simulations in the domain $x=3/7$ is
the configuration of the pentagonal phase \textrm{P}$_1$ as identified
in the EA investigations for $\eta=0.410$ and $A=0.86$
(cf. Fig. \ref{fig:runs-MC}); details of this configuration and the
function $A_{x=3/7}(\eta)$, that specifies the asymmetry parameter as
a function of $\eta$ in this domain are collected in Table
\ref{tab:path-MC} of Appendix \ref{appendix:MC}.  In
Fig. \ref{fig:Boop-3d7}, we represent the most significant order
parameters; the values of $x=N_2/N$, obtained in MC simulations, are
shown in the inset of panel (c).
The order parameters calculated along the curve $A_{x=3/7}(\eta)$
allow to identify five different phases with decreasing $\eta$: phases
\textrm{V}$_x$, \textrm{DV}$_x$, and \textrm{P}$_1$, a disordered
phase, and eventually phase \textrm{I\kern -0.3ex I}$_x$.

According to the criteria put forward in Table \ref{tab:table_order},
phases \textrm{V}$_x$ are found for $0.75 \lesssim \eta$.
An enlarged snapshot of this phase is shown in panel (A5) of
Fig. \ref{fig:selected_snapshots}; the panel displays a Moir\'e type
pattern composed of equilateral triangles.  In the range $0.51
\lesssim \eta \lesssim 0.75$, a phase emerges that we denote as the
distorted phase \textrm{V}$_x$ (i.e., phase \textrm{DV}$_x$): in this
$\eta$-interval, the six-fold order parameter reaches for both layers
still significant values, while $\langle \Psi^{(1)}_8 \rangle$ for
layer 1 and $\langle \Psi^{(2)}_{10} \rangle$ for layer 2 deviate
significantly from 0. Further, as can be seen from
Fig. \ref{fig:gr-3d7}, the correlation function $g_{2}(s)$ differs
(for the state point $\eta=0.7$ and $A = 0.814$) distinctively from
the correlation functions obtained for the \textrm{V}$_x$ phases
(depicted for the state point $\eta = 0.80$ and $A = 0.811$): the
second and third peaks are both split into two secondary peaks,
reflecting the distortion of the triangular lattice. In panels (C4)
and (C5) of Fig. \ref{fig:selected_snapshots}, we represent enlarged
snapshots of a typical \textrm{DV}$_x$ structures, along with the
related Voronoi constructions for both layers. Further details are
provided in Appendix \ref{appendix:MC}.
The long-range bond orientational function $G_{6, \alpha}(s)$ is shown
in panel (a) of Fig. \ref{fig:G-3d7}. In the subsequent $\eta$-range,
i.e., $0.345 \lesssim \eta \lesssim 0.51$ the pentagonal phase
\textrm{P}$_1$ is stable, in full agreement with the EA-based
analysis. The corresponding correlation function $g_{2}(s)$ is shown
in Fig. \ref{fig:gr-3d7}: again we find significant differences with
respect to the corresponding correlation functions of the previously
discussed phases \textrm{V}$_x$ and \textrm{DV}$_x$. The long-range
bond orientational correlation function $G_{8, \alpha}(s)$ is
displayed in panel (b) of Fig. \ref{fig:G-3d7}.

Within the interval $0.255 \lesssim \eta \lesssim 0.345$, no long-range 
order could be identified at all, as evidenced by the order
parameters displayed in Fig. \ref{fig:Boop-3d7}; likewise, the intra-
and inter-layer correlation functions exhibit no structure (not
shown).
We encounter here a disordered phase.
In this region, EA calculations indicate the close vicinity of the
boundaries that separate the regions of stability of several competing
phases, namely structures \textrm{H}, \textrm{I\kern -0.3ex I}$_x$,
\textrm{I\kern -0.3ex I}, \textrm{I\kern -0.3ex I \kern -0.3ex},
\textrm{P}$_1$, and \textrm{P}$_2$. We interpret the lack of
orientational and long-range order observed in the MC simulations at
this small temperature as a result of a strong competition between
these phases. The abrupt changes in the order parameters close to the
boundaries of this $\eta$-regime (i.e., for $\eta \sim 0.345$ and
$\eta \sim 0.255$; see Fig. \ref{fig:Boop-3d7}) are indications of the
coexistence between various phases and of first-order transitions
between them. A more quantitative statement on this issue requires
significantly larger computational efforts.

Eventually for $\eta \lesssim 0.255$, the Voronoi construction for all
particles projected onto the same plane (not displayed) indicates the
occurrence of a regular hexagonal tiling with thermal distortions,
which we identify as a \textrm{I\kern -0.3ex I}$_x$ type phase.

\subsection{The domain $x=1/3$}
\label{subsec:x=1/3}

\begin{figure}[htbp]
\begin{center}
\includegraphics[width=8.cm]{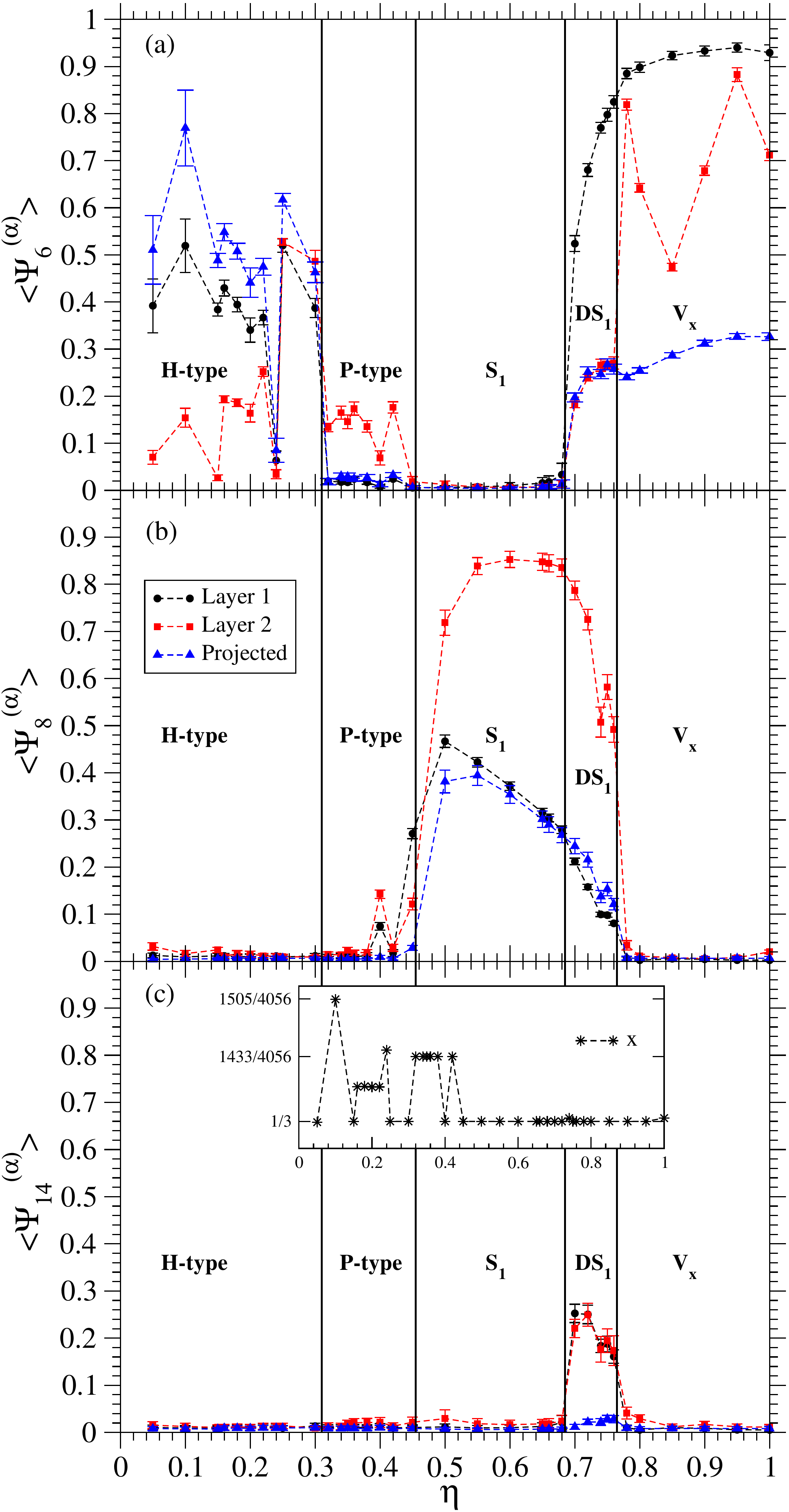}
\caption{(color online) Same as Fig. \ref{fig:Boop-3d7}, for the
  domain $x = 1/3$, and (a) $\langle \Psi_6^{(\alpha)} \rangle $, (b)
  $\langle \Psi_8^{(\alpha)} \rangle$, and (c) $\langle
  \Psi_{14}^{(\alpha)}\rangle$; $\alpha=$1, 2 or 3.
}
\label{fig:Boop-1d3}
\end{center}
\end{figure}

\begin{figure}[htbp]
\begin{center}
\includegraphics[width=8.5cm]{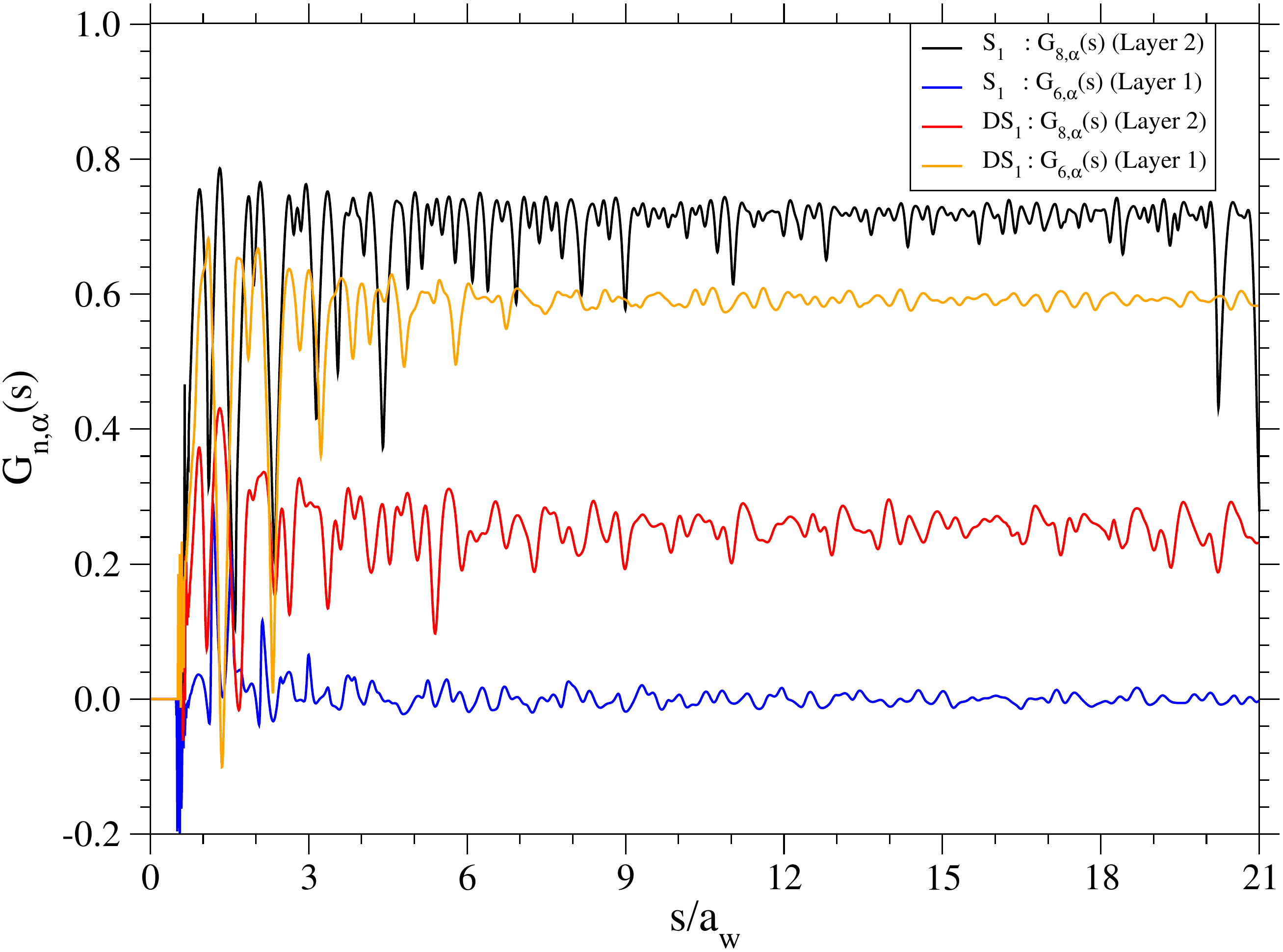}
\caption{Bond orientational correlation functions $G_{6, \alpha}(s)$
  and $G_{8, \alpha}(s)$ (defined in Eq. (\ref{Boop-gr})) as obtained
  in MC simulations along the domain $x = 1/3$ for phases
  \textrm{S}$_1$ ($\eta=0.65$ and $A=0.646$) and \textrm{DS}$_1$
  ($\eta=0.74$ and $A=0.629$), as labeled. The lengths used to rescale
  distances are $a_{i; {\rm w}}= (2/\sqrt{3} n_i)^{1/2}$, $n_i$ being
  the density of the respective layer.}
\label{fig:G-1d3}
\end{center}
\end{figure}

The initial configuration for MC simulations in the domain $x=1/3$ is
the snub square phase \textrm{S}$_1$ as identified in the EA
investigations for $\eta=0.629$ and $A=0.65$
(cf. Fig. \ref{fig:runs-MC}); details of this configuration and the
function $A_{x=1/3}(\eta)$, that specifies the asymmetry parameter as
a function of $\eta$ in this domain are collected in Table
\ref{tab:path-MC} of Appendix \ref{appendix:MC}. We note that within
the range $0.31 \lesssim \eta \lesssim 0.455$, the curve
$A_{x=1/3}(\eta)$ slightly passes through the domain $x=1/3$. In our
MC simulations we find in this $\eta$-range $x=1433/4056$ while EA
calculations predict $x=6/17$; the two values differ thus by $\sim$
0.1\%.

In Fig. \ref{fig:Boop-1d3}, we represent the most significant averaged
order parameters for the runs $x = 1/3$; the values of $x = N_2/N$ are
shown in the inset of panel (c).  The order parameters calculated
allow to identify five different phases with decreasing $\eta$:
\textrm{V}$_x$, \textrm{DS}$_1$, and \textrm{S}$_1$, as well as the
\textrm{P}-type and the \textrm{H}-type phases.  For $0.765 \lesssim
\eta$, phase \textrm{V}$_x$ is found to be stable.
In layer 2 the BOOP $\langle \Psi_6^{(2)} \rangle$ is significantly
smaller than one, which is due to different orientations of grains in
the triangular lattice.

As we decrease the distance between the layers below $\eta \sim
0.765$, structure \textrm{V}$_x$ transforms into a distorted structure
with order parameters $\langle \Psi^{(\alpha)}_6 \rangle$, $\langle
\Psi^{(\alpha)}_{10} \rangle$ (not shown), and $\langle
\Psi^{(\alpha)}_{14} \rangle$ that differ in their values from those
characteristic for phase \textrm{S}$_1$.  For $0.685 \lesssim \eta
\lesssim 0.765$, layer 2 has a rhombic structure, an enlarged snapshot
of this layer along with the respective Voronoi construction is
represented in panels (B2) and (B3) of
Fig. \ref{fig:selected_snapshots}. Particles in layer 1 form a
distorted triangular lattice and the Voronoi construction indicates a
structure similar to the one formed by isohedrally-tiled hexagons with
the symmetry group ${\it p31m}$ (see respective panel in
Fig. \ref{fig:selected_snapshots}). In this phase, the long-range
orientational order, expressed via the bond-orienational correlation
functions $G_{6, \alpha}(s)$ and $G_{8, \alpha}(s)$, distinctively
differs from the order of structure \textrm{S}$_1$, as demonstrated by
the results shown in Fig. \ref{fig:G-1d3}. Since the orientational
order for layer 2 resembles that of phase \textrm{S}$_1$, this phase
is termed distorted \textrm{S}$_1$ (i.e., \textrm{DS}$_1$). The abrupt
changes in the order parameters $\langle \Psi^{(\alpha)}_6 \rangle$
and $\langle \Psi^{(\alpha)}_8 \rangle$ for $\eta \simeq 0.76$ are
indications of a first-order transition between phases \textrm{DS}$_1$
and \textrm{V}$_x$; we note that for the {\it symmetric} bilayer, a
similar mechanism has been identified for the transition between
phases \textrm{I\kern -0.3ex V} and \textrm{V}, see
Refs. \cite{Weis:01,Mazars:08}.

The snub square phase \textrm{S}$_1$ (see panels (A2) and (A3) in
Fig. \ref{fig:selected_snapshots}) is found to be stable at our chosen
finite temperature for $0.455 \lesssim \eta \lesssim 0.685$. Further
structural details about this phase are shown in Fig. \ref{fig:G-1d3}
where the bond orientational correlation functions $G_{6, \alpha}(s)$
and $G_{8, \alpha}(s)$ are displayed. The Voronoi construction for
particles in layer 1 in the ground state configuration of this phase
leads to a so-called Cairo pentagonal tiling \cite{Gruenbaum87} with a
{\it p4g} symmetry. This also holds for layer 2, where we recover a
square regular tiling with a {\it p4m} symmetry. In
Ref. \cite{Leipold:15}, the distribution of the number of neighbors
for a square lattice under an infinitesimal perturbation of the
lattice position was computed; the numerical evaluation of these
analytical results yield: $p(4)=p(8)\simeq 0.044$, $p(5)=p(7)\simeq
0.2435$ and $p(6)\simeq 0.4249$. The corresponding values extracted
from our MC simulations are in good agreement with these predictions:
$p(4) = p(8) \simeq 0.01$, $p(5) = p(7) \simeq 0.22$.

As announced above, in the range $0.31 \lesssim \eta \lesssim 0.455$,
the $A_{x=1/3}(\eta)$ curve goes through the domain $x=1/3$. The
Voronoi construction for all particles projected onto the same plane
(not shown) indicates that the majority of the particles in layer 2
have five neighbors in layer 1. These observations fit very well with
the definitions of the pentagonal phases (\textrm{P}$_1$,
\textrm{P}$_2$ and \textrm{P}$_3$) as found in the EA investigations
-- see discussion in Subsection \ref{subsec:pentagonal}; however, we
were not able to identify the phases encountered in the MC simulations
with either of the three pentagonal phases, specified via the EA
approach.

Eventually, for $\eta \lesssim 0.31$ the EA investigations predict the
stability of phase \textrm{H}; within this $\eta$-range MC simulations
recover correctly this hexagonal phase, with only a few isolated
grains having different orientations. An enlarged snapshot taken from
the MC simulations is shown in panel (C1) of
Fig. \ref{fig:selected_snapshots}: the displayed configuration
(obtained for $\eta = 0.25$) exhibits the characteristic features of
the phase \textrm{H}.  As $\eta$ is further decreased along the line
$A_{x = 1/3}(\eta)$ (and in particular if $\eta \lesssim 0.15$), the
hexagonal structures formed by all particles is well preserved while
some of the specific features of the sublattices on the individual
layers do no longer correspond to the ideal features of phase
\textrm{H}, see also Appendix \ref{appendix:MC}.
It should be emphasized, that for such small $\eta$-values and at
finite temperatures, the difference between intralayer and interlayer
energies are too small to allow a precise observation of the ideal
ground state structure of phase \textrm{H} in MC simulations.

\subsection{The domain $x=1/4$}
\label{subsec:x=1/4}

\begin{figure}[htbp]
\begin{center}
\includegraphics[width=8.cm]{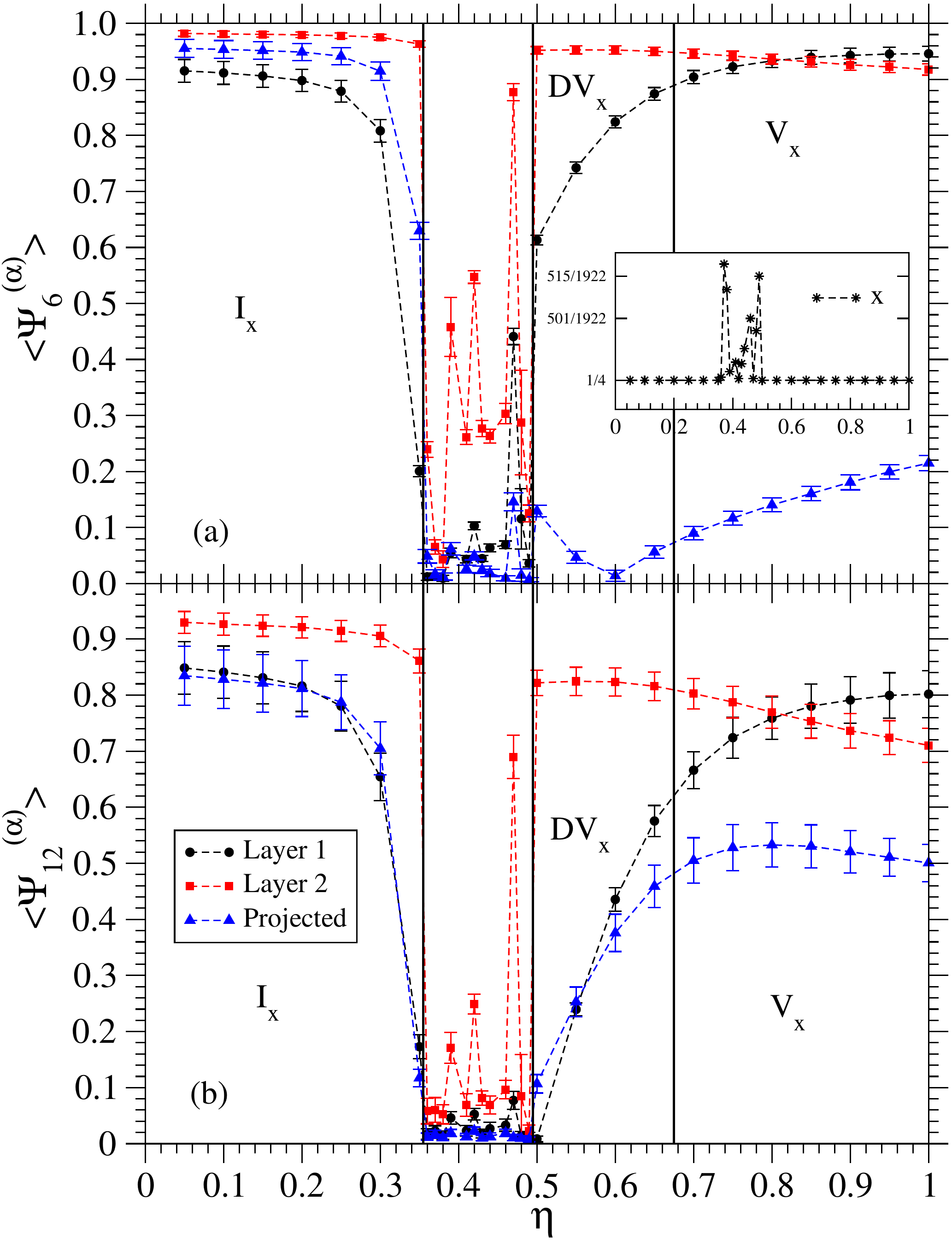}
\caption{(color online) Same as Fig. \ref{fig:Boop-3d7} for the
    domain $x = 1/4$: (a) $\langle \Psi_6^{(\alpha)} \rangle$ and (b)
    $\langle \Psi_{12}^{(\alpha)} \rangle$.
  }
\label{fig:Boop-1d4}
\end{center}
\end{figure}

The initial configuration for MC simulations in the domain $x=1/4$ is
the configuration of the phase \textrm{V}$_x$ as identified in the EA
investigations for $\eta=1.00$ and $A=0.45$
(cf. Fig. \ref{fig:runs-MC}); see also Table \ref{tab:path-MC} of
Appendix \ref{appendix:MC}.
Fig. \ref{fig:Boop-1d4} shows the order parameters which allow to
identify three different phases with decreasing $\eta$:
\textrm{V}$_x$, \textrm{DV}$_x$, and eventually phase
\textrm{I}$_x$. Between the two latter ones, there is an $\eta$-range
(i.e., for $0.32 \lesssim \eta \lesssim 0.49$), where no stable, 
long-range order could be identified; this observation can be related to
the fact that this section of the pathway corresponds to a region
along which several ordered structures coexist
(cf. Figs. \ref{fig:runs-MC} and \ref{fig:diagram_of_states_rgb}.)

Within the domain $x=1/4$, the phase \textrm{V}$_x$ occurs for $0.67
\lesssim \eta$, which is built up by commensurate triangular
sublattices in layers 1 and 2; consequently the long-range
orientational orders in both layers are particularly stable as the
spacing between the layers decreases. The order parameters $\langle
\Psi_{6p}^{(\alpha)} \rangle$ ($p$ being a positive integer) reache
rather large values (see Fig.  \ref{fig:Boop-1d4}) and the limiting
case embodied in Eq. (\ref{Boop-gr-lim}) is well fulfilled. In
contrast, all other order parameters $\langle \Psi_n^{(\alpha)}
\rangle$ with $n \neq 6p$ vanish. As a consequence of the
commensurability of the structures on the two sublattices (which can
be observed at even large distances between layers, i.e. for $\eta
\gtrsim 0.8$), the interlayer correlation function $g_3(s)$ exhibits
strong correlations for long distances $s$; this fact is in striking
contrast to the observations made for other \textrm{V}$_x$ phases
(i.e., for other values of $x$).

Decreasing the distance $\eta$, we observe that the hexagonal order in
layer 2 is still well preserved for $\eta$-values down to $\simeq
0.5$; in contrast, the hexagonal order in layer 1 is rather strongly
distorted for $0.67 \lesssim \eta \lesssim 0.7$. The snapshots in
panels (B4) and (B5) of Fig. \ref{fig:selected_snapshots} reveal the
distortions of the hexagonal cells in layer 1 via the related Voronoi
constructions: these distortions have the same vertex topology as the
regular hexagonal tiling, but with three different orientations of the
distorted hexagons similar to the monohedral convex hexagonal tilings
with a {\it p3} symmetry. Because of these continuous distortions of
the lattice in layer 1 in this $\eta$-range, we coin this structure
again as a distorted \textrm{V}$_x$ phase (\textrm{DV}$_x$).

For $0.32 \lesssim \eta \lesssim 0.49$, no ordered structures with
$x=1/4$ could be identified in the EA-based investigations. To gain a
better understanding of the transition \textrm{V}$_x$ $\rightarrow$
\textrm{I}$_x$ (which we identify for $\eta \lesssim 0.355$) within
the domain $x = 1/4$, simulations have been carried out within this
particular $\eta$-interval. As shown in the inset of
Fig. \ref{fig:Boop-1d4}-(a), the extracted values of $x$ deviate in
this $\eta$-range significantly from the ideal value of $x=1/4$, an
observation similar to the one made in the EA-based
investigations. However, no significant stable and long-range order
could be identified within the interval $0.32 \lesssim \eta \lesssim
0.49$; instead, we have observed some metastabilities of the structure
\textrm{DV}$_x$ and some grains of phase I$_{x=1/4}$.

For $\eta \lesssim 0.355$ we find -- in full agreement with the EA
computations (see panel (B1) in Fig. \ref{fig:selected_snapshots} and
panel (a) of Fig. \ref{fig:Kagome-1d4} in Appendix \ref{appendix:MC}),
-- that phase \textrm{I}$_x$ is stable. As shown in
Fig. \ref{fig:Kagome-1d4}-(a), the Kagom{\'e} lattice for particles in
layer 1 and the triangular lattice for particles in layer 2 are
perfectly recovered.

\subsection{The domain $x=1/5$}
\label{subsec:x=1/5}

\begin{figure}[htbp]
\begin{center}
\includegraphics[width=8.cm]{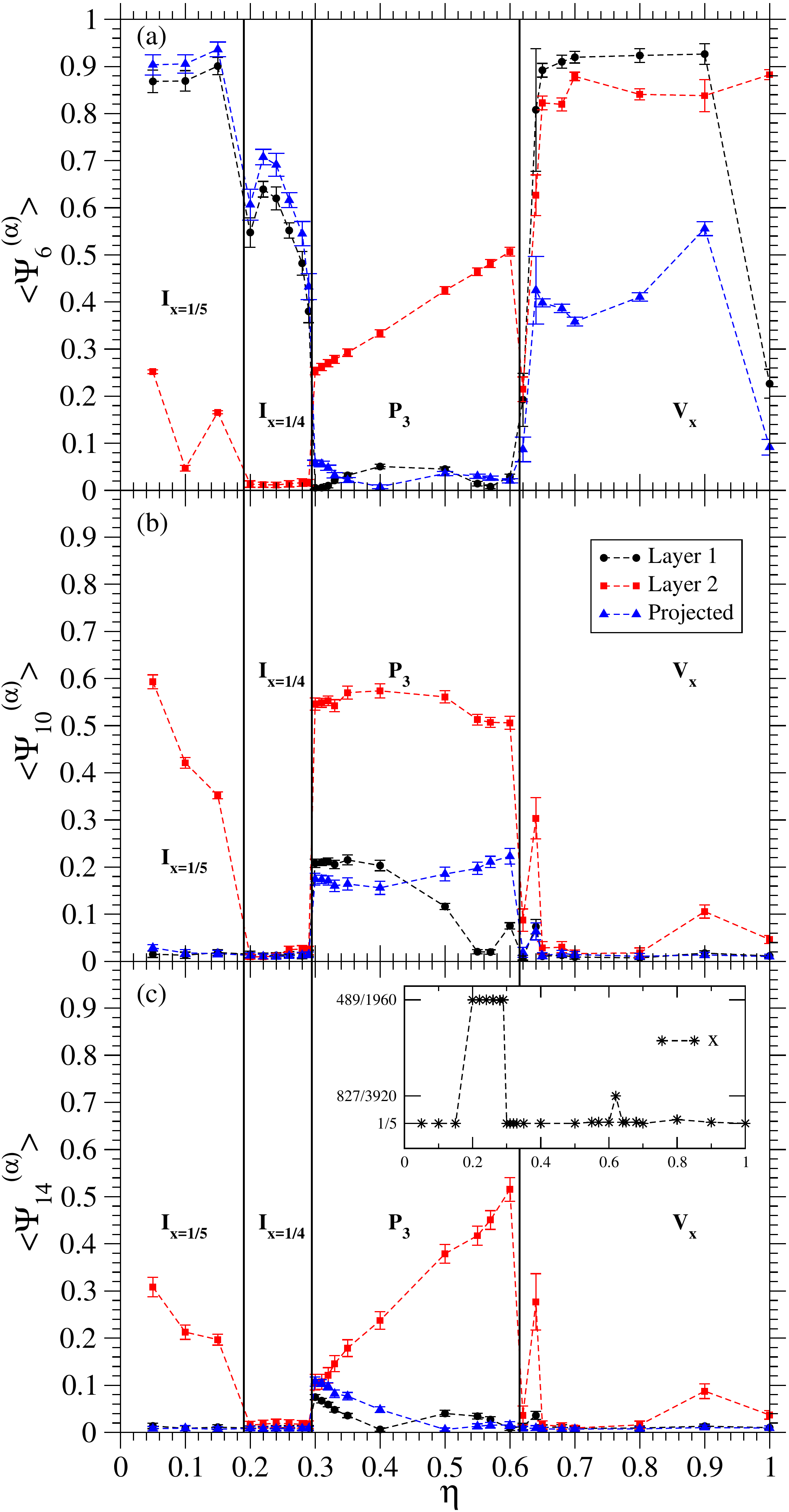}
\caption{(color online) Same as Fig. \ref{fig:Boop-3d7}, for $x =
  1/5$: (a) $\langle \Psi_6^{(\alpha)} \rangle$ , (b) $\langle
  \Psi_{10}^{(\alpha)} \rangle$, and (c) $\langle \Psi_{14}^{(\alpha)}
  \rangle$.
  }
\label{fig:Boop-1d5}
\end{center}
\end{figure}

Here, the ``seed'' for MC simulations is the pentagonal phase
\textrm{P}$_3$ as identified with the EA for $\eta=0.410$ and $A=0.86$
(see details in Table \ref{tab:path-MC} of Appendix \ref{appendix:MC}.
Fig. \ref{fig:Boop-1d5} allows us to identify four different phases
with decreasing $\eta$: \textrm{V}$_x$, \textrm{P}$_3$,
\textrm{I}$_{x=1/4}$, and \textrm{I}$_{x=1/5}$.  Phase \textrm{V}$_x$
is stable for $0.62 \lesssim \eta$.  For a snapshot, see panel (A4) of
Fig. \ref{fig:selected_snapshots} and Fig. \ref{fig:snapshots-sup}-(c)
in Appendix \ref{appendix:MC}.
The formation of grains of phase \textrm{V}$_x$ and the abrupt change
in the order parameters for $\eta \simeq 0.61$ (see
Fig. \ref{fig:Boop-1d5}) are strong indications of a first-order phase
transition that occurs as we pass from structure \textrm{V}$_x$ to the
subsequent structure \textrm{P}$_3$.

In the interval $0.295 \lesssim \eta \lesssim 0.61$, we encounter the
pentagonal phase \textrm{P}$_3$ (identified by the EA investigations)
as a stable phase at finite temperatures. Snapshots of this structure
along with the related Voronoi constructions for particles in layer 2
and for all particles projected onto the same plane are presented in
panels (C2) and (C3) of Fig. \ref{fig:selected_snapshots}. Almost all
Voronoi cells with five sides (cells represented in green in (C2))
host particles that belong to layer 2; this observation is in
agreement with the definitions specified for phase \textrm{P}$_3$. The
Voronoi construction for particles in layer 2 has a vertex
configuration $3^6$ and a symmetry {\it pg}. Rather large values of
the order parameter $\langle \Psi_{10}^{(2)} \rangle \simeq 0.55$ for
layer 2 (see panel (b) of Fig. \ref{fig:Boop-1d5}) stem from the
weights $W_{ij}$ (defined in Eq. (\ref{W-BOOP})) and the pentagonal
shape of the Voronoi cells.

Proceeding to smaller $\eta$-values ($0.19\lesssim \eta \lesssim
0.295$), we observe that the MC sampling favors the features of
commensurability of phase \textrm{I}$_{x=1/4}$; this is corroborated
via the following facts: (i) as shown in the inset of
Fig. \ref{fig:Boop-1d5}-(c), we obtain $x=489/1960$, which differs
only by 0.2\% from 1/4; (ii) in Fig. \ref{fig:Kagome-1d4}-(b), we
present a corresponding snapshot as obtained for $\eta = 0.28$ and $A
= 0.67$. Grains of phase \textrm{I}$_{x=1/4}$ are detected (with
Kagom{\' e} and triangular lattices formed by particles of layers 1 and 2,
respectively). This observation, a fingerprint of thermal effects at
{\it finite} temperature, can be interpreted as a strong competition
between the various structures within the family of \textrm{I}$_x$
phases, found as ground states within EA.

Finally, we have identified phase \textrm{I}$_{x=1/5}$ for $\eta
\lesssim 0.19$ within the domain $x =1/5$; a snapshot of this
structure is shown in panel (A1) of
Fig. \ref{fig:selected_snapshots}. This particular structure
represents an impressive example of the large diversity of phases
encountered within the family of \textrm{I}$_x$ structures (see
Subsection \ref{app:I_11}): its primitive cell hosts 20 particles
and  particles in layer 2 do not form a hexagonal lattice.


                        
\section{Conclusions} 
\label{sec:conclusions}

We have investigated in a comprehensive manner the ordered ground
state configurations of a system of identical point charges, immersed
into the space confined between two parallel plates (labeled 1 and 2)
of opposite charge and separated by a dimensionless distance $\eta$. A
state point is defined by a value of $\eta \in [0, \infty]$ and of the
asymmetry parameter $A$, which is the ratio of the surface charge
densities of the plates ($A = \sigma_2/\sigma_1$); for symmetry
reasons $A$ varies in the range $[0, 1]$.

Investigations were predominantly carried out at vanishing
temperature, using two approaches: (i) an analytical method and (ii) a
highly specialized and reliable optimization technique that is based
on ideas of evolutionary algorithms (EA). The methods are
complementary in the sense that they mutually compensate for their
respective deficiencies: (i) the analytic approach provides
essentially exact numerical results for the energy of some candidate
structures. However, the method is limited -- due to the rapidly
increasing complexity of the formalism -- to ordered structures that
are ideal (i.e., they are free from distortions) and that are not too
complicated in their internal architecture; (ii) in contrast, the
optimization technique is able to treat highly complex and also
possibly distorted structures which are by far out of reach for the
analytic approach: unit cells with up to 40 particles could be
considered; however, the approach is not able to provide a rigorous
proof that a particular structure is indeed the ground state
configuration for a specific state point.  These investigations were
completed by comprehensive Monte Carlo (MC) simulations, carried out
at small, but finite temperatures; here we have tested the thermal
stability of the particle configurations, which were predicted in the
preceding investigations as ground state configurations.

The {\it symmetric} Wigner bilayer problem (i.e., $A = 1$) is
meanwhile well documented in literature: five different ground state
configurations (labeled \textrm{I} to \textrm{V}) have been identified
in previous investigations. Our generalization of the problem to the
{\it asymmetric} case (i.e., $0 \leq A < 1$) leads to a plethora of
ordered bilayers whose features can be summarized as follows:

\begin{itemize} 
\itemsep=0pt
\item[(i)] Even for $A < 1$, the phase \textrm{I} (i.e., the hexagonal
  monolayer at plate 1) remains stable up to an $A$-dependent distance
  $\eta_c(A)$, which monotonously increases with decreasing $A$.

\item[(ii)] This monolayer phase competes for $\eta \simeq \eta_c(A)$
  with the newly emerging phases $\textrm{I}_x$ and $\textrm{V}_x$:
  both are bilayer structures where a fraction $x$ of particles has
  been moved from layer 1 to layer 2. Within high numerical accuracy,
  there is evidence that the transitions \textrm{I} $\to$
  \textrm{I}$_x$ and \textrm{I} $\to$ \textrm{V}$_x$ are both of
  second order; along the respective critical lines, $\eta_c(A)$, the
  critical exponents acquire non-classical values (as, for instance,
  $\beta = 2/3$). The stability regions of the three phases meet at
  the bi-critical point, located at $(\eta_{\rm bi} \simeq 0.47,
  A_{\rm bi} \simeq 0.408)$.

\item[(iii)] The region of stability of phase \textrm{V}$_x$ dominates
  for intermediate and large $\eta$-values (i.e., for $0.7 \lesssim
  \eta$) in the diagram of states. In contrast, for small
  $\eta$-values we find variations of phase \textrm{I}$_x$, which
  transform with monotonously increasing $A$ into the highly stable
  honeycomb structure \textrm{H} (characterized by $x = 1/3$); when
  further increasing $A$, phases {I\kern -0.3ex I}$_x$, {I\kern -0.3ex
    I}, and {I\kern -0.3ex I\kern -0.3ex I} emerge. In the range of
  intermediate $\eta$-values (i.e., up to $\eta \simeq 0.7$) we have
  identified -- in addition to the aforementioned phase \textrm{V}$_x$
  -- a broad variety of complex, sometimes exotic structures, some of
  which show a pronounced five-fold symmetry: the snub square phase
  (which is related to the Archimedean tiling) or pentagonal
  structures. At the moment, it is difficult to provide a decisive
  answer if they can be considered as precursors of quasi-crystalline
  lattices. Work along these lines is in progress.

\item[(iv)] In large parts of the $(\eta, A)$-plane, the diagram of
  states is characterized by rather thin, stripe-shaped regions where
  the occupation parameter $x$ attains rational values (i.e., 1/2,
  1/3, etc.). These $x$-values are imposed by commensurability
  requirements of the emerging sublattices on the two plates, hence
  the corresponding structures are rather simple (i.e. they have
  relatively small unit cells) and are characterized by a high degree
  of stability.

\item[(v)] Eventually, for $0.9 \lesssim A$ we could identify regions
  in the $(\eta, A)$-plane where the ground state configurations of
  the {\it symmetric} case (phases \textrm{I\kern -0.3ex I} to
  \textrm{V}), are found to be stable.  The transitions between phases
  \textrm{I\kern -0.3ex I} $\to$ \textrm{I\kern -0.3ex I\kern -0.3ex
    I} and \textrm{I\kern -0.3ex I\kern -0.3ex I} $\to$ \textrm{I\kern
    -0.3ex V} are continuous with the critical exponents belonging to
  the mean-field universality class (i.e., $\beta = 1/2$, etc.). Thus
  we conclude that (at least) two second-order phase transitions can
  be identified in the diagram of states of our system, pertaining to
  two different universality classes.
\end{itemize}

The results obtained via the analytic and the numerical approaches
agree remarkably well in those $(\eta, A)$-regions where both methods
are applicable (i.e., for not too complex and undistorted particle
arrangements). Discrepancies in the boundaries of stability regions
are observed in the case that the emerging ordered ground state
configurations are highly complex (e.g., lattices with a five-fold
symmetry) and/or where the structures deviate via distortions from
their idealized version.

The plethora of (sometimes highly complex) structures emerging in the
asymmetric Wigner bilayer problem is the result of the competition
between two driving forces:

\begin{itemize}
\item[(i)] on one side there is the system's desire for local charge
  neutrality, separately on each of the two plates (note that global
  electroneutrality always holds).  This principle is dominant either
  for the archetypical structures \textrm{I} to \textrm{V}, which were
  identified for the symmetric case (where $x (A = 1) = x _{\rm neutr}
  = 1/2$), or for large interplate distances $\eta$ (where the
  effective interactions between the layers becomes weak, the layers
  do not ``feel' each other anymore and thus the two sublattices are
  essentially uncorrelated).
\item[(ii)] On the other side there is the attempt of the particles to
  self-assemble in commensurate sublattices on the two plates, a
  strategy which is in particular in place for small and intermediate
  interplate distances $\eta$: here the two layers are strongly
  correlated. The commensurability requirement imposes discrete,
  rational values of $x$.
\end{itemize}

In the transition region, where neither of the two competing forces is
dominant (i.e., at intermediate $\eta$-values), the formerly discrete
$x$-values become essentially continuous. The competition between
preferred {\it discrete} values of $x$ and a {\it continuous} increase
$x(\eta)$ can serve as an explanation for the intricate shape of the
boundary separating structures $\textrm{I}_x$ and $\textrm{P}_3$ (see
Fig. \ref{fig:diagram_of_states_rgb}), which undulates back and forth
with decreasing $A$. The emerging structures represent a compromise
between the two disparate driving forces.  Except for the symmetric
case ($A = 1$) and the limiting case $\eta \to \infty$, the system is
in general not able to realize this compromise between these two
requirements. Thus charge-neutrality is in general violated in either
of the two following senses: (i) in an overwhelming portion of the
$(\eta, A)$-plane the charged particles attached to plate 1
overcompensate for the surface charge, while the other plate is
``underpopulated''; this case is termed {\it undercharging}; (ii) yet,
for $A \gtrsim 0.9$ we could identify regions in the $(\eta, A)$-plane
where this trend is inverted and where we observe {\it overcharging};
this occurs in particular in regions where the pure phases {I\kern
  -0.3ex I} to \textrm{V} are stable.

Finally, another consequence of the local violation of
charge-neutrality should be mentioned: (i) in the symmetric case ($A =
1$) where both plates (along with the attached charged particles) are
neutral, the effective interaction between the layers decreases
exponentially with distance; (ii) however, as soon as $A < 1$ and
local charge-neutrality is violated in either sense, we observe a
long-range attraction between the plates which decays as $1/\eta^2$.

Extensive, complementary Monte Carlo (MC) simulations have been
performed at small, but finite temperatures and in selected regions of
the $(\eta, A)$-plane where promising features could be
expected. Based on several structural observables (such as bond
orientational order parameters or different types of correlation
functions) the simulations confirm that the ground state
configurations (as they were predicted by the analytic and the
numerical approaches) remain in general stable even at small, finite
temperatures. This statement does not hold for state points that are
located close to phase boundaries that separate competing structures:
in these regions, the limiting ensemble size (with typically up to
4000 particles) and the limited simulation time prevent us from making
more decisive conclusions on the stability of the emerging structures.

The model at hand is a simple, yet striking example that demonstrates
that the emergence of highly complex particle arrangements can easily
be triggered via changes in solely two parameters (the interplate
distance $\eta$ and the charge asymmetry parameter $A$).

Acknowledgments.
M.A. and G.K. gratefully acknowledge financial support by the Austrian
Science Foundation (FWF) under projects Nos. P23910-N16 and F41 (SFB
ViCoM) and by E-CAM, an e-infrastructure center of excellence for
software, training and consultancy in simulation and modelling funded
by the EU (Proj. No. 676531). L.\v{S}. acknowledges support from grant
VEGA 2/0003/18. All authors acknowledge financial support from the
projects PHC-Amadeus-2012 and 2015 (project numbers 26996UC and
33618YH), Projekt Amad\'ee (project numbers FR 10/2012 and FR
04/2015), and funding by Investissement d'Avenir LabEx PALM (grant
ANR-10-LABX-0039).





\begin{appendix}

\section{From non-neutralized to neutralized plates}
\label{app:A}


The ground state energy $E$ is a function of both $\eta$ and $A$.
Likewise, the occupation parameter $x$ is a function of both of these
variables.  We can extend the definition of the ground state energy to
$E(\eta,A;x)$ with {\it a priori} independent variables $\eta$, $A$
{\em and} $x$.  In our investigations we retain for a given state
point ($\eta$, $A$) only the configuration with the lowest reduced
energy per particle $E(\eta, A)/N$ as the ground state, which is given
by
\begin{equation}
E (\eta, A) = \min_{x \in [0,1]} E(\eta, A; x) .
\end{equation}

Let us consider the general situation when each plate as a whole
(i.e. including the ions in residence) has a nonzero charge, i.e.,
$\sigma_1 \ne n_1$ and $\sigma_2 \ne n_2$, under the overall
electro-neutrality constraint (\ref{neutrality}).  Is there a relation
between the energy of this configuration of charged plates and the
energy of neutralized plates with the surface charge densities
$\sigma_1=n_1$ and $\sigma_2=n_2$?  In connecting the two situations,
the ionic configurations is fixed, meaning that $n_1$ and $n_2$ are
fixed. Only the surface charges $\sigma_1$ and $\sigma_2$ are allowed
to change, fulfilling electroneutrality.  Microscopically, the total
energy of the system (i.e., plates {\it and} charges) is given by
\begin{equation}
E(\eta,A;x) = E_{pp} + E_{ps} + E_{ss} ,
\end{equation}
where $E_{pp}$ describes the particle-particle interactions, $E_{ps}$
stands for the interaction of particles with the fixed surface charges
on the plates and $E_{ss}$ is the interaction energy of the surface
charge densities on the two plates.  The particle-particle energy
reads as
\begin{equation}
E_{pp} = N_1 \frac{e^2}{2} \left( \sum_{j\ne 1} \frac{1}{R_{1j}^{\alpha\alpha}}
+  \sum_j \frac{1}{R_{1j}^{\alpha\beta}} \right) 
+ N_2 \frac{e^2}{2} \left( \sum_{j\ne 1} \frac{1}{R_{1j}^{\beta\beta}}
+  \sum_j \frac{1}{R_{1j}^{\beta\alpha}} \right) .
\end{equation}
Here, $N_1$ particles on plate 1 form sublattice $\alpha$ and $N_2$
particles on plate 2 form sublattice $\beta$; $R_{1j}^{\nu\nu'}$ means
the distance between a reference particle 1 belonging to lattice $\nu$
and particle $j$ belonging to lattice $\nu'$.

We now introduce the particle-particle interactions renormalized by
the neutralizing background charge density, $\sigma'_1=n_1$ for plate
1 and $\sigma'_2=n_2$ for plate 2:
\begin{equation}
\frac{e^2}{2} \sum_{j\ne 1} \left( \frac{1}{R_{1j}^{\alpha\alpha}} - n_1
\int_0^R {\rm d}^2r \frac{1}{\vert {\bf r}\vert} \right) , \qquad
\frac{e^2}{2} \sum_j \left( \frac{1}{R_{1j}^{\alpha\beta}} - n_2
\int_0^R {\rm d}^2r \frac{1}{\vert {\bf r}+{\bf d}\vert}  \right) ,
\end{equation} 
and so on.  Here, the cutoff $R$ is the radius of the disk within
which the background is considered; at the end of calculations one
should take the limit $R \to \infty$.  Thus $E_{pp}$ can be rewritten in
terms of the neutralized particle-particle energy $E_{pp}^{\rm
  neutr}(\eta;x)$ as follows
\begin{equation} \label{pp}
E_{pp} \sim E_{pp}^{\rm neutr} + S e^2 n_1^2 \pi R  + S e^2 n_2^2 \pi R 
+ S e^2 n_1 n_2 2 \pi (R-d) ,    
\end{equation} 
where we have used that
\begin{equation}
\int_0^R {\rm d}^2r \frac{1}{\vert {\bf r}\vert} = 2\pi R , \qquad
\int_0^R {\rm d}^2r \frac{1}{\vert {\bf r}+{\bf d}\vert} = 2\pi \left[
  \sqrt{R^2+d^2} - d \right] \sim 2\pi (R-d)
\end{equation}
for large $R$.  The interaction of particles with the fixed surface
charge densities $\sigma_1$ and $\sigma_2$ on the plates 1 and 2,
respectively, is given by
\begin{eqnarray}
E_{ps} & = & - e^2 \left( N_1 \sigma_1 + N_2 \sigma_2 \right) 
\int_0^R {\rm d}^2r \frac{1}{\vert {\bf r}\vert} 
- e^2 \left( N_1 \sigma_2 + N_2 \sigma_1 \right) 
\int_0^R {\rm d}^2r \frac{1}{\vert {\bf r}+{\bf d}\vert} \nonumber \\ 
& \sim & - S e^2 (n_1+n_2)(\sigma_1+\sigma_2) 2 \pi R 
+ S e^2 (n_1\sigma_2 + n_2\sigma_1) 2 \pi d . \label{ps}  
\end{eqnarray}
The mutual interaction energy of the surface charge densities
$\sigma_1$ and $\sigma_2$ on the plates 1 and 2, respectively, is
expressible as
\begin{eqnarray}
E_{ss} & = & \frac{e^2}{2} \left( \sigma_1^2 + \sigma_2^2 \right) S 
\int_0^R {\rm d}^2r \frac{1}{\vert {\bf r}\vert} 
+ e^2 \sigma_1 \sigma_2 S 
\int_0^R {\rm d}^2r \frac{1}{\vert {\bf r}+{\bf d}\vert} \nonumber \\ 
& \sim & S \frac{e^2}{2} (n_1+n_2)(\sigma_1+\sigma_2) 2 \pi R - 
S e^2 \sigma_1\sigma_2 2 \pi d . \label{ss}   
\end{eqnarray}
Combining Equs. (\ref{pp}), (\ref{ps}), and (\ref{ss}) and using the
overall electro-neutrality condition (\ref{neutrality}), the
dependence on the background cutoff $R$ disappears and we finally
arrive at
\begin{equation} \label{renormenergy}
E = E_{pp}^{\rm neutr} - S e^2 (\sigma_1-n_1)(\sigma_2-n_2) 2\pi d 
=  E_{pp}^{\rm neutr} + S e^2 (\sigma_1-n_1)^2 2\pi d = 
E_{pp}^{\rm neutr} + S e^2 (\sigma_2-n_2)^2 2\pi d .
\end{equation}
The dimensionless version of this relation reads 
\begin{equation} \label{resequation}
\frac{E(\eta,A;x)}{N e^2 \sqrt{\sigma_1+\sigma_2}} = \frac{E_{pp}^{\rm
    neutr}(\eta;x)}{N e^2 \sqrt{\sigma_1+\sigma_2}} + 2^{3/2} \pi \eta
\left( x - \frac{A}{1+A} \right)^2 .
\end{equation}

The obtained formula is useful also for the simplified case of neutral
plates $\sigma_1=n_1$ and $\sigma_2=n_2$, when the last term in
Eq. (\ref{resequation}) disappears due to the equality $x=x^*$, see
Eq. (\ref{x_neutr}). The relation tells us that the {\em total} energy
of the charged system (i.e., plates plus particles) is equal
exclusively to the sum of particle-particle interactions appropriately
renormalized by the neutralizing background charge densities.  In the
general case of $\sigma_1\ne n_1$ and $\sigma_2\ne n_2$, an additional
positive term emerges. It simply stems from the fact that the electric
field in the slab is non vanishing.

Since $E_{pp}^{\rm neutr}$ is by definition a function of only $\eta$
and $x$, the dependence of the energy $E$ on $\sigma_1$ and $\sigma_2$
is solely encoded in this explicit additional contribution.  Like for
instance, writing explicitly the relation (\ref{resequation}) for the
case $A=0$,
\begin{equation} 
\frac{E(\eta,A=0;x)}{N e^2 \sqrt{\sigma_1+\sigma_2}} =
\frac{E_{pp}^{\rm neutr}(\eta;x)}{N e^2 \sqrt{\sigma_1+\sigma_2}}
+ 2^{3/2} \pi \eta x^2 ,
\end{equation} 
and subtracting this equality from (\ref{resequation}) leads to the
relation
\begin{equation} \label{crucial}
\frac{E(\eta,A;x)}{N e^2 \sqrt{\sigma_1+\sigma_2}} 
= \frac{E(\eta,A=0;x)}{N e^2 \sqrt{\sigma_1+\sigma_2}} 
+ 2^{3/2} \pi \eta \frac{A}{(1+A)^2} \left[ A-2x(1+A) \right] 
\end{equation}
which has been used in Subsection \ref{subsec:methods_EA}.

\section{Computation of the total energy with the Ewald method}
\label{appendix:ewald}

For the asymmetric bilayers, the charge density distribution of the
electric point charges (with nominal value $-e$) and of the
neutralizing background on layers 1 and 2 [$e\sigma_1(\bm{r})$ and
  $e\sigma_2(\bm{r})$] (which for the moment still can be
$r$-dependent) is given by
\begin{equation}
\displaystyle \rho(\bm{r}) =-e \sum_{i\in L_1} \delta(\bm{r}_i-\bm{r}) 
\delta(z_i) - e \sum_{i\in L_2} \delta(\bm{r}_i-\bm{r}) \delta(z_i-d) + e 
\sigma_1(\bm{r})\delta(z) + e \sigma_2(\bm{r})\delta(z-d) ,
\end{equation}
with $\delta(x)$ the Dirac distribution.  The total energy of the
system can be computed as a sum of Coulomb interactions via
\begin{equation} 
\label{E_Coulomb}
\begin{array}{ll}
\displaystyle E = & \displaystyle \frac{e^2}{2} 
\sum_{i \in L} \sum_{j \in L} \sum_{\mbox{\small $\bm{S}_{\bm{n}}$}}
\mbox{}^{\prime}\frac{1}{\mid \bm{r}_{ij}+\bm{S}_{\bm{n}}\mid}
- \frac{e^2}{2} \sum_{i\in L} \int_{L}
{\rm d} \bm{r} \sum_{\mbox{\small $\bm{S}_{\bm{n}}$}}
\frac{\sigma(\bm{r})}{\mid \bm{r}_i-\bm{r}+
  \bm{S}_{\bm{n}}\mid} \\ & \\ & \displaystyle
+ \frac{e^2}{2} \int_{L} {\rm d} \bm{r}' \int_{L} {\rm d} \bm{r}
\sum_{\mbox{\small $\bm{S}_{\bm{n}}$}
}\frac{\sigma(\bm{r})\sigma(\bm{r}')}{\mid \bm{r}-\bm{r}'+
  \bm{S}_{\bm{n}}\mid} .
\end{array}
\end{equation}
Here, $\sigma(\bm{r}) = \sigma_1(\bm{r}) + \sigma_2(\bm{r})$ and $L =
L_1 \bigcup L_2$; the prime in the first term of the rhs excludes
contributions where $i = j$. The ${\bf r}_i$ and ${\bf r}_j$ are the
particle positions with ${\bf r}_{ij} = {\bf r}_i - {\bf r}_j$; $L_1$
and $L_2$ denote the two layers. Further, $\bm{S}_{0}$ denotes the
simulation box (with the primitive vectors $\bm{a}$ and $\bm{b}$) and
the periodic images of $\bm{S}_{0}$ are defined by $\bm{S}_{\bm{n}} =
n_a \bm{a} +n_b\bm{b}$ with $\bm{n} = (n_a,n_b) \in \mathbb{Z}^2$; the
prime in the above summation indicates that contributions with $i=j$
are excluded from the summations in $\bm{S}_{0}$. In the EA approach,
the actual size of the unit cell, which contains only a few particles,
is rather small; thus, for the sake of efficiency, several images
$\bm{S}_{\bm{n}}$ of the simulation box are included in the real-space
contribution to the Ewald sum.

We now split the total energy into intralayer (index $a$) and
interlayer (index $e$) contributions:
\begin{equation} 
\label{E_Ewald}
\displaystyle E = E_1^{(a)}+E_2^{(a)}+E_{12}^{(e)} .
\end{equation}
With the Ewald method we obtain \cite{Mazars:11} for the {\it
  intralayer} energy for layer $\nu = (1, 2)$
\begin{equation}
\label{U_2D_a}
E_\nu^{(a)} = 
\frac{e^2}{2} \sum_{i,j\in L_\nu} \sum_{\mbox{\small $\bm{S}_{\bm{n}}$} }
\mbox{}^{\prime}\frac{\mbox{erfc}(\alpha\mid \bm{s}_{ij}+\bm{S}_{\bm{n}}\mid)}
{\mid \bm{s}_{ij}+\bm{S}_{\bm{n}}\mid} +
\frac{\pi e^2}{S}\sum_{\bf{G}\neq0} \frac{\mbox{erfc}(G/2\alpha)}{G}
\left|\sum_{i\in L_\nu}\exp\left(j\bf{G} \bf{s}_i\right)\right|^2 -
\frac{\sqrt{\pi}N_\nu^2 e^2}{\alpha S}-\frac{\alpha N_\nu e^2}{\sqrt{\pi}} ;
\end{equation}
here $N_\nu$ is the number of point charges in layer $\nu$, $S$ is the
area of $\bm{S}_{0}$, $\alpha$ stands for the Ewald damping parameter
and the ${\bf G}$ (with $G = | {\bf G}|$) are the wave vectors in
reciprocal space \cite{Mazars:11}. The vectors ${\bf s}$ (with or
without layer index) are the projections of the vectors ${\bf r}$
(again, with or without layer index) projected onto the respective
plane.  Finally, the {\it interlayer} energy is given by
\begin{equation}
\label{U_2D_e}
\begin{array}{ll}
E_{12}^{(e)} & = e^2\sum_{i\in L_1}\sum_{i\in
  L_2}\sum_{\mbox{\small $\bm{S}_{\bm{n}}$}
}\mbox{}^{\prime}\frac{\mbox{ erfc}\left(\alpha\sqrt{\mid
    \bm{s}_{ij}+\bm{S}_{\bm{n}}\mid^2+d^2}\right)}{\sqrt{\mid
    \bm{s}_{ij}+\bm{S}_{\bm{n}}\mid^2+d^2}}+\frac{\pi
  e^2}{S}\sum_{\bf{G}\neq0} F(G,\alpha ; d)\mbox{
}\mathcal{R}\left[\left(\sum_{i\in
    L_1}e^{i\bf{G}\bf{s}_i}\right)\left(\sum_{i\in
    L_2}e^{-i\bf{G}\bf{s}_i}\right)\right]\\ &\\ &\displaystyle
-\frac{\pi N_1 N_2 e^2}{S}\left[\frac{e^{-\alpha^2
      d^2}}{\alpha\sqrt{\pi}}+d\mbox{ erf}(\alpha d)\right]-\pi
e^2\sigma_2 d\left[\sigma_1 S-2N_1\right]-\pi e^2\sigma_1
d\left[\sigma_2 S-2N_2\right] ,
\end{array}
\end{equation}
introducing $d$ the distance between the layers, ${\cal R}(z)$
the real part of $z$, and
\begin{equation}
\label{def_F}
F(G,\alpha ; d) = 
\frac{1}{G}
\left[\exp(Gd) \mbox{ erfc} \left( \frac{G}{2\alpha}+\alpha d \right) +
\exp(-Gd)\mbox{ erfc} \left( \frac{G}{2\alpha}-\alpha d \right) \right] .
\end{equation}

\section{Series representations of lattice sums} 
\label{app:B}

The rhs of Eq. (\ref{border}) can be rewritten as $\left[ K(\Delta =
  \sqrt{3},\eta_c) + 4 \pi \eta_c \right]$, where we define
\begin{equation} \label{K}
K(\Delta,\eta) = \frac{1}{\sqrt{\pi}} \int_0^{\infty} \frac{{\rm d}t}{\sqrt{t}} 
\left( 1 - {\rm e}^{-\eta^2 t}\right) \left\{
\left[ \theta_3({\rm e}^{-\Delta t}) \theta_3({\rm e}^{-t/\Delta}) - 1 
- \frac{\pi}{t} \right] + 
\left[ \theta_2({\rm e}^{-\Delta t}) \theta_2({\rm e}^{-t/\Delta}) 
- \frac{\pi}{t} \right] \right\} .
\end{equation}
In terms of the functions
\begin{eqnarray}
I_2(\Delta;x,y) & \equiv & \int_0^{\pi} \frac{dt}{\sqrt{t}} 
{\rm e}^{-x t/\pi^2} {\rm e}^{-y \pi^2/t} \left[ 
\theta_2({\rm e}^{-\Delta t}) \theta_2({\rm e}^{-t/\Delta}) - \frac{\pi}{t} \right] 
\nonumber \\ & = & 2 \sum_{j=1}^{\infty} (-1)^j 
\left[ z_{3/2}(x,y+j^2/\Delta) + z_{3/2}(x,y+j^2\Delta) \right] \nonumber \\
& & + 4 \sum_{j,k=1}^{\infty} (-1)^j (-1)^k z_{3/2}(x,y+j^2/\Delta+k^2\Delta) , \\
I_3(\Delta;x,y) & \equiv & \int_0^{\pi} \frac{dt}{\sqrt{t}} 
{\rm e}^{-x t/\pi^2} {\rm e}^{-y \pi^2/t} \left[ 
\theta_3({\rm e}^{-\Delta t}) \theta_3({\rm e}^{-t/\Delta}) -1-\frac{\pi}{t} \right] 
\nonumber \\ & = & 2 \sum_{j=1}^{\infty}
\left[ z_{3/2}(x,y+j^2/\Delta) + z_{3/2}(x,y+j^2\Delta) \right] \nonumber \\ 
& & + 4 \sum_{j,k=1}^{\infty} z_{3/2}(x,y+j^2/\Delta+k^2\Delta) - \pi z_{1/2}(x,y) , 
\\ I_4(\Delta;x,y) & \equiv & \int_0^{\pi} \frac{dt}{\sqrt{t}} 
{\rm e}^{-x t/\pi^2} {\rm e}^{-y \pi^2/t} \left[ 
\theta_4({\rm e}^{-\Delta t}) \theta_4(e^{-t/\Delta}) -1 \right] 
\nonumber \\ & = & 4 \sum_{j,k=1}^{\infty} 
z_{3/2}(x,y+(j-1/2)^2/\Delta+(k-1/2)^2\Delta) - \pi z_{1/2}(x,y) ,
\end{eqnarray}
$K(\Delta,\eta)$ can be expressed as
\begin{eqnarray}
K(\Delta,\eta) & = & \frac{1}{\sqrt{\pi}} \big[ 2 I_3(\Delta;0,0) 
- I_3(\Delta;(\pi\eta)^2,0) - I_3(\Delta;0,\eta^2) \nonumber \\
& & + I_2(\Delta;0,0) - I_2(\Delta;(\pi\eta)^2,0) 
+ I_4(\Delta;0,0) - I_4(\Delta;0,\eta^2) \big] .
\end{eqnarray}

From the expression for the energy of phase $\textrm{V}_x$, Eq.
(\ref{phaseVx}), the difference in the energies of phases
$\textrm{V}_x$ and \textrm{I} can be expressed as
\begin{eqnarray}
\frac{E_{\rm V_x}(\eta,A;x)-E_{\rm I}(\eta,A)}{N e^2 \sqrt{\sigma_1+\sigma_2}} 
& = & 2^{3/2}\pi\eta x^2 - 2^{5/2} \pi \frac{A}{1+A} \eta x
+ c \left[ (1-x)^{3/2} + x^{3/2} - 1 \right] \nonumber \\ & &
- \frac{x\sqrt{1-x}}{2^{3/2}\sqrt{\pi}} \Big\{
I_3[(\pi\eta)^2(1-x),0] + I_3[0,\eta^2(1-x)] \nonumber \\ & &
+ I_2[(\pi\eta)^2(1-x),0] + I_4[0,\eta^2(1-x)] 
\Big\} \nonumber \\ & &
+ \frac{\sqrt{3} x \sqrt{1-x}}{2^{3/2}\sqrt{\pi}} \Big\{
I_3[3(\pi\eta)^2(1-x),0] + I_3[0,3\eta^2(1-x)] \nonumber \\ & &
+ I_2[3(\pi\eta)^2(1-x),0] + I_4[0,3\eta^2(1-x)] \Big\} , \label{energydif}  
\end{eqnarray} 
where $I_{\nu}(x,y)\equiv I_{\nu}(\sqrt{3};x,y)$ and $\nu=2,3,4$.

\section{More on critical features}
\label{app:crit}

We can derive two further critical indices by adding to the energy
difference (\ref{xinvolved}) the term $-h x$ where $h\to 0^+$ is a
small positive external field (or chemical potential), that couples
linearly to the order parameter.  For $\eta\ge \eta_c$, the extremum
condition for the energy leads to
\begin{equation}
x(\eta,h) \simeq \left( \frac{\lambda}{5\sqrt{2}\pi\eta_c^2} \right)^{2/3} 
\left[ h + g(\eta-\eta_c) \right]^{2/3} . 
\end{equation} 
At the critical point $\eta=\eta_c$, we find that
\begin{equation}
x(\eta_c,h) \propto h^{1/\delta} , \qquad {\rm i.e.} ~~~ \delta = \frac{3}{2} .
\end{equation}
For the susceptibility, we have
\begin{equation}
\frac{\partial x(\eta,h)}{\partial h}\Big\vert_{h=0} 
\propto \frac{1}{(\eta-\eta_c)^{\gamma}} , \qquad {\rm i.e.} ~~~ 
\gamma = \frac{1}{3} . 
\end{equation}
In the the region $\eta < \eta_c$ and including a small $h$-field
within the range $0 < h < g (\eta_c-\eta)$ we obtain the trivial
solution $x=0$.  The derivative of $x$ with respect to $h$ vanishes,
so the critical index $\gamma'$ has no meaning.

Our critical indices, i.e.,
\begin{equation} 
\alpha = \frac{1}{3} , \qquad \beta = \frac{2}{3} , 
\qquad \gamma = \frac{1}{3} , \qquad \delta = \frac{3}{2}
\end{equation}
differ from the standard mean-field (``MF'') critical indices:
\begin{equation} \label{MFindices}
\alpha_{\rm MF} = 0 , \qquad \beta_{\rm MF} = \frac{1}{2} , \qquad 
\gamma_{\rm MF} = 1 , \qquad \delta_{\rm MF} = 3 ; \qquad
\alpha'_{\rm MF} = \alpha_{\rm MF} , \qquad \gamma'_{\rm MF} = \gamma_{\rm MF} .
\end{equation}
In this context it should be mentioned that a simple Ginzburg-Landau
expansion is not able to yield an exponent $\beta=2/3$ (as we have
found); instead it necessarily leads to $\beta=1/n$ where $n$ is some
positive integer \cite{Chaikin:2013}.


In the region $\eta \ge \eta_c$ and under the influence of an external
field $h$, the difference in the energies of phases \textrm{I}$_x$ and
\textrm{I} reads for small $\eta - \eta_c$ and small $h$
(cf. Eq. (\ref{xinvolved}))
\begin{equation}
\frac{E_{\textrm{I}_x}(\eta;x(\eta))-E_{\textrm{I}}(\eta)}{
e^2 N \sqrt{\sigma_1+\sigma_2}}
\simeq g (\eta_c - \eta) x(\eta, h) + 
\frac{2^{3/2} \pi}{\lambda} \eta_c^2 x^{5/2} (\eta, h) - h x(\eta, h) + \dots .
\end{equation} 
%
This function is a homogeneous function of the parameters $(\eta -
\eta_c)$ and $h$, as it should be for any expression for the (free)
energy close to a critical point \cite{Ma}.  In particular, rescaling
$(\eta - \eta_c) \to \xi(\eta-\eta_c)$ and $h\to \xi h$ with some
parameter $\xi$, the energy scales like $E_{\textrm{I}_x}\to
\xi^{5/2}E_{\textrm{I}_x}$.  This feature guarantees that our critical
indices $\alpha=1/3$, $\beta=2/3$, $\gamma=1/3$ and $\delta=3/2$
fulfill two scaling relations \cite{Ma}
\begin{equation} 
2-\alpha = 2\beta + \gamma = \beta (\delta+1) .
\end{equation}
The critical indices $\eta$ and $\nu$, which typically describe the
large-distance behavior of the pair correlation function close to the
critical point are not available in our model due to the absence of
spatial fluctuations.

\section{Large distance analysis} 
\label{app:C}

We study the large distance asymptotic behaviour (i.e.,
$\eta\to\infty$) of the integral of the rhs of Eq. (\ref{integral}).
Using the substitution $t=t'/\eta^2$, this integral can be rewritten
as
\begin{eqnarray}
J(x,\eta) & = & x\sqrt{1-x} \frac{1}{2^{3/2}\sqrt{\pi}\eta}
\int_0^{\infty} \frac{{\rm d}t}{\sqrt{t}} 
\left[ - {\rm e}^{-t(1-x)} + \sqrt{3} {\rm e}^{-3 t(1-x)} \right]
\nonumber \\ & & \times \left\{ 
\left[ \theta_3({\rm e}^{-\sqrt{3}t/\eta^2}) \theta_3({\rm e}^{-t/(\sqrt{3}\eta^2)}) 
- 1 - \frac{\pi\eta^2}{t} \right] \right. \nonumber \\ & & \left.
+ \left[ \theta_2({\rm e}^{-\sqrt{3}t/\eta^2}) 
\theta_2({\rm e}^{-t/(\sqrt{3}\eta^2)}) -\frac{\pi\eta^2}{t} \right] \right\} .
\label{largedist}
\end{eqnarray}
Next we use the Poisson summation formula (\ref{PSF}) to derive the
small-$t$ expansions of the Jacobi theta functions
\begin{eqnarray}
\theta_3({\rm e}^{-t}) & \simeq & \sqrt{\frac{\pi}{t}}
\left[ 1 + 2 {\rm e}^{-\pi^2/t} + 2 {\rm e}^{-4\pi^2/t} + \cdots \right] ,
\nonumber \\
\theta_2({\rm e}^{-t}) & \simeq & \sqrt{\frac{\pi}{t}}
\left[ 1 - 2 {\rm e}^{-\pi^2/t} + 2 {\rm e}^{-4\pi^2/t} + \cdots \right] .
\end{eqnarray}
Thus
\begin{eqnarray}
\left[ \theta_3({\rm e}^{-\sqrt{3}t/\eta^2}) \theta_3({\rm
    e}^{-t/(\sqrt{3}\eta^2)}) - 1 - \frac{\pi\eta^2}{t} \right] +
\left[ \theta_2({\rm e}^{-\sqrt{3}t/\eta^2}) \theta_2({\rm
    e}^{-t/(\sqrt{3}\eta^2)}) -\frac{\pi\eta^2}{t} \right] \nonumber
\\ \mathop{\simeq}_{\eta\to\infty} -1 + 12 \frac{\pi\eta^2}{t}
\exp\left( - \frac{4 (\pi\eta)^2}{\sqrt{3}t} \right) + \cdots.
\end{eqnarray}
Inserting this expansion into Eq. (\ref{largedist}) we obtain
\begin{equation} \label{eqqq}
J(x,\eta) \mathop{\sim}_{\eta\to\infty} 3 \sqrt{2\pi} \eta x
\sqrt{1-x} \left[ - J_1(x,\eta) + J_2(x,\eta) \right] ,
\end{equation}
where
\begin{eqnarray}
J_1(x,\eta) & = & \int_0^{\infty} \frac{{\rm d}t}{t^{3/2}} \exp\left(
- \frac{4 (\pi\eta)^2}{\sqrt{3}t} - t(1-x) \right) , \nonumber
\\ J_2(x,\eta) & = & \sqrt{3} \int_0^{\infty} \frac{{\rm d}t}{t^{3/2}}
\exp\left( - \frac{4 (\pi\eta)^2}{\sqrt{3}t} - 3 t (1-x) \right) .
\end{eqnarray}

Let us first evaluate $J_1(x,\eta)$ by using the saddle-point method.
The ``action'' function
\begin{equation}
S(t) = - \frac{3}{2} \ln t - \frac{4 (\pi\eta)^2}{\sqrt{3}t} - t (1-x) ,
\end{equation}
has its maximum at $t^*$, given by the extremum condition $\partial
S(t)/\partial t \vert_{t=t^*}=0$.  For $\eta\to\infty$, we find
\begin{equation}
t^* = \frac{2\pi}{3^{1/4}} \frac{1}{\sqrt{1-x}} \eta  + O(1) .
\end{equation}
The expansion of $S(t)$ around $t^*$ then takes the form
\begin{equation}
S(t) = S(t^*) - \frac{3^{1/4} (1-x)^{3/2}}{2 \pi\eta} (t-t^*)^2 + \cdots .
\end{equation}
Since $t^*\to\infty$ as $\eta\to\infty$, we find the asymptotic expansion 
of the form 
\begin{eqnarray} 
J_1(x,\eta) = 
\simeq  {\rm e}^{S(t^*)} \int_{-t^*}^{\infty} {\rm d}t\, 
\exp \left( - \frac{3^{1/4} (1-x)^{3/2}}{2 \pi\eta} t^2 \right) \nonumber \\
\mathop{\simeq}_{\eta\to\infty} \frac{3^{1/4}}{2\sqrt{\pi}\eta} 
\exp\left( - \frac{4\pi\sqrt{1-x}}{3^{1/4}} \eta\right) .  \label{i1}
\end{eqnarray}
The same procedure can be applied to $J_2(x, \eta)$, with the asymptotic
result
\begin{equation}
J_2(x,\eta) \mathop{\simeq}_{\eta\to\infty}  \frac{3^{3/4}}{2\sqrt{\pi}\eta} 
\exp\left( - 3^{1/4} 4 \pi\sqrt{1-x}\, \eta \right) ,
\end{equation}
i.e. $J_2(x, \eta)$ is sub-leading with respect to $J_1(x, \eta)$ in
the large-$\eta$ limit.  Substituting relation (\ref{i1}) into
Eq. (\ref{eqqq}), we end up with the asymptotic representation
(\ref{exponential}).

\section{Details of the Monte Carlo simulations}
\label{appendix:MC}

In this Appendix, we provide some additional information related to
the MC simulations performed (see also Section
\ref{sec:results_finite}).

\subsection{Computational pathways and initial configurations for the MC simulations}

Results obtained via the EA approach provide evidence of well-defined,
stripe-shaped regions in the ($\eta$, $A$)-plane that are
characterized by a constant value of $x=N_2/N$ (see
Fig. \ref{fig:runs-MC}). In an effort to focus on these regions and to
improve thereby the sampling efficiency of parameter space via MC
simulations, we have defined pathways in the $(A, \eta)$-plane which
are characterized -- according to the EA predictions -- by
(essentially) constant $x$-values. Each of these pathways passes (i)
through $(\eta, A) = (0, 1)$ and (ii) through a selected state point
which serves as an initial configuration for all subsequent
simulations along this particular pathway; this point is specified by
$(\eta_{\rm ic}, A_{\rm ic})$. These pathways, i.e., $A_{x = {\rm
    const.}}  (\eta)$, have been parameterized in a heuristic manner
by polynomials of order four in $\eta$ with suitably defined
coefficients $a_1$ to $a_4$:
\begin{equation}
\label{path-MC}
\displaystyle A_{x = {\rm const.}}(\eta) = 1+a_1 \eta+a_2 \eta^2+a_3
  \eta^3+a_4 \eta^4 .
\end{equation}
All relevant data that specify the four $x$-domains, the respective
pathways, and the initial configurations for the MC simulations are
summarized in Table \ref{tab:path-MC}.

\begin{table}
\begin{tabular}{|c|cccc|c|cc|c|cc|}
\hline
\multicolumn{5}{|c|}{$x$-domains, pathways}  & \multicolumn{6}{c|}{specific parameters for the initial configurations} \\
      $x$   & $a_1$ & $a_2$ & $a_3$ & $a_4$ & phase & $A_{\mbox{\small ic}}$  & $\eta_{\mbox{\small ic}}$ & $N_0$ & $n\times m$ & $N$ \\
\hline
3/7  & $\mbox{   }-0.2132\mbox{   }$ & $\mbox{   }-0.8947\mbox{   }$ & $\mbox{   }1.773\mbox{   }$ & $\mbox{   }-0.8562\mbox{   }$ & \textrm{P}$_1$ & $\mbox{   }0.86\mbox{   }$ & $\mbox{   }0.41\mbox{   }$ & $\mbox{   }28\mbox{   }$ & $\mbox{   }14\times 10\mbox{   }$& $\mbox{   }3920\mbox{   }$ \\                   
1/3  & $\mbox{   }-1.0075\mbox{   }$ & $\mbox{   }0.9058\mbox{   }$ & $\mbox{   }-0.2997\mbox{   }$ & $\mbox{   }0.\mbox{   }$ & \textrm{S}$_1$ & $\mbox{   }0.65\mbox{   }$ & $\mbox{   }0.629\mbox{   }$ & $\mbox{   }6\mbox{   }$ & $\mbox{   }26\times 26\mbox{   }$& $\mbox{   }4056\mbox{   }$ \\    
1/4  & $\mbox{   }-1.384\mbox{   }$ & $\mbox{   }1.240\mbox{   }$ & $\mbox{   }-0.4061\mbox{   }$ & $\mbox{   }0.\mbox{   }$ & \textrm{V}$_x$ & $\mbox{   }0.45\mbox{   }$ & $\mbox{   }1.0\mbox{   }$ & $\mbox{   }4\mbox{   }$ & $\mbox{   }31\times 31\mbox{   }$& $\mbox{   }3844\mbox{   }$ \\    
1/5  & $\mbox{   }-1.512\mbox{   }$ & $\mbox{   }1.306\mbox{   }$ & $\mbox{   }-0.4151\mbox{   }$ & $\mbox{   }0.\mbox{   }$ & \textrm{P}$_3$ & $\mbox{   }0.6\mbox{   }$ & $\mbox{   }0.368\mbox{   }$ & $\mbox{   }20\mbox{   }$ & $\mbox{   }14\times 14\mbox{   }$& $\mbox{   }3920\mbox{   }$ \\    
\hline
\end{tabular}
\caption{{\bf Left half:} definition of the domains of constant
  $x$-values, along which extended MC simulations have been carried
  out: $x$-value specifying the domain and parameters $a_1$ to $a_4$
  which define via the polynomial $A_{x = {\rm const.}}(\eta)$ -- see
  Eq. (\ref{path-MC}) -- the respective domain in the $(\eta,
  A)$-plane. {\bf Right half:} parameters specifying the structures
  which served as initial configurations for all subsequent MC runs
  within the respective domains: specification of the phase (as
  predicted by EA calculations), $A$- and $\eta$-parameters defining
  the initial configurations (index 'ic') for the MC-runs within the
  respective domains, $N_0$, the number of particles within the
  primitive cell (as predicted by EA calculations), $n \times m$ the
  number of replications of this primitive cell, creating thereby the
  simulation cell, and $N$, the total number of particles in the
  simulation cell (with $N = N_0 \times n \times m)$.}
  \label{tab:path-MC}
\end{table}

\subsection{Scaling behaviour of the correlation functions}

\begin{figure}[htbp]
\begin{center}
\includegraphics[width=7.2cm]{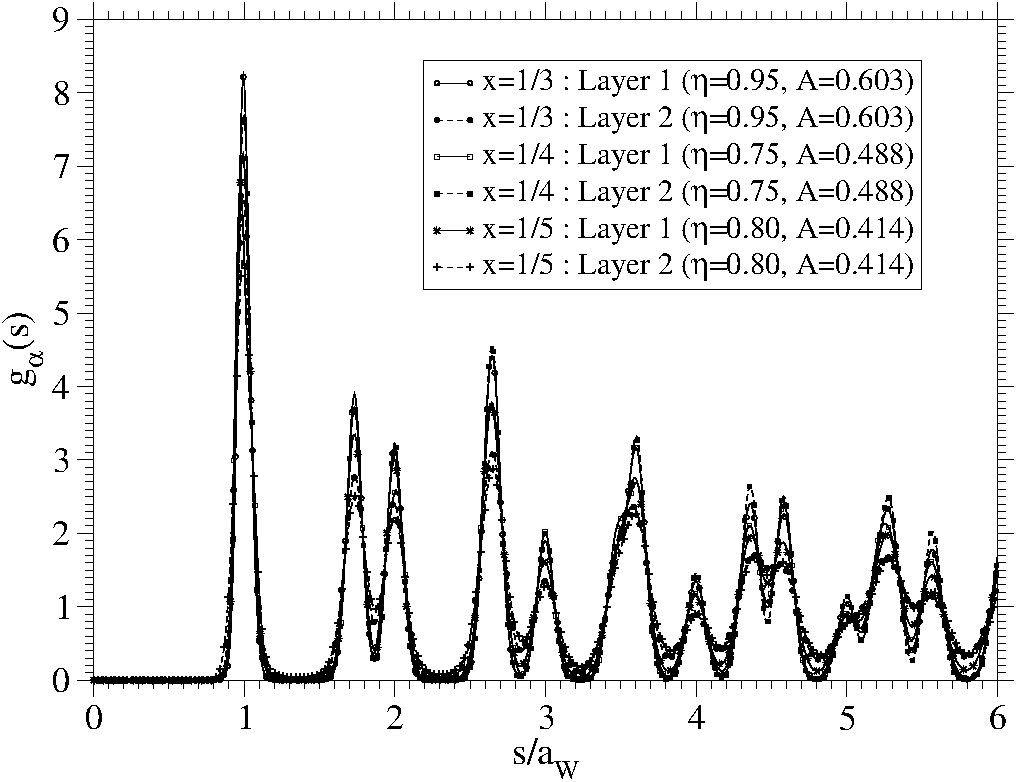}
\caption{Correlation functions $g_\alpha(s)$, $\alpha = 1, 2$, for
  phases \textrm{V}$_x$ as obtained in MC simulations at finite
  temperature, with the distances $s$ scaled by the respective $a_{i;
    {\rm W}}$-values. The correlation functions have been calculated
  for different state points and selected layers (as labeled). }
\label{fig:scaling-gr}
\end{center}
\end{figure}

We analyze here the scaling behaviour of the correlation functions
$g_\alpha(s)$ with $\alpha = 1, 2$ (see the definition
in. Eq. (\ref{g_of_r})), as observed in phase \textrm{V}$_x$.  A
relevant and appropriate length scale to represent the distance
dependence of these correlations functions -- independently of the
particle densities $n_1$ and $n_2$ in each layer -- are the respective
lattice spacings of the hexagonal 2D Wigner crystal, $a_{i; {\rm W}} =
(2/\sqrt{3}n_i)^{1/2}$, $i = 1, 2$; for each layer the respective
surface density is given by $n_1 = (\sigma_1+\sigma_2)(1-x)$ and $n_2
= (\sigma_1+\sigma_2)x$, while the total density for all particles
projected onto the same plane is given by $n = n_1 + n_2$.  As
explained in Section \ref{sec:largeeta}, particles arrange in phase
\textrm{V}$_x$ in both layers as hexagonal 2D Wigner crystals,
therefore the intralayer correlation functions have to fulfill in each
of the layers a scaling law, imposed by the respective surface
densities. More precisely, the rescaled intralayer correlation
functions fulfill in phases \textrm{V}$_x$ the relation
\begin{equation}
\label{scale-gr}
\displaystyle g_{1}\left(\frac{s}{a_{1; {\rm W}}} ; (1-x)^{1/2}\Gamma \right)
\simeq g_{2}\left( \frac{s}{a_{2; {\rm W}}} ; x^{1/2} \Gamma\right)
\end{equation}
where we have included the dependence of the correlation functions on
the coupling constant $\Gamma$ (see Subsection
\ref{subsec:methods_MC}) to emphasize the surface density dependence
in both layers.

Results shown in Fig. \ref{fig:scaling-gr} verify the scaling law
specified in Eq. (\ref{scale-gr}): in this figure the correlation
function $g_\alpha(s)$, with the distance $s$ appropriately scaled,
are shown for several state points of phase \textrm{V}$_x$ and
selected layers (as labeled). For all these correlations functions,
the first three peaks are located at $s = a_{i; {\rm W}}$, $s =
\sqrt{3}\mbox{ }a_{i; {\rm W}}$, and $s = 2a_{i; {\rm W}}$; the
differences in the height and in the width of the peaks are due to the
$\Gamma$-dependence. We could verify this scaling law of the
intralayer correlation functions in all of our MC simulations
performed for the phases \textrm{V}$_x$.

\subsection{Further structural details and additional snapshots}

This subsection contains a few snapshots of the full systems simulated
in Monte Carlo simulations (Figs. \ref{fig:snapshots-sup} and
\ref{fig:Kagome-1d4}) which do not belong into the main text; for the
discussion we refer to Section \ref{sec:results_finite}. 
\begin{figure}[htbp]
\begin{center}
(a)\includegraphics[width=6cm]{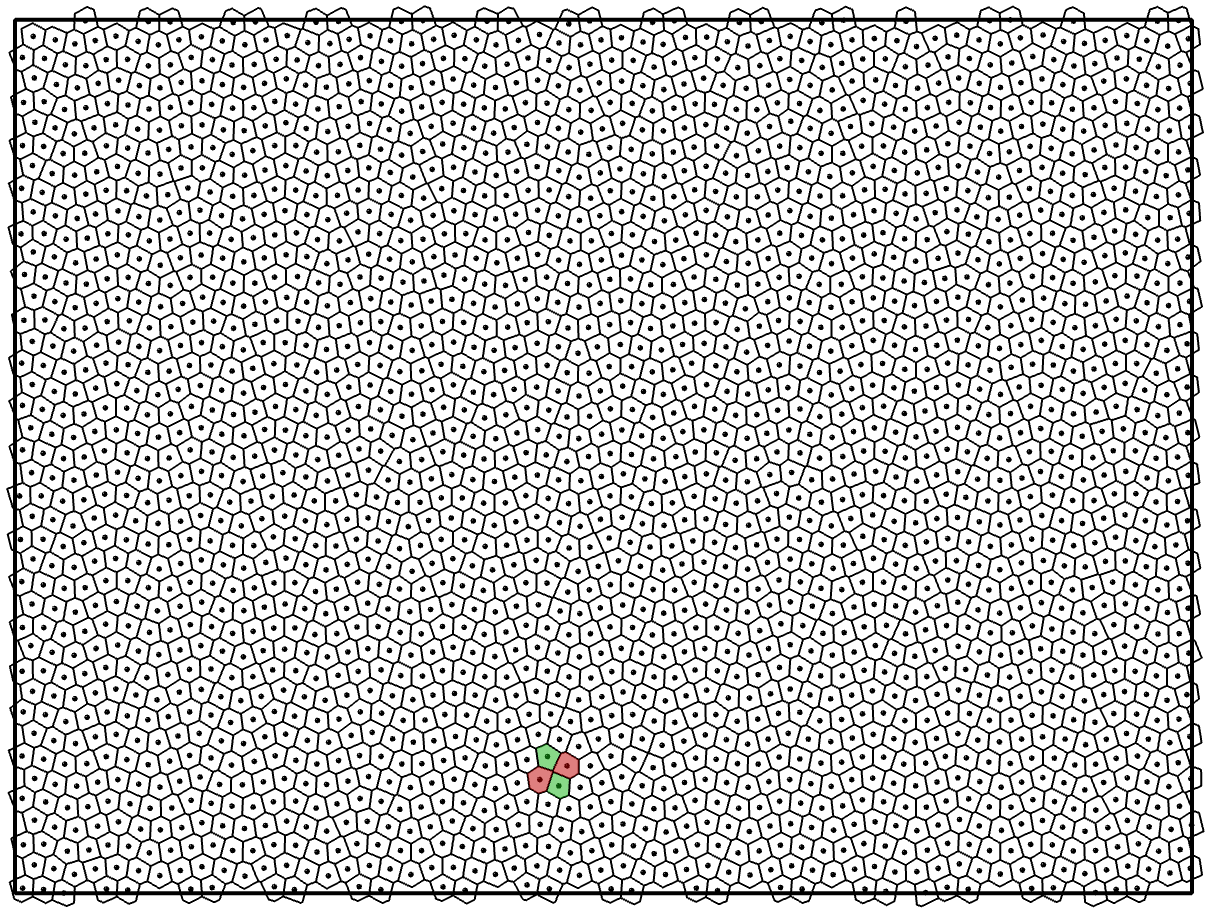}
(b)\includegraphics[width=5cm]{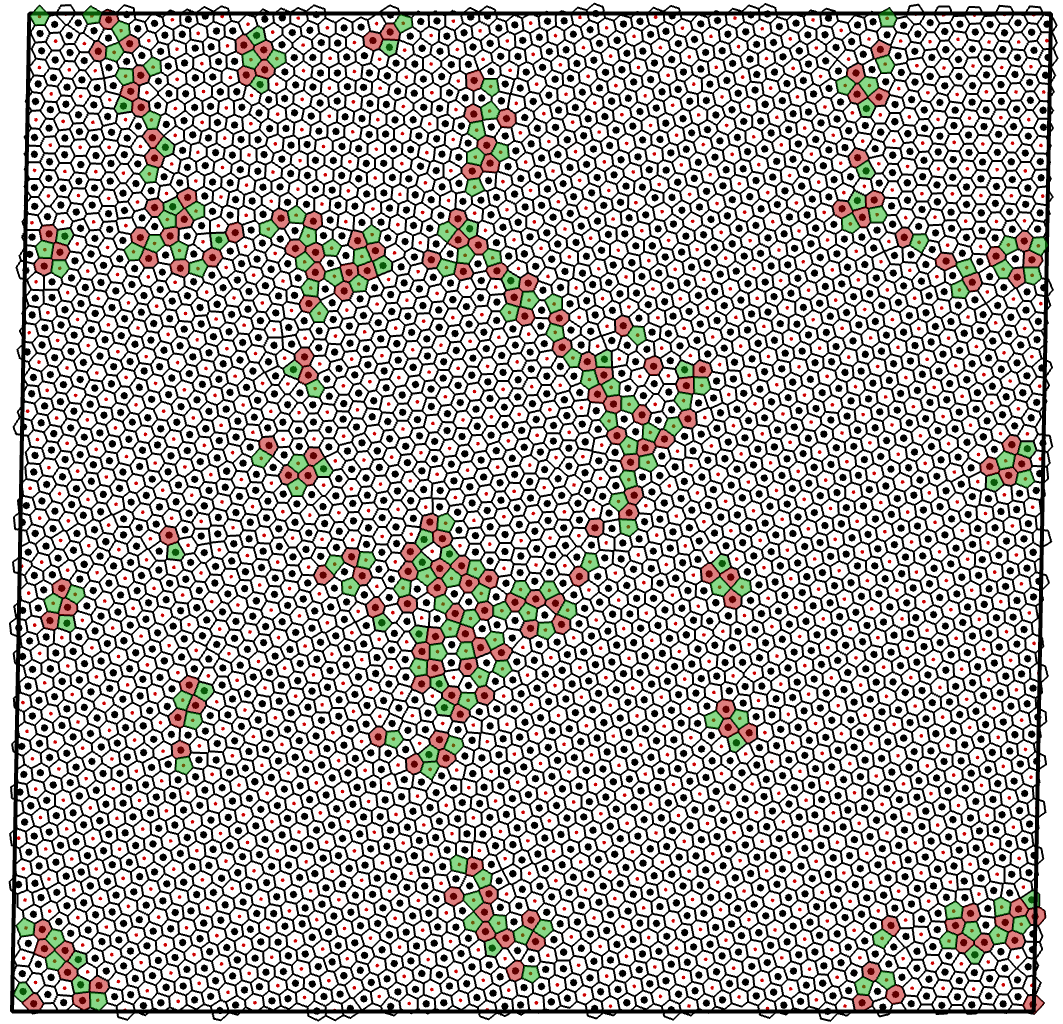}
(c)\includegraphics[width=5cm]{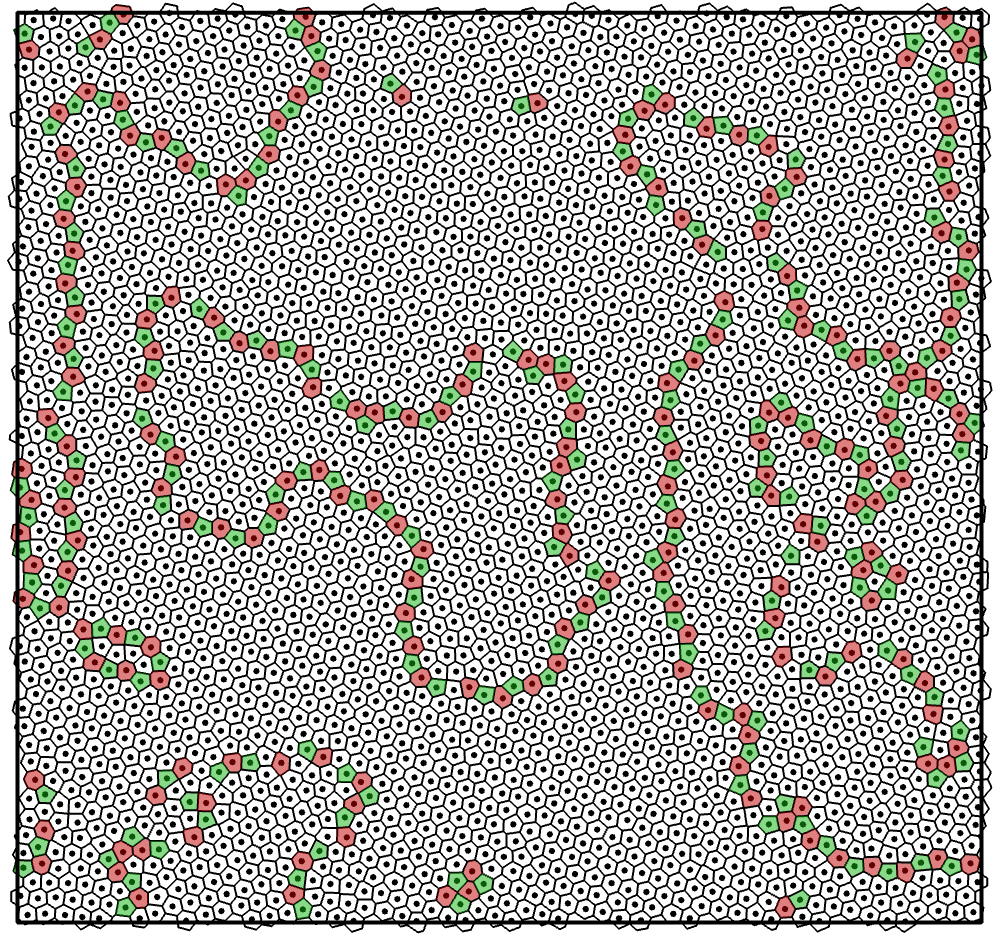}
\caption{(Color online) Selected snapshots of the full systems as
  obtained in MC simulations, along with the related Voronoi
  constructions.  The color code for the Voronoi cells is the
  following (color and number of edges): yellow (four), green (five),
  white (six), red (seven), and blue (eight). (a) Layer 1 of the
  \textrm{DV}$_x$ phase ($x=3/7$, $\eta=0.68$, $A= 0.814$); the
  Voronoi constructions have been performed only for the particles in
  layer 1 (in black). (b) phase \textrm{H} ($x=1/3$, $\eta=0.3$ and
  $A=0.771$); the Voronoi constructions have been performed for all
  particles projected onto the same plane, particles in layer 2 are
  shown in red. (c) Layer 1 of phase \textrm{V}$_x$, close to the
  transition to phase \textrm{P}$_3$ ($x=1/5$, $\eta=0.62$ and
  $A=0.466$).}
\label{fig:snapshots-sup}
\end{center}
\end{figure}

\begin{figure}[htbp]
\begin{center}
(a)\includegraphics[width=5.cm]{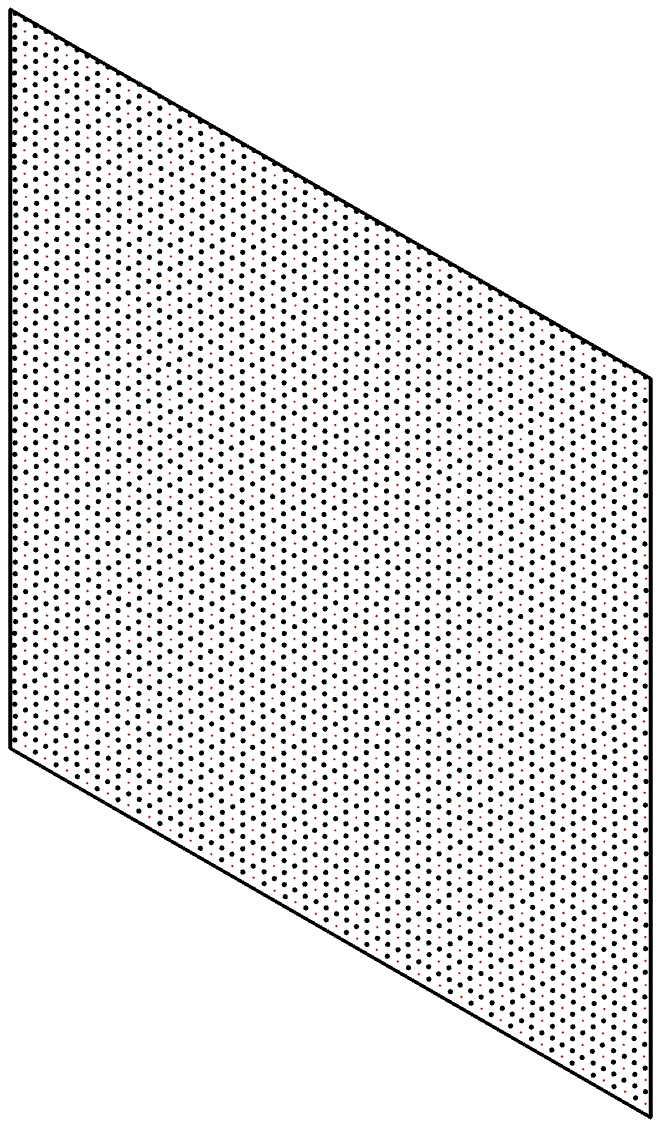}
(b)\includegraphics[width=7.5cm]{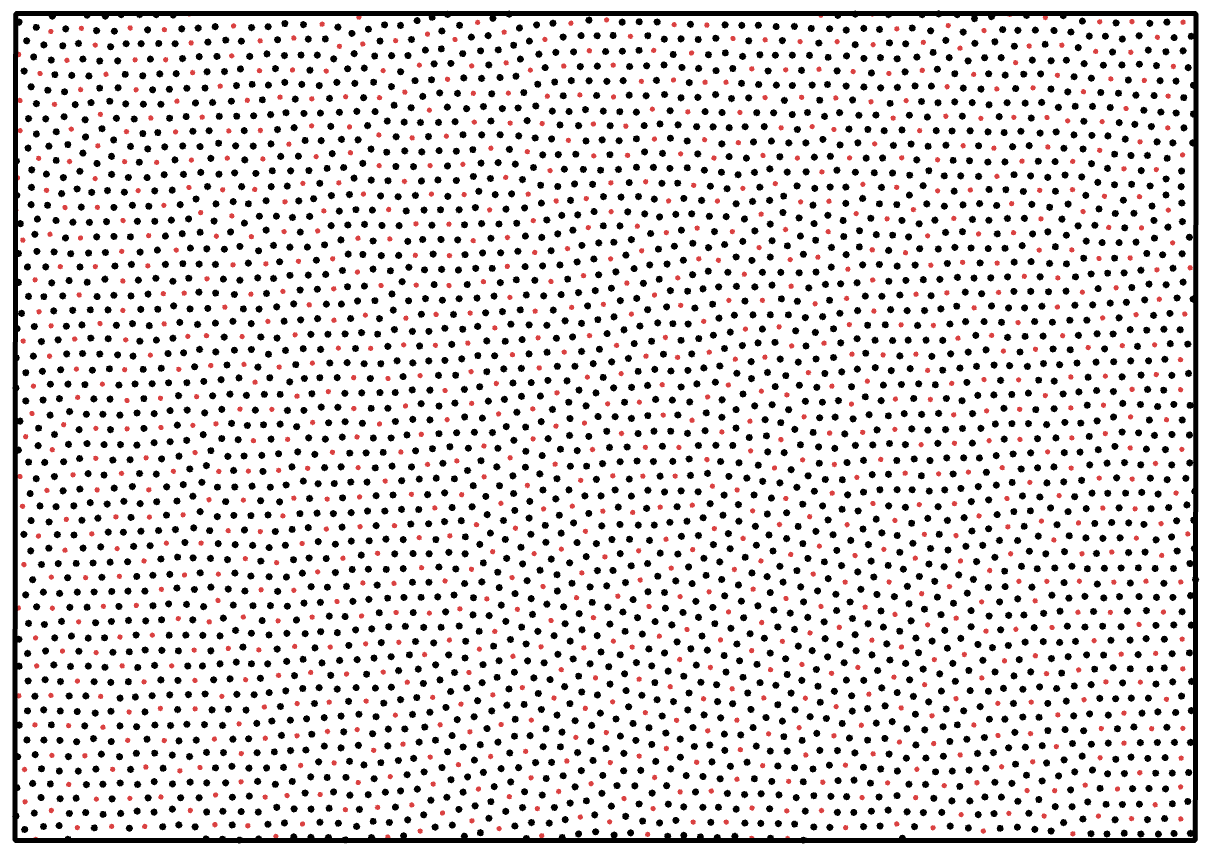}
\caption{(color online) Structures I$_{x=1/4}$ observed in
  MC simulations within the domains $x = 1/4$ and $x = 1/5$. (a)
  Kagom{\' e} lattice of layer 1 in the phase I$_{x=1/4}$ ($\eta=0.10$ and
  $A=0.874$) as identified in the domain $x = 1/4$. (b) Grains of
  phase I$_{x=1/4}$ ($\eta=0.28$, $A=0.67$) as identified in the
  domain $x = 1/5$.}
\label{fig:Kagome-1d4}
\end{center}
\end{figure}

\end{appendix}


\begin{thebibliography}{1}
\bibitem{rque10} The case $\sigma_1\sigma_2 < 0$ is somewhat simple
  and of little interest, because we deal with ions that are all of
  the same charge. In this case, one plate repels them while the other
  is attractive, and structure I ensues. The interesting cases under
  study here are for $\sigma_1$ and $\sigma_2$ of the same sign,
  opposite to that of the point ions. Without loss of generality, we
  consider $A=\sigma_2/\sigma_1<1$, and both $\sigma_1$, $\sigma_2$,
  to be positive.
\bibitem{rque20} With our code and for the choice of parameters for
  the Ewald method used in this study, CPU-times that are required to
  perform $4.0 ~ 10^4$ MC-cycles for a systems with $N \sim 4000$
  particles typically amount to 48 CPU-hours on IBM IDataplex DX360
  processors; this computer time also covers the numerical effort
  required for the Voronoi constructions \cite{Voronoi:Book} which are
  performed after each MC-cycle.
\bibitem{moire} Moir\'e patterns are created by superposing two simple
  lattices. Depending on the commensurability and the relative
  orientation of the two lattices, highly complex patterns can emerge.
\bibitem{polynomials} In order to define these pathways, $A_{x = {\rm
    const.}}(\eta)$, suitable interpolating polynomials have been
  fixed, which are specified in Appendix \ref{appendix:MC}.
\end{thebibliography}
\end{document}